\documentclass[10pt,journal,compsoc]{IEEEtran}
\usepackage{amsthm}
\usepackage{amsmath,amssymb,amsfonts}
\usepackage{algorithm}
\usepackage{algpseudocode}
\usepackage{graphicx}
\usepackage{textcomp}
\usepackage{xcolor}
\usepackage{soul}
\usepackage{makecell}
\usepackage{ctable}
\usepackage{booktabs} 
\usepackage{multirow}
\usepackage{thm-restate}
\usepackage{sistyle}
\SIthousandsep{,}
\usepackage{calligra}
\usepackage{enumitem}
\usepackage{subfigure}
\usepackage{footmisc} 
\usepackage{url}
\usepackage{makecell}
\usepackage{bigstrut}
\usepackage{multicol}
\usepackage{adjustbox}
\usepackage{makecell}
\usepackage{color}
\usepackage[nocompress]{cite}

\ifCLASSINFOpdf
\else
\fi

\begin{document}
\title{A Comparison of Image Denoising Methods}

\author{Zhaoming~Kong, Fangxi Deng, Haomin Zhuang, Jun Yu, \\
Lifang He, ~\IEEEmembership{Member,~IEEE}, and Xiaowei~Yang
\IEEEcompsocitemizethanks{\IEEEcompsocthanksitem Z. Kong, H. Zhuang and X. Yang are with the School of Software Engineering, South China University of Technology, Guangzhou, China (e-mail: kong.zm@mail.scut.edu.cn; semzm@mail.scut.edu.cn; xwyang@scut.edu.cn). Corresponding author: Zhaoming Kong.
\IEEEcompsocthanksitem F. Deng is with Tencent Technology (email: fangxideng@outlook.com).
\IEEEcompsocthanksitem J. Yu and L. He are with the Department of Computer Science and Engineering, Lehigh University, PA, USA (email: juy220@lehigh.edu; lih319@lehigh.edu).
}
}

%
%


\IEEEtitleabstractindextext{%
\begin{abstract}

The advancement of imaging devices and countless images generated everyday pose an increasingly high demand on image denoising, which still remains a challenging task in terms of both effectiveness and efficiency. To improve denoising quality, numerous denoising techniques and approaches have been proposed in the past decades, varying from different regularization terms, algebraic representations and especially to advanced deep neural network (DNN) architectures. Despite their sophistication, many methods may fail to achieve desirable results for simultaneous noise removal and fine detail preservation. In this paper, to investigate the applicability of existing denoising techniques, we compare a variety of denoising methods on both synthetic and real-world datasets for different applications. A new dataset is introduced for benchmarking, and evaluations are performed from four different perspectives including quantitative metrics, visual effects, human ratings and computational cost. Throughout extensive experiments, we witness the astonishing success brought by DNN methods, and also recognize the competitive performance of traditional denoisers. In spite of the tremendous progress in recent years, we discuss shortcomings and possible extensions of existing techniques. Datasets, code and results are available and will be continuously updated at https://github.com/ZhaomingKong/Denoising-Comparison.

\end{abstract}

\begin{IEEEkeywords}
Image denoising, nonlocal self-similarity, block-matching filters, deep neural network, real-world images.
\end{IEEEkeywords}}

\maketitle

\IEEEdisplaynontitleabstractindextext

\IEEEpeerreviewmaketitle

\IEEEraisesectionheading{\section{Introduction}\label{sec:introduction}}

\IEEEPARstart{I}mage denoising enjoys a long history and pioneering works \cite{elad2023image} may date back for decades. The primary goal of denoising is to enhance image quality by estimating underlying clean images from noisy observations. As a simple and special form of inverse problems \cite{jin2017deep}, it has drawn extensive attention from both academia and industry. In real-world applications, image denoising can be integrated into many different tasks such as visual tracking \cite{isard1998condensation}, image segmentation \cite{pal1993review} and classification \cite{qiao2016effective}. \\
\indent These years, the rapid development of modern imaging systems and technologies has largely enriched the information preserved and presented by an image, which can deliver more faithful representation for real scenes. A good example is that the rise and spread of advanced mobile phones facilitates the production of high-quality images and videos. In practice, noise removal has become a necessity for various imaging sensors and techniques such as multispectral/hyperspectral imaging (MSI/HSI) \cite{peng2014decomposable}, magnetic resonance imaging (MRI) \cite{mohan2014survey} and computed topography (CT) \cite{diwakar2018review}. Meanwhile, the increase of image size and dimension also puts forward higher requirements for denoising in terms of both effectiveness and efficiency. Therefore, the interest in the realm of denoising grows consistently with a large quantity of approaches \cite{elad2023image, Collection_denoising_methods}, which may be roughly divided into two categories, namely \textit{traditional denoisers} and \textit{DNN methods}, depending on whether neural network architectures are utilized. \\
\indent Briefly, traditional denoisers normally filter out noise based solely on the input noisy observation by taking advantage of different regularization terms and image priors \cite{katkovnik2010local, zoran2011learning}. A particular and powerful solution that deserves our specific attention is the block-matching 3D (BM3D) framework \cite{dabov2007image}, which integrates the nonlocal self-similarity (NLSS) characteristic of natural images \cite{buades2005review}, sparse representation \cite{elad2006image} and transform domain techniques \cite{yaroslavsky2001transform} into a subtle paradigm. Since the birth of BM3D, there is no shortage of extensions originating from different disciplines. To name a few, Dabov et al. \cite{dabov2009bm3d} improve BM3D by exploiting shape-adaptive image patches and principal component analysis (PCA). Maggioni et al. \cite{maggioni2012nonlocal} and Rajwade et al. \cite{rajwade2012image} introduce 3D cubes of voxels for high-dimensional data. Zhang et al. \cite{gu2014weighted} replace the sparsity constraint with the low-rank assumption. Xu et al. \cite{xu2018trilateral} employ the Maximum A-Posterior (MAP) estimation technique \cite{murphy2012machine} and propose a trilateral weighted sparse coding scheme. \\
\indent Despite the steady improvements brought by classic algorithms, they suffer from several drawbacks \cite{lucas2018using} such as the need for solving complex optimization problems in the test phase, manual setting parameters and failure to exploit auxiliary information. To address these issues, DNN methods have been given an exceptionally large attention and shown promising results in image denoising \cite{tian2019deep}. The arrival of the deep learning era has significantly broadened the scope of denoising and infused new insights into the design of effective denoisers. For example, Zhang et al. \cite{zhang2017beyond} incorporate batch normalization (BN) \cite{ioffe2015batch}, rectified linear unit (ReLU) \cite{nair2010rectified} and residual learning \cite{he2016deep} into the convolutional neural network (CNN) model. Chen et al. \cite{chen2018image} introduce generative adversarial networks (GANs) \cite{radford2015unsupervised} to resolve the problem of unpaired noisy images. Lefkimmiatis et al. \cite{lefkimmiatis2017non} and Davy \cite{davy2018non} et al. combine NLSS and CNN to efficiently remove noise. \\
\indent Accompanying the significant and inspiring progress of denoising algorithms, concerns may arise about their practical applicability, as a large proportion of approaches are verified on a limited number (often less than three) of datasets. Besides, with a considerable amount of existing methods \cite{Collection_denoising_methods, yaroslavsky2001transform, buades2005review, milanfar2012tour, thanh2019review, shao2013heuristic, mohan2014survey, schmidhuber2015deep, tian2019deep, izadi2022image, elad2023image}, there still lacks a study on their performance for different image denoising tasks and applications. In this paper, we intend to narrow the gap by collecting and comparing various denoisers to investigate their effectiveness, efficiency, applicability and generalization ability with both synthetic and real-world experiments. \\
\indent The main contributions of the paper are as follows. \\
\indent (1) We construct a real-world dataset for image and video denoising tasks with a variety of digital cameras and smartphones. The dataset is composed of images and video sequences of both indoor and outdoor scenes under different lighting conditions, which serves as a good complement to current benchmark datasets. \\
\indent (2) We compare a variety of methods and perform extensive experiments in both synthetic and real-world scenarios for different denoising tasks and applications, including images, video sequences, 3D MRI volumes and MSI/HSI data. We adopt both objective and subjective metrics and evaluate the denoising quality of compared methods with quantitative results and human visual ratings. \\
\indent (3) We make several interesting observations based on experimental results. First, representative traditional denoisers such as the BM3D family\cite{dabov2007image, dabov2007color, kostadin2007video, maggioni2012video, maggioni2012nonlocal} still demonstrate very competitive performance in several denoising applications. In addition, we argue that a simple modified singular value decomposition (M-SVD) method is able to produce similar results with tensor-based approaches in image denoising. For DNN methods, advanced network architectures gain significant improvements over traditional denoisers when fine-tuned with given training/validation samples, but the pretrained models and predefined settings may not generalize well to other datasets. Nevertheless, we intend to identify models that exhibit impressive generalizability. For example, FCCF \cite{yue2019high}, DRUNet \cite{zhang2020plug} and PNGAN \cite{cai2021learning} produce state-of-the-art results on a number of real-world image datasets. FastDVDNet \cite{Tassano_2020_CVPR}, FloRNN \cite{li2022unidirectional} and RVRT \cite{liang2022recurrent} show outstanding performance in the video denoising task in terms of both effectiveness and efficiency. Interestingly, many of these effective models such as FCCF, RVRT and FastDVDNet take advantage of Gaussian noise modeling and denoisers.\\
\indent The rest of this paper is organized as follows. Section \ref{section_background} introduces background knowledge. Section \ref{section_related_works} gives a brief review on related denoising techniques and datasets. Section \ref{section_experiments} provides experimental results and discussions. Section \ref{section_conclusion} concludes the paper.
\vspace{-10pt}

\section{Background} \label{section_background}
\subsection{Symbols and Notations}
\indent In this paper, we mainly adopt the mathematical notations and preliminaries of tensors from \cite{kolda2009tensor} for image representation. Vectors and matrices are first- and second- order tensors which are denoted by boldface lowercase letters $\mathbf{a}$ and capital letters $\mathbf{A}$, respectively. A higher order tensor (the tensor of order three or above) is denoted by calligraphic letters, e.g., $\mathcal{A}$. An $N$th-order tensor is denoted as $\mathcal{A} \in \mathbb{R}^{I_1\times I_2\times\cdots\times I_N}$. The $n$-mode product of a tensor $\mathcal{A}$ by a matrix $\mathbf{U}\in \mathrm{R}^{P_n\times I_n}$, denoted by $\mathcal{A}\times _n\mathbf{U} \in \mathbb{R}^{I_1 \cdots I_{n-1} P_n I_{n+1} \cdots I_N}$ is also a tensor. The mode-$n$ matricization or unfolding of $\mathcal{A}$, denoted by $\mathbf{A}_{(n)}$, maps tensor elements $(i_1,i_2,\ldots,i_N)$ to matrix element $(i_n,j)$ where $j=1+\sum_{k=1,k\neq n}^{N}(i_k-1)J_k$, with $J_k = \prod_{m=1,m\neq n}^{k-1}I_m$. The Frobenius norm of a tensor $\mathcal{A} \in \mathbb{R}^{I_1\times I_2\times\cdots\times I_N}$ is defined as $\|\mathcal{A}\|_F = \sqrt{\sum_{i_1=1}...\sum_{i_N=1}\mathcal{A}_{i_1...i_N}^2}$.
\vspace{-2pt}
\subsection{Noise Modeling}
Let us consider a noisy observation $\mathcal{Y}$ and its underlying clean image $\mathcal{X}$, a general assumption of noise distribution is additive white Gaussian noise (AGWN) with variance $\sigma^2$ represented by $\mathcal{N}(0, \sigma^2)$, and the degradation process is then given as
\begin{equation}\label{awgn}
  \mathcal{Y} = \mathcal{X} + \mathcal{N}
\end{equation}
Indeed, noise modeling can be more complex and challenging in that noise in real-world images may be multiplicative and signal dependent \cite{ramanath2005color}. Therefore, there are a plenty of non i.i.d Gaussian models tailored for the need of different applications, such as the mixed Gaussian impulse noise of grayscale and color images \cite{yan2013restoration, huang2017mixed}, sparse random corruptions of video data \cite{zhang2014novel}, mixture noise removal of MSI/HSI \cite{chen2017denoising, xu2022hyperspectral}, and Rician noise reconstruction of MRI \cite{awate2007feature, maggioni2012nonlocal}. In this paper, our synthetic experiments and analysis are mainly based on the AGWN model because: (i) the majority of existing methods are able to handle Gaussian noise, (ii) certain types of noise can be transformed to Gaussian distribution, and (iii) Romano et al. \cite{romano2017little} have recently pointed out that the removal of AGWN from an image is largely a solved problem, which may help explain the effectiveness of the simplified noise modeling in Eq. (\ref{awgn}).
\vspace{-3.8pt}
\subsection{Nonlocal Self-similarity}
\begin{figure}[htbp]
\graphicspath{{Figs/Fig_frameworks/}}
\centering
\subfigure[Image data]{
\label{Fig4}
\includegraphics[width=0.81in]{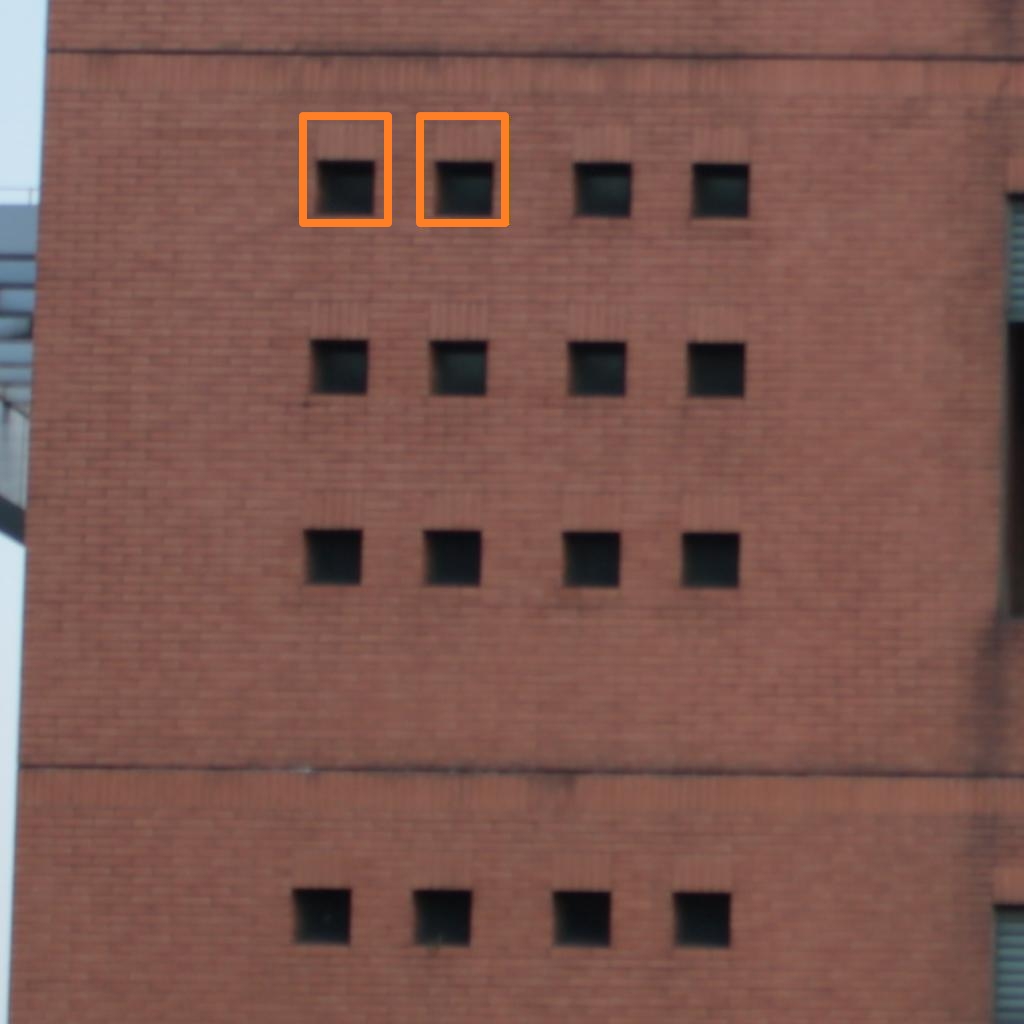}}
\subfigure[Video data]{
\label{Fig4}
\includegraphics[width=0.81in]{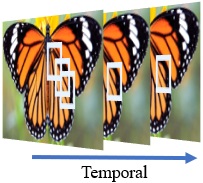}}
\subfigure[MRI data]{
\label{Fig4}
\includegraphics[width=0.81in]{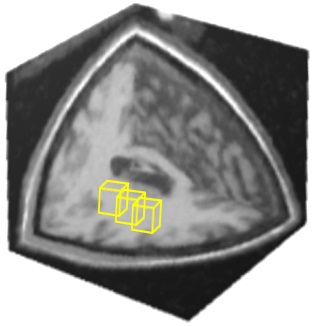}}
\subfigure[MSI data]{
\label{Fig4}
\includegraphics[width=0.81in]{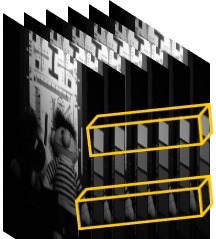}}

\caption{The nonlocal self-similarity (NLSS) prior and patch representation of different imaging sensors, techniques and applications.}
\label{Fig_NSS}
\end{figure}

\noindent The nonlocal self-similarity (NLSS) property of natural images is probably the most important prior adopted by many different denoising methods. Briefly, NLSS refers to the fact that a local image patch often has many nonlocal similar patches to it across the image \cite{xu2015patch}. Usually, the similarity between two patches $\mathcal{P}_A$ and $\mathcal{P}_B$ is measured by their Euclidean distance $d_{AB} = \|\mathcal{P}_A - \mathcal{P}_B\|$. In practice, to save some time, the search for similar patches is restricted to a local window $\Omega_{SR}$ with predefined size. As illustrated in Fig. \ref{Fig_NSS}, the patch representation and the rule of similar patch search (SPS) may vary for different types of data. For example, SPS of grayscale/color images can be conducted only on the single/luminance channel; for video sequences, SPS is performed along both temporal and spatial dimensions; and for MRI and MSI/HSI data, a patch could be represented by a 3D cube or a square tube with multiple spectral bands.
\vspace{-0.18cm}
\section{Related Works}\label{section_related_works}

In this section, we briefly introduce related denoising methods and datasets of different applications, which are summarized in Table \ref{Table_traditional_method}, Table \ref{Table_DNN_method} and Table \ref{Table_dataset_description}. More details can be found in the supplemental material and previous works\cite{yaroslavsky2001transform, buades2005review, milanfar2012tour, thanh2019review, shao2013heuristic, mohan2014survey, schmidhuber2015deep, tian2019deep, elad2023image}.

\begin{table*}[htbp]
  \centering
  \caption{Representative traditional denoisers with different noise modeling techniques and representation.}
  \scalebox{0.859}{
    \begin{tabular}{|c|c|c|c|c|c|}
    \hline
    Category & Representation & Methods & year  & Noise modeling & Key words \\
    \hline
    \multirow{56}{*}{\shortstack[l]{Traditional \\ denoisers}} & \multirow{16}[56]{*}{Matrix} & AWT \cite{donoho1995noising, huang2005color}  & 1995  & AWGN  & Adaptive wavelet thresholding methods \\
\cline{3-6}          &       & DCT \cite{yaroslavsky1996local, foi2007pointwise} & 1999  & AWGN  & Image denoising with discrete cosine transform (DCT) \\
\cline{3-6}          &       & NLM \cite{buades2005review, dai2013multichannel} & 2005  & AWGN  & nonlocal means (NLM) algorithm with Gaussian kernels \\
\cline{3-6}          &       & AMF \cite{chan2005salt} & 2005  & Impulse & Impulsive noise removal by adaptive median filter \\
\cline{3-6}          &       & K-SVD \cite{elad2006image} & 2006  & AWGN  & Over-complete dictionary learning and sparse coding \\
\cline{3-6}          &       & ONLM \cite{coupe2008optimized} & 2008  & AWGN and Rician & An optimized blockwise NLM Filter \\
\cline{3-6}          &       & ROLMMSE \cite{aja2008noise} & 2010  & Rician & A novel linear minimum mean square error estimator \\
\cline{3-6}          &       & AONLM \cite{manjon2010adaptive} & 2010  & AWGN and Rician & Adaptive NLM with spatially varying noise levels \\
\cline{3-6}          &       & TCVD \cite{liu2010high} & 2010  & AWGN  & Combination of robust optical flow with NLM \\
\cline{3-6}          &       & NLPCA \cite{zhang2010two, dong2012nonlocal, phophalia20173d} & 2010  & AWGN  & Applications of SVD transform \\
\cline{3-6}          &       & EPLL \cite{zoran2011learning} & 2011  & AWGN  & A Bayesian method for whole image restoration \\
\cline{3-6}          &       & PRI-NLM \cite{manjon2012new} & 2012  & AWGN and Rician & A rotationally invariant version of the NLM filter \\
\cline{3-6}          &       & ODCT \cite{manjon2012new} & 2012  & AWGN and Rician & A 3D DCT implementation \\
\cline{3-6}          &       & LRMR \cite{zhang2013hyperspectral} & 2013  & AWGN  & A low-rank matrix decomposition model \\
\cline{3-6}          &       & MCWNNM \cite{gu2014weighted, xu2017multi} & 2014  & AWGN  & Extension of WNNM to color images \\
\cline{3-6}          &       & PRI-NLPCA \cite{manjon2015mri} & 2015  & AWGN and Rician & A two stage filter based on NLPCA and PRI-NLM \\
\cline{3-6}          &       & NoiseClinic \cite{lebrun2015multiscale} & 2015  & AWGN and Realistic & A multiscale blind Bayes denoising algorithm \\
\cline{3-6}          &       & SPTWO \cite{buades2016patch} & 2016  & AWGN  & NLPCA with optical flow estimation \\
\cline{3-6}          &       & LSCD \cite{rizkinia2016local} & 2016  & AWGN  & Spectral component decomposition with line feature \\
\cline{3-6}          &       & Global-Search \cite{ehret2017global} & 2017  & AWGN  & Efficient approximate global patch search \\
\cline{3-6}          &       & LSM-NLR \cite{huang2017mixed} & 2017  & AWGN, Poisson and Impulse & Low-rank with laplacian scale mixture modeling \\
\cline{3-6}          &       & NMoG \cite{chen2017denoising} & 2017  & AWGN, Stripe and Impulse & Low-rank matrix model with non i.i.d Gaussian noise \\
\cline{3-6}          &       & FastHyde \cite{zhuang2018fast} & 2018  & AWGN and Poisson & Low-rank and sparse representation \\
\cline{3-6}          &       & TWSC \cite{xu2018trilateral} & 2018  & AWGN  & A trilateral weighted sparse coding scheme \\
\cline{3-6}          &       & VNLB \cite{arias2018video} & 2018  & AWGN  & A patch-based Bayesian model for video denoising \\
\cline{3-6}          &       & GID \cite{xu2018external} & 2018  & AWGN  & External data guided and internal prior learning \\
\cline{3-6}          &       & NLH \cite{hou2020nlh} & 2020  & AWGN  & Nonlocal pixel similarity with Haar transform \\
\cline{3-6}          &       & Bitonic \cite{treece2022real} & 2022  & AWGN and Realistic  & A non-learning bitonic filter with locally adaptive masks \\
\cline{2-6}          & \multirow{12}[38]{*}{Tensor} & CBM3D \cite{dabov2007color} & 2007  & AWGN  & BM3D with oppenent and YUV color mode transforms \\
\cline{3-6}          &       & LRTA \cite{renard2008denoising} & 2008  & AWGN  & A low Tucker rank tensor decomposition  model \\
\cline{3-6}          &       & PARAFAC \cite{liu2012denoising} & 2012  & AWGN  & A rank-one candecomp/parafac (CP) model \\
\cline{3-6}          &       & BM4D \cite{maggioni2012nonlocal} & 2012  & AWGN  & An extention of BM3D to MSI and MRI using 3D patch \\
\cline{3-6}          &       & 4DHOSVD \cite{rajwade2012image, zhang2015denoising, zhang2017denoise} & 2012  & AWGN  & Applications of 4DHOSVD transform \\
\cline{3-6}          &       & TDL \cite{peng2014decomposable} & 2014  & AWGN  & A Tucker based tensor dictionary learning algorithm \\
\cline{3-6}          &       & KTSVD \cite{Zhang2015KTSVD} & 2015  & AWGN  & A t-SVD based tensor dictionary learning algorithm \\
\cline{3-6}          &       & LRTV \cite{he2015total} & 2015  & AWGN  & Low-rank model with total variation (TV) regularization \\
\cline{3-6}          &       & NLTA-LSM \cite{dong2015low} & 2015  & AWGN and Poisson & Low Tucker rank with laplacian scale mixture modeling \\
\cline{3-6}          &       & ITSReg \cite{xie2016multispectral} & 2016  & AWGN  & Intrinsic Tensor Sparsity Regularization \\
\cline{3-6}          &       & LLRT \cite{chang2017hyper} & 2017  & AWGN  & low-rank tensor with hyper-laplacian regularization \\
\cline{3-6}          &       & MSt-SVD \cite{kong2017new, kong2019color} & 2017  & AWGN  & An efficient one-step t-SVD implementation \\
\cline{3-6}          &       & LLRGTV \cite{He2018LLRGTV} & 2018  & AWGN  & Low Tucker rank decomposition with total variation \\
\cline{3-6}          &       & WTR1 \cite{wu2018weighted} & 2019  & AWGN  & A weighted rank-one CP decomposition model \\
\cline{3-6}          &       & ILR-HOSVD \cite{lv2019denoising} & 2019  & AWGN  & A recursive low Tucker rank model with rank estimation \\
\cline{3-6}          &       & NGMeet \cite{he2019non} & 2019  & AWGN  & Low-rank tensor model with iterative regularization \\
\cline{3-6}          &       & LTDL \cite{gong2020low} & 2020  & AWGN  & A low-rank tensor dictionary learning method \\
\cline{3-6}          &       & OLRT \cite{chang2020hyperspectral} & 2020  & AWGN and Stripe  & A low-rank tensor model with sparse error component \\
\cline{3-6}          &       & GLF \cite{zhuang2021hyperspectral} & 2021  & AWGN  & A low-rank tensor model with global and local factorization \\
\cline{3-6}          &       & HLTA \cite{xie2022novel} & 2022  & AWGN, Stripe and Impulse  & A low-rank tensor model for denoising and completion \\
\cline{3-6}          &       & TRSTR \cite{zhang2023hyperspectral} & 2023  & AWGN, Stripe and Impulse  & A low-rank tensor-ring decomposition model \\
    \hline
    \end{tabular}}%
  \label{Table_traditional_method}%
\end{table*}%

\subsection{Traditional Denoisers}
\begin{figure}[htbp]
  \centering
  \graphicspath{{Figs/Fig_frameworks/}}
  \includegraphics[width=3.46in]{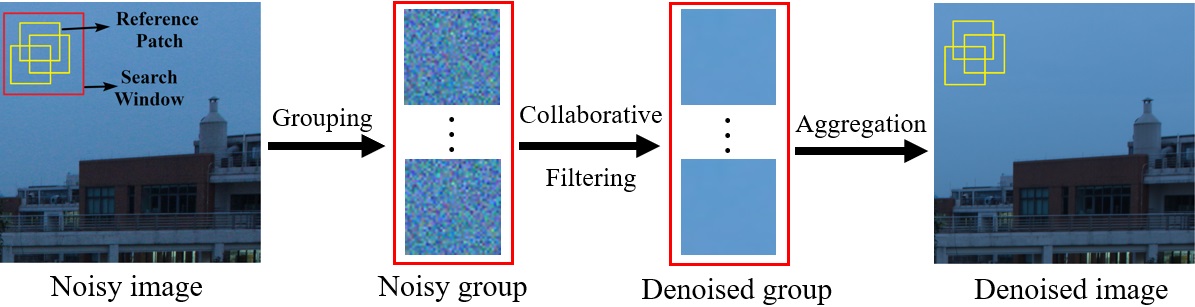}\\
  \caption{Illustration of the grouping-collaborative filtering-aggregation framework for traditional denoisers.}
  \label{Fig_traditional_framework}
\end{figure}

\noindent For traditional denoisers, learning and denoising are usually accomplished only with the noisy image by leveraging the NLSS property. The design of related algorithms can stem from the Bayesian point of view with various image priors \cite{elad2023image}. In our journey, we shall focus on the popular `grouping-collaborative filtering-aggregation' framework with different algebraic representations. The flowchart of this effective three-stage paradigm is illustrated in Fig. \ref{Fig_traditional_framework}.
\subsubsection{Grouping}
For every $d$-dimensional noisy image patch $\mathcal{P}_{n}$, based on certain patch matching criteria \cite{foi2007pointwise, rubel2014metric, buades2016patch, foi2016foveated, ehret2017global, Foi2020}, the grouping step stacks $K$ similar (overlapping) patches located within a local window $\Omega_{SR}$ into a $d+1$-dimensional group. For example, consider a 3D patch $\mathcal{P}_{n} \in \mathbb{R}^{H \times W \times N}$, where $H$, $W$ and $N$ represents height, width and the number of channels or spectral bands, respectively, the 4D group of $K$ patches can be directly represented by a fourth-order tensor $\mathcal{G}_n \in \mathbb{R} ^{H \times W \times N \times K}$, or a 2D matrix $\mathbf{G}_n \in \mathbb{R}^{HWN \times K}$ if every patch $\mathcal{P}_n$ is reshaped into a long vector $\mathbf{p}_n \in \mathbb{R}^{HWN}$.
\subsubsection{Collaborative Filtering}
Collaborative filters operate on the noisy patch group $\mathcal{G}_n$ to estimate the corresponding underlying clean group $\mathcal{G}_c$ via
\begin{equation}\label{tensor_collaborative_filtering}
  \hat{\mathcal{G}}_c = \mathop{\arg\min_{\mathcal{G}_c}} \| \mathcal{G}_n - \mathcal{G}_c \|_{F}^2 + \rho\cdot\Psi(\mathcal{G}_c)
\end{equation}
or in the matrix form
\begin{equation}\label{matrix_collaborative_filtering}
  \hat{\mathbf{G}}_c = \mathop{\arg\min_{\mathbf{G}_c}} \| \mathbf{G}_n - \mathbf{G}_c \|_{F}^2 + \rho\cdot\Psi(\mathbf{G}_c)
\end{equation}
where $\| \mathcal{G}_n - \mathcal{G}_c \|_{F}^2$ or $\| \mathbf{G}_n - \mathbf{G}_c \|_{F}^2$ indicates the conformity between the clean and noisy groups, and $\Psi(\cdot)$ is a regularization term for certain priors. For example, to model the nonlocal redundancies, the low-rank prior is adopted in \cite{xu2017multi, dong2012nonlocal, dong2015low, chang2017hyper} with $\Psi(\mathbf{G}_c) = \|\mathbf{G}_c\|_\ast$ for matrix and $\Psi(\mathcal{G}_c) = \sum_{n = 1}^4 a_{(n)}\|\mathbf{G}_{c_{(n)}}\|_\ast$ for tensor \cite{liu2012tensor}. In addition, the dictionary learning model with over-complete representations \cite{elad2006image,mairal2007sparse,xu2018trilateral} is utilized to reconstruct $\mathbf{G}_c$ with a dictionary $\mathbf{D}$ and sparse coding coefficients $\mathbf{C}$ via
\begin{equation}\label{matrix_dictionary}
  \hat{\mathbf{C}} = \mathop{\arg\min_{\mathbf{C}}}\|\mathbf{G}_n - \mathbf{D} \mathbf{C}\|_F^2 + \lambda\|\mathbf{C}\|_1
\end{equation}
where $\lambda \| \cdot \|$ is the regularization term that enforces sparsity constraint on $\mathbf{C}$. Once $\hat{\mathbf{C}}$ is computed, the latent clean patch group $\hat{\mathbf{G}}_c$ can be estimated as $\hat{\mathbf{G}}_c = \mathbf{D}\hat{\mathbf{C}}$. In \cite{peng2014decomposable} and \cite{Zhang2015KTSVD}, Eq. (\ref{matrix_dictionary}) is extended to tensor for MSI/HSI denoising with higher-order SVD (HOSVD) \cite{tucker1966some, de2000multilinear} and tensor-SVD (t-SVD) \cite{kilmer2011factorization, kilmer2013third} transforms, respectively. A simple and effective method is to model the sparsity with certain thresholding techniques \cite{donoho1994ideal, donoho1995noising} to attenuate the noise. For example, the hard-thresholding technique is adopted by the BM3D family and some state-of-the-art methods \cite{rajwade2012image, kong2019color}, which attempts to shrink the coefficients $\mathcal{T}(\mathcal{G}_n)$ in the transform-domain \cite{yaroslavsky2001transform} under a threshold $\tau$ via
 \begin{equation}\label{hard_thresholding}
  \mathcal{G}_{t}=\left\{
    \begin{aligned}
    \mathcal{T}(\mathcal{G}_n), \quad |\mathcal{T}(\mathcal{G}_n)| \geq \tau \\
    0, \quad |\mathcal{T}(\mathcal{G}_n)| < \tau
    \end{aligned}
    \right.
 \end{equation}
where $\mathcal{T}$ represents a pre-defined or learned transform. The estimated clean group $\hat{\mathcal{G}}_c$ is obtained by inverting the transform via
\begin{equation}\label{clean_group_estimate}
  \hat{\mathcal{G}}_c = \mathcal{T}^{-1} (\mathcal{G}_{t})
\end{equation}
It is noticed that SVD-based transforms are widely adopted among traditional denoisers, and such popularity largely results from the invertible orthogonal bases of SVD.
\subsubsection{Aggregation}
To further smooth out noise, the estimated clean patches of $\hat{\mathcal{G}}_c$ are averagely written back to their original location. More specifically, at the pixel level, every pixel $\hat{p}_i$ of the denoised image is the (weighted) average of all pixels at the same position of filtered group $\hat{\mathcal{G}}_c$, which can be formulated as
 \begin{equation}\label{aggregation}
   \hat{p}_i = \sum_{\hat{p}_{i_k} \in \hat{\mathcal{G}}_c} w_{i_k} \hat{p}_{i_k}
 \end{equation}
 where $w_{i_k}$ and $\hat{p}_{i_k}$ are weight and local pixel, respectively.
 \subsubsection{Discussion}
 \indent The major difference among various traditional patch-based denoisers mainly lies in the collaborative filtering step, which often varies according to the selection of regularization terms, transforms and algebraic representation. Intuitively, reshaping the 4D group $\mathcal{G}$ of Eq. (\ref{tensor_collaborative_filtering}) into the 2D matrix $\mathbf{G}$ of Eq. (\ref{matrix_collaborative_filtering}) may break the internal structure of natural images. Therefore, a typical assumption \cite{muti2008lower, rajwade2012image, peng2014decomposable, zhang2015denoising, Zhang2015KTSVD, chang2017hyper, kong2019color, gong2020low} is that tensor representation and decomposition techniques can help preserve more structure information, based on the fact that images can be naturally represented by multi-dimensional array. However, the conventional and widely-used tensor model may fall into the unbalance trap \cite{oseledets2011tensor, bengua2017efficient}, leading to unsatisfactory image restoration performance. In this paper, we further show that with slight modifications, a simple modified SVD (M-SVD) implementation may be able to produce competitive results compared with several tensor-based methods. The M-SVD approach is described in Algorithm \ref{m-svd} with more details given in the supplemental material.

\begin{algorithm}[ht]
\caption{Modified SVD (M-SVD)}
{\bf Input:} Noisy image $\mathcal{A}$, patch size $ps$, number of similar patches $K$ and search window size $\Omega_{SR}$.\\
{\bf Output:} Estimated clean image $\hat{\mathcal{A}}_c$.\\
{\bf Step 1} (Grouping): For every reference patch of $\mathcal{A}$, stack $K$ similar patches in a group $\mathcal{G}$ within $\Omega_{SR}$.\\
{\bf Step 2} (Collaborative filtering):\\
 \hspace*{0.18in}(1) Obtain the group and patch level transform $\mathbf{U}$ and $\mathbf{V}$ by performing SVD on $\mathbf{G}_{opp_{(3)}}$ and $\mathbf{G}_{(4)}^T$, respectively.\\
 \hspace*{0.18in}(2) Apply the hard-threshold technique to $\mathbf{C} = \mathbf{U}^T \mathbf{G}_{(4)} \mathbf{V}$ via $\mathbf{C}_{trun} = \text{hard-threshold}(\mathbf{C})$.\\
 \hspace*{0.18in}(3) Take the inverse transform to obtain estimated clean group via $\hat{\mathbf{G}}_{c} = \mathbf{U} \mathbf{C}_{trun} \mathbf{V}^T$.\\
{\bf Step 3} (Aggregation): Averagely write back all image patches in $\hat{\mathbf{G}}_{c}$ to their original locations.
\label{m-svd}
\end{algorithm}
 \vspace{-0.3cm}

\vspace{-0.269cm}
 \subsection{DNN Methods}
\begin{table*}[htbp]
    \renewcommand{\arraystretch}{0.098}
    \scriptsize
    \centering
    \caption{Representative DNN denoising methods with different noise modeling techniques and training strategy.}
    \scalebox{0.8468}{
    \begin{tabular}{|c|c|c|c|c|c|}
    \hline
    Category & Training srategy & Methods & Year  & Noise modeling & Key words \\
    \hline
    \multirow{198}[136]{*}{DNN Methods} & \multirow{136}[100]{*}{Supervised} & MLP\cite{burger2012image} & 2012  & AWGN  & A multilayer perceptron model \\
\cline{3-6}          &       & TNRD\cite{chen2016trainable} & 2016  & AWGN  & A trainable nonlinear diffusion model \\
\cline{3-6}          &       & DnCNN\cite{zhang2017beyond} & 2017  & AWGN  & A CNN model with batch normalization and residual learning \\
\cline{3-6}          &       & NLNet\cite{lefkimmiatis2017non} & 2017  & AWGN  & A nonlocal CNN model for grayscale/color image denoising \\
\cline{3-6}          &       & UDNet\cite{Lefkimmiatis_2018_CVPR} & 2018  & AWGN  & A robust and flexible CNN model using UNet \\
\cline{3-6}          &       & HSID-CNN\cite{yuan2018hyperspectral} & 2018  & AWGN  & A 2D and 3D combined CNN model \\
\cline{3-6}          &       & VNLNet\cite{davy2018non} & 2018  & AWGN  & The first nonlocal CNN model for video denoising\\
\cline{3-6}          &       & FFDNet\cite{zhang2018ffdnet} & 2018  & AWGN  & A flexible CNN model with tunable input noise levels \\
\cline{3-6}          &       & HSI-DeNet\cite{chang2018hsi} & 2018  & AWGN and Stripe & A CNN model for mixed noise removal \\
\cline{3-6}          &       & PRI-PB-CNN\cite{manjon2018mri} & 2018  & AWGN an Rician & A combination of sliding window scheme and 3D CNN \\
\cline{3-6}          &       & GCBD\cite{chen2018image} & 2018  & AWGN and Realistic & A GAN-based blind denoiser \\
\cline{3-6}          &       & DnGAN\cite{yeh2018image} & 2018  & AWGN and Blur & A GAN model with maximum a posteriori (MAP) framework \\
\cline{3-6}          &       & DBF\cite{chen2019real} & 2019  & AWGN and Realistic & Integration of CNN into a boosting algorithm \\
\cline{3-6}          &       & FCCF\cite{yue2019high} & 2019  & AWGN and Realistic & A deep fusion scheme of collaborative filtering (CBM3D) and CNN \\
\cline{3-6}          &       & DIDN\cite{yu2019deep} & 2019  & AWGN and Realistic & A deep iterative down-up CNN model \\
\cline{3-6}          &       & CBDNet\cite{guo2019toward} & 2019  & Realistic & A convolutional blind denoising network \\
\cline{3-6}          &       & ViDeNN\cite{claus2019videnn} & 2019  & AWGN  & A combination of spatial and temporal filtering with CNN \\
\cline{3-6}          &       & RIDNet\cite{anwar2019real} & 2019  & AWGN and Realistic & A single stage model with feature attention \\
\cline{3-6}          &       & MIFCN\cite{abbasi2019three} & 2019  & Realistic & A fully convolutional network model \\
\cline{3-6}          &       & DRDN\cite{song2019dynamic} & 2019  & Realistic & A dynamic residual dense network model \\
\cline{3-6}          &       & SGN\cite{gu2019self} & 2019  & Realistic & A self-guided network with top-down architecture \\
\cline{3-6}          &       & ADGAN\cite{Lin_2019_CVPR_Workshops} & 2019  & Realistic & An attentive GAN model with noise domain adaptation \\
\cline{3-6}          &       & QRNN3D\cite{wei20203} & 2020  & AWGN and Stripe & A deep recurrent neural network model with 3D convolution \\
\cline{3-6}          &       & CycleISP\cite{zamir2020cycleisp} & 2020  & Realistic & A GAN based framework modeling the camera pipeline \\
\cline{3-6}          &       & GCDN\cite{valsesia2020deep} & 2020  & AWGN and Realistic & A graph convolution network based denoising model \\
\cline{3-6}          &       & FastDVDNet \cite{Tassano_2020_CVPR}& 2020  & AWGN  & A real-time video denoising network without flow estimation \\
\cline{3-6}          &       & DANet\cite{yue2020dual} & 2020  & Realistic & A Bayesian framework for noise removal and generation \\
\cline{3-6}          &       & LIDIA \cite{Vaksman_2020_CVPR_Workshops} & 2020  & AWGN  & A lightweight model with instance adaptation \\
\cline{3-6}          &       & SADNet\cite{chang2020spatial} & 2020  & AWGN  & A CNN model with residual spatial-adaptive blocks \\
\cline{3-6}          &       & AINDNet\cite{Kim_2020_CVPR} & 2020  & AWGN and Realistic & A CNN model based on a transfer learning scheme \\
\cline{3-6}          &       & MIRNet\cite{Zamir2020MIRNet} & 2020  & Realistic & A multi-scale model with parallel convolution streams \\
\cline{3-6}          &       & PD-denosing\cite{zhou2020awgn} & 2020  & AWGN and Realistic & A CNN model with pixel-shuffle down-sampling adaptation \\
\cline{3-6}          &       & ADRN\cite{zhao2020adrn} & 2020  & AWGN  & A deep residual network model with channel attention scheme \\
\cline{3-6}          &       & MPRNet\cite{zamir2021multi} & 2021  & Realistic & A multi-stage network with cross-stage feature fusion and attention modules \\
\cline{3-6}          &       & DRUNet\cite{zhang2021plug} & 2021  & AWGN  & A plug-and-play method with deep denoiser prior \\
\cline{3-6}          &       & DudeNet\cite{tian2021designing} & 2021  & AWGN and Realistic & A dual network with four different blocks and sparse mechanism \\
\cline{3-6}          &       & InvDN\cite{liu2021invertible} & 2021  & Realistic & An invertible denoising network with wavelet transformation \\
\cline{3-6}          &       & Restormer\cite{liu2021invertible} & 2021  & AWGN and Realistic & An efficient Transformer model with multi-head attention networks \\
\cline{3-6}          &       & DCDicL\cite{zheng2021deep} & 2021  & AWGN  & A deep convolution dictionary learning denoising network \\
\cline{3-6}          &       & NBNet\cite{cheng2021nbnet} & 2021  & AWGN and Realistic & A UNet based model that learns subspace basis and image projection \\
\cline{3-6}          &       & DeamNet\cite{ren2021adaptive} & 2021  & AWGN and Realistic & A CNN model with adaptive consistency prior and self-attention mechanism \\
\cline{3-6}          &       & PNGAN\cite{cai2021learning} & 2021  & AWGN and Realistic & A GAN based model for real image synthesis with domain alignment \\
\cline{3-6}          &       & SwinIR\cite{liang2021swinir} & 2021 & AWGN & A shifted window (Swin) Transformer model with self-attention mechanism \\
\cline{3-6}          &       & NAFNet\cite{chen2022simple} & 2022  & Realistic & A UNet based model without nonlinear activation layers \\
\cline{3-6}          &       & MalleNet\cite{jiang2022fast} & 2022  & AWGN and Realistic & A CNN with spatially-varying convolution kernels \\
\cline{3-6}          &       & Masked\cite{chen2023masked} & 2023 & AWGN and Realistic & A novel training approach that randomly masks pixels of the input image\\
\cline{3-6}          &       & SERT\cite{li2023spectral} & 2023 & AWGN, Mixed and Realistic & A Transformer framework with rectangle self-attention modeling\\
\cline{2-6}          & \multirow{89}[22]{*}{\shortstack[l]{Self-supervised/\\Unsupervised}} & Noise2Noise\cite{lehtinen2018noise2noise} & 2018  & AWGN and Poisson & A model trained on noisy image pairs of the same scene \\
\cline{3-6}          &       & SCGAN\cite{yan2019unsupervised} & 2019  & AWGN  & Unsupervised modeling with self-consistent GAN \\
\cline{3-6}          &       & Noise2Void \cite{krull2019noise2void} & 2019  & AWGN  & A blind-spot masking strategy that excludes central pixels of the input \\
\cline{3-6}          &       & Self2Self\cite{quan2020self2self} & 2020  & AWGN and Realistic & A dropout based strategy that generates noisy pairs with the Bernoulli sampler \\
\cline{3-6}          &       & C2N\cite{jang2021c2n} & 2021  & Realistic & A GAN based model that generates real noisy images from arbitrary clean ones \\
\cline{3-6}          &       & UDVD\cite{sheth2021unsupervised} & 2021  & AWGN & An unsupervised video denoising network based on blind-spot strategy \\
\cline{3-6}          &       & R2R\cite{pang2021recorrupted} & 2021  & AWGN and Realistic & A model trained on recorrupted noisy pairs \\
\cline{3-6}          &       & Neighbor2Neighbor \cite{huang2021neighbor2neighbor} & 2021  & AWGN and Realistic & A model trained on noisy pairs generated by the downsampling strategy \\
\cline{3-6}          &       & IDR\cite{zhang2022idr} & 2022  & AWGN and Realistic & A model trained on noisy pairs with an iterative refinement strategy \\
\cline{3-6}          &       & CVF-SID\cite{neshatavar2022cvf} & 2022  & Realistic & A CNN model with a self-supervised cycle \\
\cline{3-6}          &       & Blind2Unblind\cite{wang2022blind2unblind} & 2022  & AWGN and Realistic & A blind-spot network with a global-aware mask mapper \\
\cline{3-6}          &       & AP-BSN\cite{lee2022ap} & 2022  & Realistic & A blind-spot network (BSN) with an asymmetric pixel-shuffle downsampling strategy \\
\cline{3-6}          &       & DDS2M \cite{miao2023dds2m} & 2023  & AWGN & A deep diffusion model with a variational spatio-spectral module \\
\cline{3-6}          &       & C-BSN \cite{jang2023self} & 2023  & Realistic & A conditional BSN with downsampled invariance loss \\
\cline{3-6}          &       & MM-BSN \cite{zhang2023mm} & 2023  & Realistic & A BSN method with multiple masks of different shapes \\
\cline{3-6}          &       & LGBPN \cite{wang2023lgbpn} & 2023  & Realistic & A BSN with a patch-masked convolution module and a dilated Transformer block \\
\cline{3-6}          &       & SASL \cite{li2023spatially} & 2023  & Realistic & A BSN with spatially adaptive supervision \\
    \hline
    \end{tabular}}%
  \label{Table_DNN_method}%
\end{table*}%

\subsubsection{Overview}
The most recent development of image processing stems largely from the applications of deep learning techniques, which demonstrate outstanding performance in a wide variety of tasks \cite{abdelhamed2019ntire}. Image denoising is not an exception. From the early plain networks \cite{jain2008natural, burger2012image} to recently proposed generative and diffusion models \cite{cai2021learning, miao2023dds2m}, numerous network architectures and frameworks have been developed with different training paradigms, including supervised, self-supervised and unsupervised learning\footnote{It is noticed that the terms \textit{self-supervised}, \textit{unsupervised} and \textit{blind} denoising are often used interchangeably in the literature \cite{izadi2022image}.}.

\subsubsection{Supervised Methods}
\begin{figure}[htbp]
  \centering
  \graphicspath{{Figs/Fig_frameworks/}}
  \includegraphics[width=2.866in, height=1.399in]{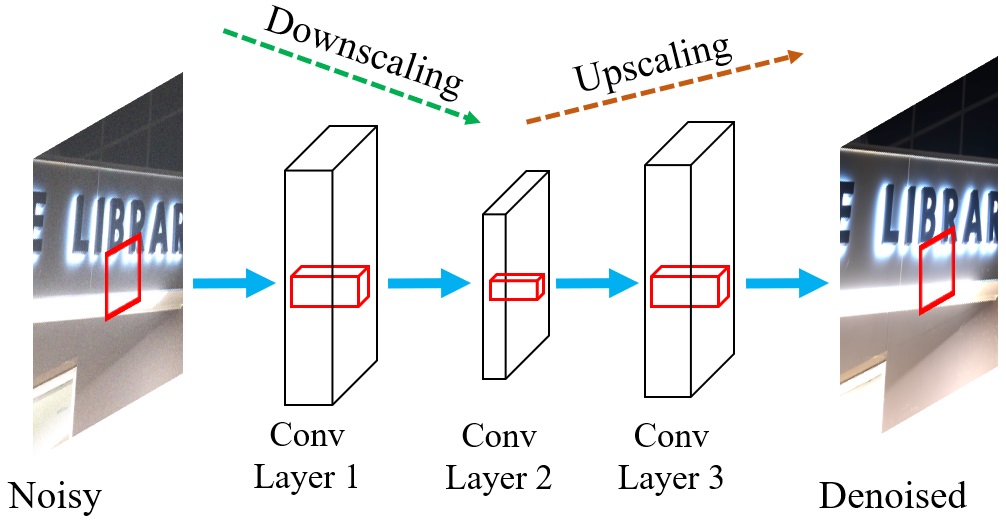}\\
  \caption{A plain CNN denoising framework with three convolutional layers.}
  \label{Fig_CNN}
\end{figure}

Different from traditional denoisers that use only internal information of the noisy observation, the supervised training strategy of DNN methods is often guided by external priors and data. The goal is to minimize the distance $\mathcal{L}$ between predicted and clean images via
\begin{equation}\label{DNN_denoising_equation}
  \underset{\theta}{\min} \sum_{i} \mathcal{L} (\mathcal{F}_{\theta}(\mathcal{Y}_i), \mathcal{X}_i) + \rho\cdot\Psi(\mathcal{F}_{\theta}(\mathcal{Y}_i))
\end{equation}
where $\mathcal{L}$ can be measured by different loss functions \cite{izadi2022image}, $\mathcal{X}_{i}$ and $\mathcal{Y}_{i}$ are clean-noisy image (patch) pairs, $\mathcal{F}_{\theta}$ with parameter $\theta$ is a nonlinear function that maps noisy patches onto predicted clean patches, and $\Psi(\cdot)$ represents certain regularizers \cite{lefkimmiatis2017non, ulyanov2018deep}. Early methods \cite{zhou1987novel, chiang1989multi} work with the known shift blur function and weighting factors. Burger et al. \cite{burger2012image} show that a simple multi-layer perceptron (MLP) network is able to compete with representative traditional denoisers at certain noise levels. To extract latent features and exploit the self-similarity property of images, a widely adopted network is CNN \cite{lecun1998gradient, jain2009natural} with flexible size of convolution filters and local receptive fields. Fig. \ref{Fig_CNN} illustrates a simple CNN denoising framework with three convolutional layers. Due to the effectiveness of CNN, its variations are quite extensive. To name a few, GCDN\cite{valsesia2020deep} utilizes graph convolution networks (GCNs) to capture self-similar information. SADNet \cite{chang2020spatial} introduces a residual spatial-adaptive block and context block to sample related features and obtain multi-scale information. QRNN3D \cite{wei20203} exploits 3D convolutions to extract structural spatio-spectral correlation of MSI/HSI data.

\vspace{-3pt}

\subsubsection{Self-supervised and Unsupervised Methods}
Collecting a large scale matched clean-noisy image pairs for training is expensive and impractical, especially in the medical imaging sector\cite{guo2020unsupervised}. Such limitation of supervised denoising prompts the development of self-supervised and unsupervised denoising networks. \\
\indent To get rid of the prerequisite on training data and leverage the power of DNNs, pioneering methods such as DIP \cite{ulyanov2018deep} trains a network on a single image to fit itself by early stopping. SURE-based methods \cite{metzler2018unsupervised, soltanayev2018training} impose regularizations on DNNs to avoid overfitting. An interesting alternative strategy is to use noisy-noisy pairs for training \cite{lehtinen2018noise2noise}. For example, Noise2Self \cite{batson2019noise2self} and Noise2Void \cite{krull2019noise2void} introduce blind-spot learning by masking pixels of the noisy input. Self2Self \cite{quan2020self2self} adopts a dropout-based scheme with noisy pairs generated by the Bernoulli sampler. Noise-As-Clean \cite{xu2020noisy} and R2R \cite{pang2021recorrupted} obtain noisy pairs by corrupting the input with certain noise distributions and noise levels. Neighbor2Neighbor \cite{huang2021neighbor2neighbor} creates noisy pairs via sub-sampling and pixel-wise independent noise assumption. \\
\indent A recent appealing tool and new trend for self-supervised denoising is the generative framework such as GANs \cite{song2020unsupervised, guo2020unsupervised, jang2021c2n} and diffusion models \cite{ fadnavis2020patch2self, kawar2022denoising, miao2023dds2m}. Briefly, thanks to the powerful image synthesis capability of GAN, it is able to learn a variety of complex noise distributions and thus generate more accurate realistic noisy images. The diffusion model can go beyond image synthesis \cite{miao2023dds2m}, which intends to approximate the score function \cite{elad2023image, song2020improved} and then adopts an iterative algorithm.
\subsubsection{Discussion}
The penetration of deep learning in image denoising has beyond doubt pushed forward the frontiers of denoising. Despite the effectiveness of DNN methods, they may not be cure-all, which enjoy three major advantages and also face the same challenges. First, by utilizing external information to guide the training process, DNN methods are not comfined to the theoretical and practical bound of traditional denoisers \cite{chatterjee2009denoising}. However, the high quality training datasets and certain prior information such as ISO, shutter speed and camera brands are not always available in practice. Second, with the aid of advanced GPU devices for acceleration, real-time denoising \cite{Tassano_2020_CVPR} is achievable for certain tasks, yet the expensive computational resources may not be accessible to ordinary users and researchers. Last but not least, the deep, powerful and sophisticated architectures are capable of capturing latent features underlying noisy images. But compared with benchmark traditional denoisers which only store several small predefined transform matrices, the complex networks with millions of parameters may drastically increase the storage cost.

\vspace{-0.18cm}
\subsection{Datasets}
\noindent In this section, we provide a short overview of popular datasets for various denoising applications and briefly introduce the proposed dataset. More information is available in the supplemental material and \cite{anaya2018renoir, nam2016holistic, xu2018real, abdelhamed2018high, Perazzi2016, brummer2019natural, yue2020supervised, CAVE_0293, arad_and_ben_shahar_2016_ECCV, chakrabarti2011statistics, cocosco1997brainweb, marcus2007open}.
\subsubsection{A Brief Overview}
%
%
%
%
%

\begin{figure}[htbp]

\graphicspath{{Figs/Fig_dataset_illus/}}

\centering
\includegraphics[width=3.39in]{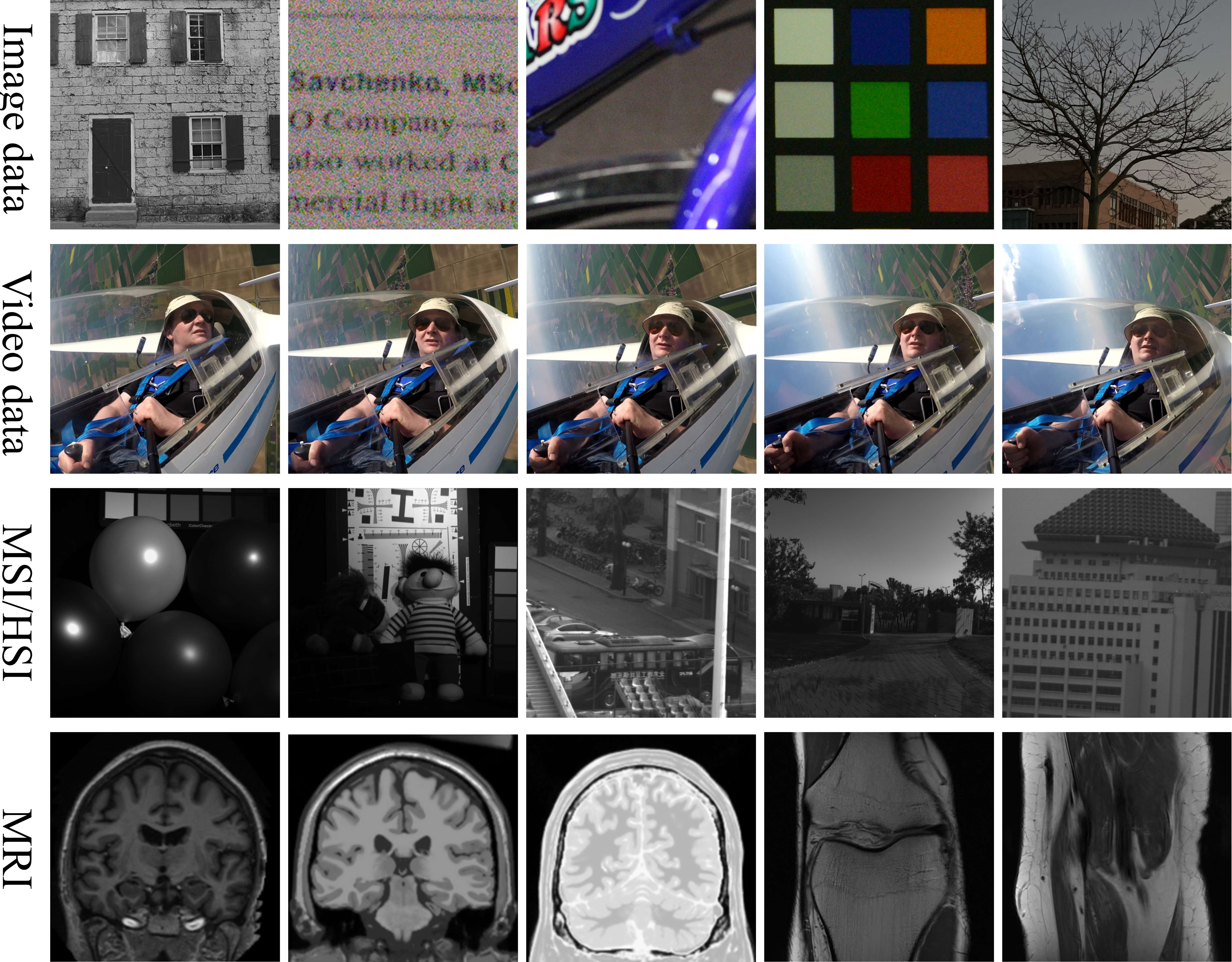}

\caption{Illustration of datasets for different applications. From the first row to the fourth row: grayscale/color image, video, MSI/HSI and MRI data.}

\label{Fig_dataset_illus}
\end{figure}

\noindent The statistics of popular datasets for different denoising applications is summarized in Table \ref{Table_dataset_description}. Some examples are illustrated in Fig. \ref{Fig_dataset_illus}. Typically, datasets used for synthetic experiments consist of noise-free (ground-truth) images acquired under ideal conditions with sufficient light and careful camera settings, while the corresponding noisy images are generated by manually adding noise of different levels and distributions to the noise-free ones. In many real-world applications, images are inevitably contaminated by noise to various degrees, often decided by the environments and imaging devices. In such cases, the image averaging strategy is often adopted to generate the ground-truth data by averaging a series of images captured based on the same, static scene. It is noteworthy that compared to grayscale/color images and videos, collecting MSI/HSI and MRI data is of greater difficulty and also more expensive.
\begin{table}[htbp]
\tiny
  \centering
  \caption{Popular datasets for synthetic and real-world experiments. 'GT': ground-truth, '$\surd$': Available, '-': Not available, 'F': number of frames.}
  \scalebox{0.93}{
    \begin{tabular}{cccccc}
    \toprule
    Applications & Name  & Experiments & GT    & Data size & \# Data \\
    \midrule
    \multicolumn{1}{c}{\multirow{10}[29]{*}{Grayscale/Color Image}} & Standard set \cite{dabov2007image} & Synthetic & $\surd$     & 512 $\times$ 512 / 256$\times$256 & 11 \\
\cmidrule{2-6}          & BSD \cite{MartinFTM01} & Synthetic & $\surd$     & 481 $\times$ 321 $\times$ 3 & 500 \\
\cmidrule{2-6}          & Kodak \cite{Kodak} & Synthetic & $\surd$     & 512 $\times$ 512 $\times$ 3 & 24 \\
\cmidrule{2-6}          & RENOIR \cite{anaya2018renoir} & Real-world & $\surd$     & 3684 $\times$ 2760 $\times$ 3 & 120 \\
\cmidrule{2-6}          & Nam-CC15 \cite{nam2016holistic} & Real-world & $\surd$     & 512 $\times$ 512 $\times$ 3 & 15 \\
\cmidrule{2-6}          & Nam-CC60 \cite{nam2016holistic} & Real-world & $\surd$     & 500 $\times$ 500 $\times$ 3 & 60 \\
\cmidrule{2-6}          & PolyU \cite{xu2018real} & Real-world & $\surd$     & 512 $\times$ 512 $\times$ 3 & 100 \\
\cmidrule{2-6}          & DnD \cite{plotz2017benchmarking} & Real-world & -   & 512 $\times$ 512 $\times$ 3 & 1000 \\
\cmidrule{2-6}          & SIDD \cite{abdelhamed2018high} & Real-world & -   & 256 $\times$ 256 $\times$ 3 & 1280 \\
\cmidrule{2-6}          & HighISO \cite{yue2019high} & Real-world & $\surd$     & 512 $\times$ 512 $\times$ 3 & 100 \\
\cmidrule{2-6}          & Our IOCI & Real-world & $\surd$     & 1024 $\times$ 1024 $\times$ 3 & 848 \\
    \midrule
    \multirow{6}[9]{*}{Video} & Set8 \cite{Set8} & Synthetic & $\surd$     & 960 $\times$ 540 $\times$ 3 $\times$ F & 8 \\
\cmidrule{2-6}          & DAVIS \cite{Perazzi2016} & Synthetic & $\surd$     & 854 $\times$ 480 $\times$ 3 $\times$ F & 30 \\
\cmidrule{2-6}          & CRVD \cite{yue2020supervised} & Real-world & $\surd$     & 1920 $\times$ 1080 $\times$ 3 $\times$ F & 61 \\
\cmidrule{2-6}          & PVDD \cite{xu2022pvdd} & Real-world & -    & - & 200 \\
\cmidrule{2-6}          & Our IOCV & Real-world & $\surd$     & 512 $\times$ 512 $\times$ 3 $\times$ F & 39 \\
    \midrule
    \multirow{6}[12]{*}{MSI/HSI} & CAVE \cite{CAVE_0293} & Synthetic & $\surd$     & 512 $\times$ 512 $\times$ 31 & 32 \\
\cmidrule{2-6}          & ICVL \cite{arad_and_ben_shahar_2016_ECCV} & Synthetic & $\surd$     & 1392 $\times$ 1300 $\times$ 31 & 201 \\
\cmidrule{2-6}          & Indian Pines \cite{PURR1947} & Synthetic & $\surd$     & 145 $\times$ 145 $\times$ 224 & 1 \\
\cmidrule{2-6}          & Urban \cite{PURR1947} & Real-world & -   & 307 $\times$ 307 $\times$ 210 & 1 \\
\cmidrule{2-6}          & HHD \cite{chakrabarti2011statistics} & Real-world & -   & 1392 $\times$ 1040 $\times$ 31 & 77 \\
\cmidrule{2-6}          & Real-HSI \cite{zhang2021hyperspectral} & Real-world & $\surd$     & 696 $\times$ 520 $\times$ 34 & 59 \\
    \midrule
    \multirow{3}[6]{*}{MRI} & BrainWeb \cite{cocosco1997brainweb} & Synthetic & $\surd$     & 181 $\times$ 217 $\times$ 181 & 3 \\
\cmidrule{2-6}          & fastMRI \cite{zbontar2018fastmri} & Synthetic & $\surd$     & 320 $\times$ 320 $\times$ 40 & 50 \\
\cmidrule{2-6}          & OASIS \cite{marcus2007open} & Real-world & -   & 256 $\times$ 256 $\times$ 128 & 2 \\
    \bottomrule
    \end{tabular}%
    }
  \label{Table_dataset_description}%
\end{table}%

\vspace{-0.2cm}
\subsubsection{The Proposed Dataset}
\begin{figure}[htbp]
\graphicspath{{Figs/Fig_dataset_illus/}}
\centering
\subfigure[The proposed IOCI dataset]{
\label{Fig4}
\includegraphics[width=3.38in]{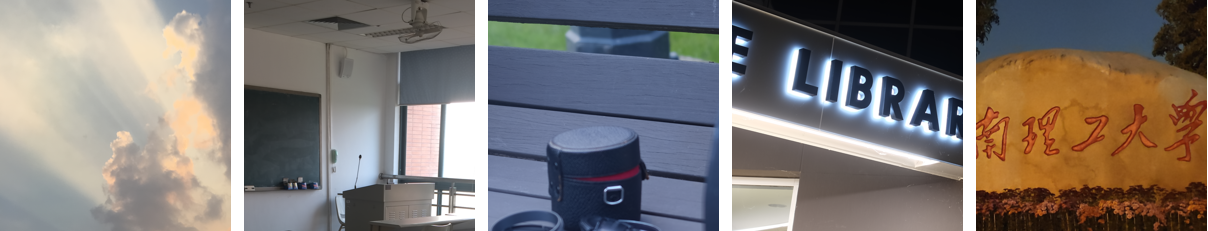}}
\subfigure[The proposed IOCV dataset]{
\label{Fig4}
\includegraphics[width=3.38in]{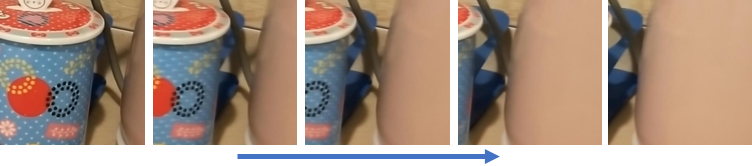}}

\caption{Illustration of the proposed IOCI and IOCV dataset.}
\label{Fig_Kong_dataset}
\end{figure}

\begin{figure}[htbp]

\graphicspath{{Figs/Fig_frameworks/}}

\centering
\includegraphics[width=3.58in]{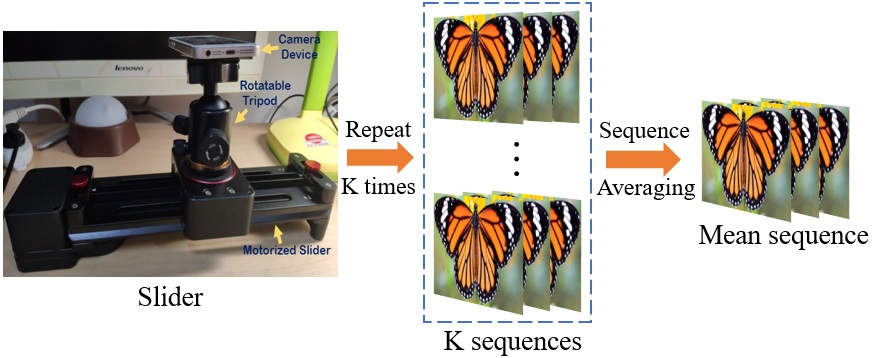}

\caption{The procedure of generating mean video sequences with a motorized slider. The camera is fixed to a rotatable tripod ball-head placed on
top of the slider.}

\label{Fig_Video_slider}
\end{figure}

\noindent In this subsection, we briefly introduce the motivation and details regarding the setup and protocol followed by our indoor-outdoor color image (IOCI) and video (IOCV) dataset, some examples are illustrated in Fig. \ref{Fig_Kong_dataset}.\\
\indent \textbf{IOCI}. To capture images of various scenes, 13 different camera devices are used to collect data in both indoor and outdoor environments. To reduce human interference and simulate daily usage, we mostly adopt the cameras' \textit{auto mode} instead of predefined settings such as ISO, shutter speed and aperture \cite{abdelhamed2018high, xu2018real, yue2019high}. In uncontrollable and dark environments, we use short exposures and increase ISO values to reduce misalignment and produce over 100 images with high noise levels. Sampled data with obvious misalignment and illumination differences are discarded.\\
\indent \textbf{IOCV}. To obtain reference videos for benchmarking, we adopt a video-by-video strategy. Instead of manually controlling static objects \cite{yue2020supervised}, we propose to move cameras automatically. The procedure of generating mean videos as ground-truth is illustrated in Fig. \ref{Fig_Video_slider}. We fix the cameras onto a rotatable tripod ball-head placed on top of a motorized slider. The slider and the tripod ball-head can be set to repeatedly move and rotate at different speeds, which simulate the movement of observed objects in more than one directions. Both the slider and cameras are controlled remotely to reduce human interference.

\vspace{-0.28cm}
\vspace{-0.28cm}
\section{Experiments} \label{section_experiments}
In this section, we report the results of different methods/models for the denoising task of images, videos, MSI/HSI and MRI data. All implementations and source codes are provided by the authors or downloaded from the authors' personal websites. For methods that require GPU devices, we resort to Google ColabPro's computing resources. All other experiments are performed on a computer equipped with Core(TM) i5-7500 @ 3.4 GHz and 16GB RAM.
 \vspace{-0.18cm}
\subsection{Results for Real-World Image Datasets}  \label{color_image_experiments}
\subsubsection{Experimental Settings} \label{color_image_settings}
We compare the performance of over 40 methods or models with real-world image datasets, including DnD, SIDD, Nam (CC15 \cite{nam2016holistic, xu2017multi} and CC60\cite{nam2016holistic, xu2018external}), PolyU, High-ISO and our IOCI, where SIDD dataset provides training and validation samples. The clean-noisy image pairs of DnD and SIDD are not available, and the test results can be obtained by uploading denoised images online. In this paper, we focus on the sRGB color space since the raw data \cite{wang2020practical} are not always accessible. \\
\indent Typically, the decisive parameters for traditional denoisers are input noise level, patch size and the number of local similar patches, etc, whereas the number of layers, learning rate and weight size are essential for DNN methods. However, many existing methods are designed for synthetic experiments, while ideally in real-world cases, the parameters of all compared methods should be fine-tuned to obtain the best possible performance. But this can be computationally expensive and the clean reference images for training and validation are always not available. Therefore, a more practical way is to selectively adopt effective pretrained models or predefined parameter settings. Since many DNN methods offer two to six pretrained models with different parameter settings for testing, for fair comparison, we select the input noise level for each Gaussian denoiser from four different values, which may be regarded as the equivalance of four pretrained models corresponding to \textit{low}, \textit{medium-low}, \textit{medium-high} and \textit{high} denoising modes. We report the best average values of all compared methods on each dataset. Peak signal-to-noise ratio (PSNR) and structural similarity (SSIM) \cite{wang2004image} are employed for objective evaluations. Normally, the higher the PSNR and SSIM values, the better the quality of denoised images.
\begin{table*}[htbp]
\centering
    \caption{Average PSNR, SSIM values on real-world color image datasets. The average time is calculated based on the PolyU dataset. `-' means the results are not available. `C: CANON', `F: FUJIFILM', `H: HUAWEI', `I: IPHONE', 'N: NIKON', `O: OPPO', `S: SONY', `X: XIAOMI'. Results marked with $\ast$ are from the original papers or the official websites.}
    \vspace{-9pt}
    \renewcommand{\arraystretch}{0.2818}
    \scalebox{0.5399}{
    \begin{tabular}{ccccccccccccccccccccccc}
    \toprule
    \multicolumn{3}{c}{\multirow{2}[4]{*}{Methods/Models}} & \multirow{2}[4]{*}{DnD} & \multirow{2}[4]{*}{SIDD} & \multirow{2}[4]{*}{CC15} & \multirow{2}[4]{*}{CC60} & \multirow{2}[4]{*}{PolyU} & \multirow{2}[4]{*}{HighISO} & \multicolumn{13}{c}{IOCI}                                                                             & \multirow{3}[6]{*}{Time (s)} \\
\cmidrule{10-22}    \multicolumn{3}{c}{}  &       &       &       &       &       &       & C-100D & C-600D & C-5DMark4 & F-X100T & H-Honor6X & H-Mate40Pro & I-5S & I-6S & I-13 & N-D5300 & O-R11s & S-A6500 & X-MI8 \\
\cmidrule{1-22}    Category     & Schemes & \# Images & 1000  & 1280  & 15    & 60    & 100   & 100   & 55    & 25    & 83    & 71    & 30    & 126   & 36    & 67    & 174   & 56    & 39    & 36    & 50    &  \\
    \midrule
    \multirow{66}[52]{*}{\shortstack[c]{Traditional \\ denoisers}} & \multirow{39}[32]{*}{Matrix} & \multirow{2}[4]{*}{Bitonic \cite{treece2022real}} & 37.85  & 36.67  & 35.22  & 35.98  & 36.64  & 37.37  & 39.38  & 39.63  & 40.76  & 41.05  & 37.71  & 37.41  & 38.98  & 38.04  & 39.09  & 39.22  & 38.87  & 43.25  & 34.92  & \multirow{2}[4]{*}{2.57} \\
\cmidrule{4-22}          &       &       & 0.936  & 0.933  & 0.925  & 0.931  & 0.940  & 0.943  & 0.959  & 0.952  & 0.965  & 0.964  & 0.940  & 0.956  & 0.947  & 0.939  & 0.952  & 0.954  & 0.959  & 0.979  & 0.941  &  \\
\cmidrule{3-23}          &       & \multirow{2}[4]{*}{GID\cite{xu2018external}} & 34.32  & 27.16  & 37.17  & 38.41  & 38.37  & 39.63  & 40.86  & 41.60  & 44.11  & 42.31  & 39.52  & 38.66  & 40.12  & 40.16  & 41.50  & 41.09  & 40.50  & 44.96  & 36.11  & \multirow{2}[4]{*}{49.99} \\
\cmidrule{4-22}          &       &       & 0.817  & 0.634  & 0.946  & 0.963  & 0.967  & 0.967  & 0.974  & 0.979  & 0.989  & 0.975  & 0.965  & 0.972  & 0.964  & 0.967  & 0.979  & 0.973  & 0.973  & 0.989  & 0.960  &  \\
\cmidrule{3-23}          &       & \multirow{2}[4]{*}{LSCD \cite{rizkinia2016local}} & -     & -     & 36.20  & 37.69  & 37.97  & 38.64  & 40.65  & 41.45  & -     & -     & 39.22  & -     & 39.84  & 39.53  & -     & 40.83  & -     & 44.34  & 35.86  & \multirow{2}[4]{*}{3.19} \\
\cmidrule{4-22}          &       &       & -     & -     & 0.942  & 0.963  & 0.964  & 0.962  & 0.973  & 0.979  & -     & -     & 0.963  & -     & 0.963  & 0.963  & -     & 0.973  & -     & 0.987  & 0.957  &  \\
\cmidrule{3-23}          &       & \multirow{2}[4]{*}{MCWNNM \cite{xu2017multi}} & 37.38$^\ast$  & -  & 37.02  & 38.54  & 38.26  & 39.89  & 41.47  & 42.07  & 44.22  & 42.48  & 39.46  & 38.78  & 39.87  & 40.18  & 41.33  & 41.74  & 40.71  & 45.38  & 35.84  & \multirow{2}[4]{*}{237.57} \\
\cmidrule{4-22}          &       &       & 0.929$^\ast$  & -  & 0.950  & 0.967  & 0.965  & 0.970  & 0.977  & 0.979  & 0.988  & 0.976  & 0.961  & 0.971  & 0.957  & 0.963  & 0.976  & 0.975  & 0.973  & 0.990  & 0.952  &  \\
\cmidrule{3-23}          &       & \multirow{2}[4]{*}{M-SVD} & 37.45  & 33.51  & 37.63  & 39.36  & 38.57  & 40.27  & 41.81  & 42.55  & 44.39  & 42.68  & 39.99  & 38.93  & 40.78  & 40.52  & 41.72  & 42.18  & 40.73  & 45.71  & 36.37  & \multirow{2}[4]{*}{104.24} \\
\cmidrule{4-22}          &       &       & 0.924  & 0.867  & 0.954  & 0.971  & 0.967  & 0.971  & 0.979  & 0.983  & 0.989  & 0.976  & 0.967  & 0.973  & 0.969  & 0.969  & 0.977  & 0.977  & 0.974  & 0.990  & 0.961  &  \\
\cmidrule{3-23}          &       & \multirow{2}[4]{*}{NLHCC\cite{hou2020nlh}} & 38.85  & 35.31  & 38.49  & 39.86  & 38.36  & 40.29  & 41.77  & 42.72  & 43.66  & {\textbf{42.80}} & 38.84  & 38.31  & 40.44  & 39.94  & 41.13  & {\textbf{43.25}} & -     & {\textbf{46.02}} & 35.73  & \multirow{2}[4]{*}{40.32} \\
\cmidrule{4-22}          &       &       & 0.953  & 0.930  & 0.965  & 0.976  & 0.965  & 0.971  & 0.979  & 0.984  & 0.986  & {\textbf{0.978}} & 0.959  & 0.968  & 0.964  & 0.963  & 0.976  & {\textbf{0.981}} & -     & {\textbf{0.991}} & 0.955  &  \\
\cmidrule{3-23}          &       & \multirow{2}[4]{*}{TWSC\cite{xu2018trilateral}} & 37.96$^\ast$  & -     & 37.90  & 39.66  & 38.62  & 40.62  & 41.65  & 42.52  & 44.94  & 42.26  & 38.71  & 38.81  & 38.77  & 40.12  & 41.71  & 42.23  & 40.65  & 45.49  & 35.40  & \multirow{2}[4]{*}{350.30} \\
\cmidrule{4-22}          &       &       & 0.942$^\ast$  & -     & 0.959  & 0.976  & 0.967  & 0.975  & 0.977  & 0.982  & 0.992  & 0.973  & 0.945  & 0.972  & 0.938  & 0.962  & 0.980  & 0.975  & 0.972  & 0.990  & 0.939  &  \\
\cmidrule{3-23}          &       & \multirow{2}[4]{*}{WTR1\cite{wu2018weighted}} & -     & -     & 37.80  & 37.32  & 38.20  & 38.19  & -     & -     & -     & -     & -     & -     & -     & -     & -     & -     & -     & -     & -     & \multirow{2}[4]{*}{600.33} \\
\cmidrule{4-22}          &       &       & -     & -     & 0.958  & 0.944  & 0.965  & 0.940  & -     & -     & -     & -     & -     & -     & -     & -     & -     & -     & -     & -     & -     &  \\
\cmidrule{2-23}          & \multirow{26}[20]{*}{Tensor} & \multirow{2}[4]{*}{4DHOSVD\cite{rajwade2012image}} & 37.58  & 34.49  & 37.52  & 39.15  & 38.54  & 40.27  & 41.40  & 42.19  & 44.49  & 42.60  & 39.82  & 38.88  & 40.59  & 40.37  & 41.75  & 41.82  & 40.71  & 45.58  & 36.27  & \multirow{2}[4]{*}{123.06} \\
\cmidrule{4-22}          &       &       & 0.929  & 0.911  & 0.956  & 0.973  & 0.968  & 0.973  & 0.977  & 0.982  & 0.990  & 0.976  & 0.966  & 0.973  & 0.965  & 0.967  & 0.980  & 0.977  & 0.974  & 0.990  & 0.961  &  \\
\cmidrule{3-23}          &       & \multirow{2}[4]{*}{CBM3D1\cite{dabov2007image}} & 38.04  & 35.00  & 37.58  & 39.21  & 38.60  & 40.11  & 41.58  & 42.40  & 44.43  & 42.53  & 39.93  & 38.84  & 40.56  & 40.41  & 41.93  & 42.29  & 40.62  & 45.58  & 36.24  & \multirow{2}[4]{*}{1.02} \\
\cmidrule{4-22}          &       &       & 0.938  & 0.925  & 0.955  & 0.971  & 0.969  & 0.972  & 0.977  & 0.982  & 0.990  & 0.976  & 0.966  & 0.973  & 0.967  & {\textbf{0.970}} & 0.982  & 0.979  & 0.973  & 0.990  & 0.959  &  \\
\cmidrule{3-23}          &       & \multirow{2}[4]{*}{CBM3D2\cite{dabov2007image}} & 37.73  & 34.74  & 37.70  & 39.41  & 38.69  & 40.35  & 41.69  & 42.54  & 44.74  & 42.65  & 39.97  & 38.97  & 40.77  & 40.55  & 42.03  & 42.20  & 40.75  & 45.72  & 36.38  & \multirow{2}[4]{*}{2.94} \\
\cmidrule{4-22}          &       &       & 0.934  & 0.922  & 0.957  & 0.975  & 0.970  & 0.974  & 0.978  & 0.984  & 0.992  & 0.977  & 0.967  & 0.974  & 0.967  & 0.969  & 0.983  & 0.979  & 0.974  & 0.990  & 0.961  &  \\
\cmidrule{3-23}          &       & \multirow{2}[4]{*}{CMSt-SVD\cite{kong2019color}} & 38.25  & 34.38  & 37.95  & 39.76  & 38.85  & 40.49  & 41.99  & 42.75  & 44.65  & 42.68  & 40.08  & 38.95  & {\textbf{40.84}} & 40.53  & 42.06  & 42.72  & {\textbf{40.88}} & 45.91  & 36.40  & \multirow{2}[4]{*}{4.95} \\
\cmidrule{4-22}          &       &       & 0.940  & 0.900  & 0.959  & 0.976  & 0.971  & 0.974  & 0.979  & 0.984  & 0.991  & 0.977  & 0.967  & 0.973  & 0.967  & 0.967  & 0.982  & 0.980  & {\textbf{0.974}} & 0.991  & 0.962  &  \\
\cmidrule{3-23}          &       & \multirow{2}[4]{*}{LLRT\cite{chang2017hyper}} & 35.45  & 30.74  & 37.77  & 39.76  & 38.28  & 39.59  & 41.60  & 42.24  & 42.68  & 42.22  & 37.91  & 38.80  & 38.01  & 39.87  & 42.02  & 41.76  & 38.69  & 45.17  & 35.71  & \multirow{2}[4]{*}{285.62} \\
\cmidrule{4-22}          &       &       & 0.897  & 0.766  & 0.957  & 0.977  & 0.970  & 0.972  & 0.977  & 0.983  & 0.992  & 0.975  & 0.969  & 0.973  & 0.965  & 0.971  & 0.984  & 0.979  & 0.972  & 0.989  & 0.962  &  \\
    \midrule
    \midrule
    \multirow{218}[148]{*}{\shortstack[c]{DNN \\ methods}} & \multirow{216}[124]{*}{Supervised} & \multirow{2}[4]{*}{AINDNet\cite{Kim_2020_CVPR}} & 39.77  & 39.08  & 36.14  & 37.19  & 37.33  & 38.00  & 38.54  & 39.33  & 39.49  & 38.50  & 36.53  & 36.24  & 36.93  & 36.83  & 37.27  & 38.11  & 37.44  & 40.17  & 34.65  & \multirow{2}[4]{*}{0.33} \\
\cmidrule{4-22}          &       &       & 0.959  & 0.953  & 0.935  & 0.949  & 0.954  & 0.946  & 0.975  & 0.976  & 0.979  & 0.966  & 0.954  & 0.965  & 0.958  & 0.953  & 0.964  & 0.961  & 0.968  & 0.981  & 0.953  &  \\
\cmidrule{3-23}          &       & \multirow{2}[4]{*}{BRDNet\cite{tian2020image}} & 33.80  & -     & 37.27  & 39.16  & 38.04  & 39.64  & 39.85  & 40.95  & 43.54  & 40.27  & 39.48  & 38.66  & 38.19  & 40.07  & 41.92  & 41.34  & 39.66  & 42.76  & 36.08  & \multirow{2}[4]{*}{14.66} \\
\cmidrule{4-22}          &       &       & 0.897  & -     & 0.953  & 0.974  & 0.960  & 0.969  & 0.965  & 0.974  & 0.989  & 0.958  & 0.958  & 0.973  & 0.943  & 0.969  & 0.981  & 0.972  & 0.965  & 0.979  & 0.953  &  \\
\cmidrule{3-23}          &       & \multirow{2}[4]{*}{CBDNet\cite{guo2019toward}} & 38.06  & 33.26  & 36.20  & 37.67  & 37.81  & 38.18  & 41.43  & 42.41  & 42.55  & 41.88  & 38.35  & 38.17  & 39.80  & 39.07  & 40.63  & 40.92  & 39.54  & 44.38  & 35.54  & \multirow{2}[4]{*}{0.04} \\
\cmidrule{4-22}          &       &       & 0.942  & 0.869  & 0.919  & 0.940  & 0.956  & 0.942  & 0.977  & 0.981  & 0.980  & 0.971  & 0.946  & 0.964  & 0.960  & 0.951  & 0.968  & 0.963  & 0.965  & 0.981  & 0.950  &  \\
\cmidrule{3-23}          &       & \multirow{2}[4]{*}{CycleISP\cite{zamir2020cycleisp}} & 39.57  & 39.42  & 35.40  & 36.87  & 37.61  & 37.70  & 41.26  & 41.84  & 42.64  & 41.59  & 38.47  & 38.01  & 38.79  & 38.67  & 40.61  & 40.03  & 39.86  & 44.75  & 35.54  & \multirow{2}[4]{*}{0.26} \\
\cmidrule{4-22}          &       &       & 0.955  & 0.956  & 0.916  & 0.939  & 0.955  & 0.936  & 0.977  & 0.977  & 0.978  & 0.969  & 0.952  & 0.966  & 0.961  & 0.952  & 0.970  & 0.953  & 0.968  & 0.987  & 0.949  &  \\
\cmidrule{3-23}          &       & \multirow{2}[4]{*}{DANet\cite{yue2020dual}} & 39.55  & 39.43  & 37.17  & 38.63  & {\textbf{41.49}} & 38.92  & 41.59  & 42.59  & 42.83  & 41.70  & 37.91  & 38.17  & 39.23  & 38.74  & 41.58  & 41.54  & 39.58  & 44.03  & 36.21  & \multirow{2}[4]{*}{0.02} \\
\cmidrule{4-22}          &       &       & 0.953  & 0.956  & 0.953  & 0.969  & {\textbf{0.977}} & 0.960  & 0.979  & 0.985  & 0.979  & 0.977  & 0.963  & 0.970  & 0.964  & 0.962  & 0.980  & 0.976  & 0.972  & 0.989  & 0.961  &  \\
\cmidrule{3-23}          &       & \multirow{2}[4]{*}{DBF\cite{chen2019real}} & -     & -     & 38.48  & 40.59  & 38.87  & 40.09  & 41.56  & 42.87  & 44.85  & 42.36  & 39.98  & 38.86  & 40.17  & 40.26  & 42.12  & 43.17  & 40.67  & 44.75  & 36.45  & \multirow{2}[4]{*}{0.34} \\
\cmidrule{4-22}          &       &       & -     & -     & 0.960  & 0.981  & 0.970  & 0.970  & 0.979  & 0.984  & 0.992  & 0.975  & 0.967  & 0.972  & 0.966  & 0.966  & 0.982  & 0.982  & 0.973  & 0.987  & 0.963  &  \\
\cmidrule{3-23}          &       & \multirow{2}[4]{*}{DCDicL\cite{zheng2021deep}} & 35.86  & -     & 36.61  & -     & 37.45  & 38.12  & 37.77  & 38.24  & 38.39  & 38.62  & 35.87  & 36.07  & 35.83  & 36.25  & 39.84  & 39.38  & 37.04  & 40.24  & 34.72  & \multirow{2}[4]{*}{1.24} \\
\cmidrule{4-22}          &       &       & 0.918  & -     & 0.945  & -     & 0.972  & 0.965  & 0.970  & 0.977  & 0.990  & 0.968  & 0.968  & 0.975  & 0.956  & 0.971  & 0.984  & 0.976  & 0.969  & 0.980  & 0.955  &  \\
\cmidrule{3-23}          &       & \multirow{2}[4]{*}{DeamNet\cite{ren2021adaptive}} & 39.63  & 39.35  & 36.63  & -     & 37.70  & 36.93  & 40.90  & 40.86  & 41.22  & 39.72  & 33.67  & 36.40  & 37.87  & 35.07  & 40.25  & 38.92  & 39.56  & 40.52  & 34.61  & \multirow{2}[4]{*}{0.18} \\
\cmidrule{4-22}          &       &       & 0.953  & 0.955  & 0.936  & -     & 0.958  & 0.944  & 0.977  & 0.978  & 0.979  & 0.967  & 0.953  & 0.964  & 0.961  & 0.948  & 0.966  & 0.957  & 0.969  & 0.980  & 0.952  &  \\
\cmidrule{3-23}          &       & \multirow{2}[4]{*}{DIDN\cite{yu2019deep}} & 39.64  & 39.78  & 36.06  & -     & 37.36  & 38.24  & 40.74  & 41.68  & 42.35  & 41.17  & 38.15  & 37.87  & 38.27  & 38.58  & 40.03  & 40.32  & 39.73  & 44.07  & 35.28  & \multirow{2}[4]{*}{7.25} \\
\cmidrule{4-22}          &       &       & 0.953  & 0.958  & 0.946  & -     & 0.953  & 0.950  & 0.975  & 0.977  & 0.975  & 0.966  & 0.951  & 0.965  & 0.954  & 0.950  & 0.963  & 0.959  & 0.966  & 0.985  & 0.949  &  \\
\cmidrule{3-23}          &       & \multirow{2}[4]{*}{DnCNN\cite{zhang2017beyond}} & 37.90$^\ast$  & 37.73$^\ast$  & 37.47  & 39.32  & 38.51  & 40.01  & 40.81  & 41.91  & 44.16  & 41.57  & 39.92  & 38.72  & 39.34  & 40.22  & 42.03  & 42.12  & 40.26  & 43.97  & 36.30  & \multirow{2}[4]{*}{3.23} \\
\cmidrule{4-22}          &       &       & 0.943$^\ast$  & 0.941$^\ast$  & 0.954  & 0.974  & 0.966  & 0.971  & 0.972  & 0.979  & 0.991  & 0.969  & 0.962  & 0.973  & 0.964  & 0.969  & 0.983  & 0.977  & 0.970  & 0.984  & 0.960  &  \\
\cmidrule{3-23}          &       & \multirow{2}[4]{*}{DRUNet\cite{zhang2020plug}} & -     & -     & 38.30  & 40.33  & 38.93  & 40.77  & 41.94  & 42.62  & {\textbf{45.36}} & 42.78  & {\textbf{40.35}} & {\textbf{39.17}} & 40.80  & {\textbf{40.80}} & {\textbf{42.86}} & 42.86  & {\textbf{40.88}} & 45.76  & 36.61  & \multirow{2}[4]{*}{0.17} \\
\cmidrule{4-22}          &       &       & -     & -    & 0.961  & 0.979  & 0.970  & 0.974  & 0.978  & 0.982  & {\textbf{0.993}} & 0.977  & {\textbf{0.969}} & {\textbf{0.975}} & 0.966  & {\textbf{0.970}}  & {\textbf{0.985}} & 0.980  & {\textbf{0.974}} & 0.990  & 0.961  &  \\
\cmidrule{3-23}          &       & \multirow{2}[4]{*}{DudeNet\cite{tian2021designing}} & 33.18  & 27.55  & 36.03  & 37.44  & 39.92  & 38.18  & 39.94  & 41.13  & 42.97  & 41.06  & 38.63  & 37.83  & 39.09  & 38.58  & 40.73  & 40.63  & 39.71  & 43.34  & 35.44  & \multirow{2}[4]{*}{0.18} \\
\cmidrule{4-22}          &       &       & 0.810  & 0.645  & 0.933  & 0.953  & 0.973  & 0.955  & 0.972  & 0.977  & 0.988  & 0.971  & 0.962  & 0.969  & 0.962  & 0.961  & 0.976  & 0.969  & 0.972  & 0.985  & 0.955  &  \\
\cmidrule{3-23}          &       & \multirow{2}[4]{*}{FCCF\cite{yue2019high}} & 36.30  & 23.32  & {\textbf{39.02}} & {\textbf{40.89}} & 38.87  & {\textbf{41.20}} & {\textbf{42.03}} & 43.07  & 45.15  & {\textbf{42.80}} & 40.14  & 39.04  & 40.61  & 40.66  & 42.31  & 43.05  & {\textbf{40.88}} & 45.47  & {\textbf{36.72}} & \multirow{2}[4]{*}{11.20} \\
\cmidrule{4-22}          &       &       & 0.885  & 0.461  & {\textbf{0.968}} & {\textbf{0.982}} & 0.971  & {\textbf{0.978}} & {\textbf{0.980}} & {\textbf{0.985}} & 0.993  & {\textbf{0.978}} & 0.966  & 0.974  & 0.967  & 0.969  & 0.984  & 0.982  & {\textbf{0.974}} & 0.989  & {\textbf{0.964}} &  \\
\cmidrule{3-23}          &       & \multirow{2}[4]{*}{FFDNet\cite{zhang2018ffdnet}} & 37.61$^\ast$  & 38.27$^\ast$  & 37.67  & 39.73  & 38.76  & 40.28  & 41.67  & 42.55  & 44.77  & 42.44  & 40.05  & 38.96  & 40.60  & 40.50  & 42.43  & 42.44  & 40.75  & 45.71  & 36.47  & \multirow{2}[4]{*}{0.02} \\
\cmidrule{4-22}          &       &       & 0.942$^\ast$  & 0.948$^\ast$  & 0.956  & 0.977  & 0.970  & 0.973  & 0.977  & 0.982  & 0.992  & 0.976  & 0.967  & 0.974  & 0.964  & 0.971  & 0.984  & 0.979  & 0.973  & 0.990  & 0.961  &  \\
\cmidrule{3-23}          &       & \multirow{2}[4]{*}{InvDN\cite{liu2021invertible}} & 39.57  & 39.28  & 34.55  & 36.33  & 35.94  & 38.01  & 38.95  & 40.09  & 42.03  & 40.02  & 33.32  & 37.00  & 36.64  & 37.66  & 40.06  & 40.98  & 39.10  & 40.74  & 34.15  & \multirow{2}[4]{*}{0.74} \\
\cmidrule{4-22}          &       &       & 0.952  & 0.955  & 0.937  & 0.953  & 0.947  & 0.952  & 0.965  & 0.973  & 0.978  & 0.960  & 0.930  & 0.965  & 0.939  & 0.955  & 0.966  & 0.971  & 0.964  & 0.969  & 0.935  &  \\
\cmidrule{3-23}          &       & \multirow{2}[4]{*}{IRCNN\cite{zhang2017learning}} & 34.78  & 33.50  & 37.26  & 37.25  & 37.49  & 39.66  & 41.24  & 41.85  & 43.89  & 41.78  & 39.91  & 38.63  & 40.14  & 40.12  & 41.87  & 41.86  & 40.26  & 44.38  & 36.22  & \multirow{2}[4]{*}{2.15} \\
\cmidrule{4-22}          &       &       & 0.907  & 0.887  & 0.954  & 0.956  & 0.956  & 0.970  & 0.978  & 0.981  & 0.991  & 0.973  & 0.969  & 0.972  & 0.965  & 0.969  & 0.983  & 0.977  & 0.972  & 0.988  & 0.961  &  \\
\cmidrule{3-23}          &       & \multirow{2}[4]{*}{MIRNet\cite{Zamir2020MIRNet}} & 39.88  & 39.55  & 36.06  & 37.25  & 37.49  & 38.10  & 40.72  & 41.68  & 42.76  & 41.05  & 38.28  & 37.80  & 38.60  & 38.55  & 40.28  & 40.46  & 39.73  & 43.66  & 35.42  & \multirow{2}[4]{*}{0.71} \\
\cmidrule{4-22}          &       &       & 0.956  & 0.957  & 0.942  & 0.956  & 0.956  & 0.952  & 0.975  & 0.978  & 0.980  & 0.968  & 0.954  & 0.967  & 0.959  & 0.952  & 0.966  & 0.964  & 0.967  & 0.983  & 0.953  &  \\
\cmidrule{3-23}          &       & \multirow{2}[4]{*}{MPRNet\cite{zamir2021multi}} & 39.82  & 39.57  & 35.94  & -     & 37.50  & 38.04  & 40.59  & 41.30  & 42.59  & 41.23  & 38.35  & 37.92  & 38.95  & 38.62  & 40.21  & 40.20  & 39.77  & 44.05  & 35.49  & \multirow{2}[4]{*}{35.97} \\
\cmidrule{4-22}          &       &       & 0.954  & 0.958  & 0.935  & -     & 0.954  & 0.946  & 0.974  & 0.974  & 0.978  & 0.969  & 0.952  & 0.966  & 0.960  & 0.949  & 0.965  & 0.960  & 0.967  & 0.986  & 0.951  &  \\
\cmidrule{3-23}          &       & \multirow{2}[4]{*}{NAFNet\cite{chen2022simple}} & 38.36  & {\textbf{40.15}} & 34.39  & -     & 36.38  & 37.88  & 40.12  & 40.23  & 40.66  & 40.01  & 36.13  & 35.87  & 36.66  & 36.79  & 36.53  & 39.93  & 39.32  & 40.45  & 34.82  & \multirow{2}[4]{*}{0.35} \\
\cmidrule{4-22}          &       &       & 0.943  & {\textbf{0.960}} & 0.923  & -     & 0.947  & 0.954  & 0.973  & 0.943  & 0.936  & 0.962  & 0.919  & 0.927  & 0.948  & 0.927  & 0.884  & 0.951  & 0.966  & 0.924  & 0.948  &  \\
\cmidrule{3-23}          &       & \multirow{2}[4]{*}{NBNet\cite{cheng2021nbnet}} & 39.89  & 39.64  & 35.89  & -     & 37.37  & 38.32  & 40.95  & 41.91  & 42.92  & 41.17  & 38.02  & 37.87  & 37.79  & 38.45  & 40.51  & 40.45  & 39.73  & 43.96  & 35.37  & \multirow{2}[4]{*}{0.12} \\
\cmidrule{4-22}          &       &       & 0.955  & 0.958  & 0.935  & -     & 0.957  & 0.955  & 0.977  & 0.980  & 0.982  & 0.971  & 0.955  & 0.967  & 0.956  & 0.955  & 0.972  & 0.967  & 0.969  & 0.984  & 0.953  &  \\
\cmidrule{3-23}          &       & \multirow{2}[4]{*}{NLNet\cite{lefkimmiatis2017non}} & -     & -     & 37.56  & 39.32  & 38.58  & 40.17  & 41.43  & 42.18  & 44.53  & 42.19  & 39.96  & 38.89  & 40.06  & 40.61  & 42.14  & 42.16  & 40.63  & 44.87  & 36.34  & \multirow{2}[4]{*}{39.06} \\
\cmidrule{4-22}          &       &       & -     & -     & 0.955  & 0.974  & 0.968  & 0.973  & 0.976  & 0.981  & 0.991  & 0.974  & {\textbf{0.969}} & 0.973  & 0.964  & {\textbf{0.971}} & 0.983  & 0.978  & 0.973  & 0.988  & 0.962  &  \\
\cmidrule{3-23}          &       & \multirow{2}[4]{*}{PD-Denoising\cite{zhou2020awgn}} & 38.40  & 33.99  & 35.85  & 37.13  & 37.13  & 37.12  & 40.85  & 41.38  & 41.13  & 41.05  & 37.72  & 37.57  & 39.03  & 38.01  & 39.18  & 39.64  & 38.80  & 43.73  & 35.05  & \multirow{2}[4]{*}{0.61} \\
\cmidrule{4-22}          &       &       & 0.943  & 0.896  & 0.924  & 0.944  & 0.944  & 0.921  & 0.974  & 0.974  & 0.966  & 0.959  & 0.935  & 0.956  & 0.948  & 0.929  & 0.948  & 0.946  & 0.955  & 0.981  & 0.940  &  \\
\cmidrule{3-23}          &       & \multirow{2}[4]{*}{PNGAN\cite{cai2021learning}} & 39.38  & 39.81  & 38.44  & -     & 39.57  & 40.51  & 41.79  & {\textbf{43.13}} & 44.73  & 42.55  & 39.84  & 38.78  & 40.63  & 40.20  & 41.85  & 43.15  & {\textbf{40.88}} & 45.32  & 36.40  & \multirow{2}[4]{*}{1.53} \\
\cmidrule{4-22}          &       &       & 0.953  & 0.959  & 0.963  & -     & 0.974  & 0.974  & 0.979  & {\textbf{0.985}} & 0.992  & 0.977  & 0.964  & 0.970  & {\textbf{0.969}} & 0.962  & 0.981  & 0.982  & \textbf{0.974} & 0.988  & 0.962  &  \\
\cmidrule{3-23}          &       & \multirow{2}[4]{*}{Restormer\cite{zamir2022restormer}} & {\textbf{40.03}} & 40.02  & 36.33  & -     & 37.66  & 38.29  & 41.10  & 41.84  & 42.49  & 41.47  & 38.42  & 38.08  & 39.05  & 38.77  & 40.13  & 40.53  & 39.56  & 44.19  & 35.65  & \multirow{2}[4]{*}{0.77} \\
\cmidrule{4-22}          &       &       & {\textbf{0.956}} & 0.960  & 0.941  & -     & 0.956  & 0.948  & 0.977  & 0.979  & 0.976  & 0.968  & 0.952  & 0.966  & 0.960  & 0.949  & 0.963  & 0.962  & 0.963  & 0.986  & 0.953  &  \\
\cmidrule{3-23}          &       & \multirow{2}[4]{*}{RIDNet\cite{anwar2019real}} & 39.26  & 38.70  & 36.84  & 38.11  & 38.57  & 38.60  & 41.09  & 42.15  & 42.39  & 41.59  & 38.88  & 38.21  & 38.67  & 39.02  & 40.91  & 41.01  & 40.07  & 44.40  & 35.76  & \multirow{2}[4]{*}{0.11} \\
\cmidrule{4-22}          &       &       & 0.953  & 0.950  & 0.941  & 0.961  & 0.970  & 0.956  & 0.977  & 0.980  & 0.989  & 0.975  & 0.962  & 0.970  & 0.963  & 0.959  & 0.978  & 0.977  & 0.972  & 0.987  & 0.957  &  \\
\cmidrule{3-23}          &       & \multirow{2}[4]{*}{SADNet\cite{chang2020spatial}} & 39.59$^\ast$  & -     & 37.76  & 38.11  & 39.22  & 38.90  & 40.82  & 42.02  & 44.25  & 41.85  & 39.53  & 38.33  & 39.83  & 39.69  & 39.81  & 41.64  & 40.70  & 44.61  & 36.36  & \multirow{2}[4]{*}{0.04} \\
\cmidrule{4-22}          &       &       & 0.952$^\ast$  & -     & 0.952  & 0.961  & 0.972  & 0.965  & 0.976  & 0.980  & 0.992  & 0.975  & 0.965  & 0.970  & 0.965  & 0.963  & 0.965  & 0.979  & {\textbf{0.974}} & 0.988  & 0.962  &  \\
\cmidrule{3-23}          &       & \multirow{2}[4]{*}{SCUNet\cite{zhang2022practical}} & -     & -     & 37.44  & 39.29  & 38.08  & 39.47  & 40.96  & 42.24  & 44.02  & 41.93  & 40.74  & 38.75  & 39.97  & 40.37  & 41.79  & 42.61  & 40.64  & 44.24  & 36.13  & \multirow{2}[4]{*}{0.62} \\
\cmidrule{4-22}          &       &       & -     & -     & 0.958  & 0.972  & 0.962  & 0.971  & 0.972  & 0.980  & 0.991  & 0.966  & 0.974  & 0.971  & 0.967  & 0.975  & 0.970  & 0.978  & 0.973  & 0.981  & 0.956  &  \\
\cmidrule{3-23}          &       & \multirow{2}[4]{*}{Uformer\cite{Wang_2022_CVPR}} & 40.05  & 39.74  & 35.97  & -     & 37.44  & 38.12  & 40.83  & 41.37  & 42.45  & 41.35  & 38.36  & 37.83  & 38.95  & 38.69  & 40.30  & 40.36  & 39.69  & 43.93  & 35.50  & \multirow{2}[4]{*}{1.46} \\
\cmidrule{4-22}          &       &       & 0.956  & 0.958  & 0.941  & -     & 0.956  & 0.950  & 0.976  & 0.977  & 0.978  & 0.968  & 0.954  & 0.966  & 0.960  & 0.953  & 0.966  & 0.964  & 0.967  & 0.985  & 0.953  &  \\
\cmidrule{3-23}          &       & \multirow{2}[4]{*}{UDNet\cite{lefkimmiatis2018universal}} & -     & -     & 36.95  & 38.96  & 38.05  & 39.62  & 40.40  & 41.36  & 43.58  & 41.12  & 39.35  & 38.25  & 38.94  & 39.95  & 41.32  & 41.70  & 39.51  & 43.17  & 36.09  & \multirow{2}[4]{*}{3.68} \\
\cmidrule{4-22}          &       &       & -     & -     & 0.943  & 0.971  & 0.962  & 0.967  & 0.968  & 0.977  & 0.989  & 0.968  & 0.963  & 0.971  & 0.959  & 0.968  & 0.979  & 0.974  & 0.966  & 0.981  & 0.958  &  \\
\cmidrule{3-23}          &       & \multirow{2}[4]{*}{VDIR\cite{soh2022variational}} & 39.63  & 39.26  & 34.90  & 36.51  & 37.01  & 36.84  & 40.84  & 41.40  & 41.29  & 40.81  & 37.65  & 37.84  & 39.09  & 38.17  & 39.20  & 39.41  & 38.98  & 44.40  & 35.04  & \multirow{2}[4]{*}{2.03} \\
\cmidrule{4-22}          &       &       & 0.953  & 0.955  & 0.894  & 0.925  & 0.940  & 0.911  & 0.973  & 0.970  & 0.965  & 0.953  & 0.936  & 0.960  & 0.951  & 0.931  & 0.946  & 0.938  & 0.956  & 0.985  & 0.939  &  \\
\cmidrule{3-23}          &       & \multirow{2}[4]{*}{VDNet\cite{yue2019variational}} & -     & -     & 35.86  & -     & 37.58  & 38.43  & 40.97  & 42.37  & 43.08  & 41.19  & 38.23  & 37.95  & 37.88  & 38.60  & 40.81  & 41.21  & 39.93  & 44.27  & 35.49  & \multirow{2}[4]{*}{0.05} \\
\cmidrule{4-22}          &       &       & -     & -     & 0.945  & -     & 0.957  & 0.955  & 0.976  & 0.983  & 0.983  & 0.970  & 0.956  & 0.968  & 0.956  & 0.956  & 0.973  & 0.970  & 0.970  & 0.984  & 0.953  &  \\
\cmidrule{2-23}          & \multirow{66}[40]{*}{Self-supervised} & \multirow{2}[4]{*}{AP-BSN\cite{lee2022ap}} & 37.29  & 35.97  & 35.44  & 36.76  & 36.99  & 38.26  & 40.23  & 41.29  & 42.72  & 41.40  & 37.87  & 37.62  & 37.47  & 38.29  & 40.83  & 40.86  & 39.55  & 43.73  & 34.37  & \multirow{2}[4]{*}{3.56} \\
\cmidrule{4-22}          &       &       & 0.932  & 0.925  & 0.936  & 0.956  & 0.956  & 0.965  & 0.973  & 0.979  & 0.986  & 0.970  & 0.949  & 0.966  & 0.939  & 0.960  & 0.977  & 0.973  & 0.969  & 0.983  & 0.937  &  \\
\cmidrule{3-23}          &       & \multirow{2}[4]{*}{Blind2Unblind\cite{wang2022blind2unblind}} & -     & -     & 36.51  & 38.18  & 38.25  & 39.03  & 40.71  & 41.72  & 43.11  & 41.17  & 39.35  & 38.27  & 39.39  & 39.04  & 40.88  & 40.90  & 39.76  & 43.78  & 35.75  & \multirow{2}[4]{*}{0.74} \\
\cmidrule{4-22}          &       &       & -     & -     & 0.935  & 0.958  & 0.968  & 0.962  & 0.974  & 0.981  & 0.990  & 0.970  & 0.967  & 0.971  & 0.962  & 0.958  & 0.976  & 0.970  & 0.971  & 0.986  & 0.956  &  \\
\cmidrule{3-23}          &       & \multirow{2}[4]{*}{C2N\cite{jang2021c2n}} & 37.28  & -     & 37.02  & -     & 37.69  & 38.86  & 41.12  & 41.95  & 42.61  & 41.33  & 38.73  & 38.15  & 38.95  & 38.67  & 40.50  & 40.78  & 40.05  & 44.64  & 35.53  & \multirow{2}[4]{*}{88.00} \\
\cmidrule{4-22}          &       &       & 0.924  & -     & 0.945  & -     & 0.958  & 0.960  & 0.976  & 0.980  & 0.980  & 0.966  & 0.954  & 0.965  & 0.956  & 0.945  & 0.977  & 0.964  & 0.970  & 0.987  & 0.952  &  \\
\cmidrule{3-23}          &       & \multirow{2}[4]{*}{CVF-SID\cite{neshatavar2022cvf}} & 36.31  & 34.43  & 29.13  & 31.02  & 33.08  & 33.35  & 36.68  & 39.08  & 39.73  & 36.40  & 33.13  & 33.16  & 32.07  & 33.33  & 36.92  & 38.40  & 35.22  & 38.56  & 29.89  & \multirow{2}[4]{*}{0.12} \\
\cmidrule{4-22}          &       &       & 0.923  & 0.912  & 0.853  & 0.898  & 0.912  & 0.929  & 0.933  & 0.974  & 0.976  & 0.929  & 0.895  & 0.920  & 0.857  & 0.917  & 0.946  & 0.961  & 0.927  & 0.952  & 0.849  &  \\
\cmidrule{3-23}          &       & \multirow{2}[4]{*}{IDR\cite{zhang2022idr}} & -     & -     & 34.26  & 36.18  & 36.28  & 36.44  & 39.33  & 40.31  & 40.94  & 39.74  & 37.38  & 37.39  & 38.30  & 38.15  & 39.68  & 38.22  & 38.01  & 43.15  & 34.64  & \multirow{2}[4]{*}{0.32} \\
\cmidrule{4-22}          &       &       & -     & -     & 0.877  & 0.923  & 0.925  & 0.910  & 0.955  & 0.962  & 0.961  & 0.935  & 0.927  & 0.957  & 0.938  & 0.944  & 0.951  & 0.913  & 0.941  & 0.978  & 0.929  &  \\
\cmidrule{3-23}          &       & \multirow{2}[4]{*}{LGBPN\cite{wang2023lgbpn}} & 38.43$^\ast$   & 37.28$^\ast$   & -  & -  & -    & -  & -     & -     & -     & -     & -     & -     & -     & -     & -     & -     & -     & -     & -     & \multirow{2}[4]{*}{-} \\
\cmidrule{4-22}          &       &       & 0.942$^\ast$     & 0.936$^\ast$     & -  & -  & -     & -  & -     & -     & -     & -     & -     & -     & -     & -     & -     & -     & -     & -     & -     &  \\
\cmidrule{3-23}          &       & \multirow{2}[4]{*}{Neigh2Neigh\cite{huang2021neighbor2neighbor}} & -     & -     & 34.47  & 35.65  & 37.10  & 36.59  & 40.22  & 40.41  & 41.29  & 41.24  & 38.13  & 37.94  & 39.71  & 38.50  & 39.63  & 38.05  & 39.17  & 44.49  & 35.30  & \multirow{2}[4]{*}{2.48} \\
\cmidrule{4-22}          &       &       & -     & -     & 0.883  & 0.907  & 0.942  & 0.908  & 0.967  & 0.960  & 0.966  & 0.958  & 0.940  & 0.961  & 0.956  & 0.936  & 0.952  & 0.919  & 0.958  & 0.984  & 0.944  &  \\
\cmidrule{3-23}          &       & \multirow{2}[4]{*}{Noise2Noise\cite{lehtinen2018noise2noise}} & -     & -     & 33.55  & 34.67  & 35.91  & 35.28  & 38.88  & 39.07  & 39.34  & 39.26  & 36.81  & 36.99  & 38.04  & 37.21  & 38.00  & 36.35  & 37.67  & 42.59  & 34.44  & \multirow{2}[4]{*}{0.03} \\
\cmidrule{4-22}          &       &       & -     & -     & 0.850  & 0.874  & 0.913  & 0.875  & 0.950  & 0.940  & 0.938  & 0.931  & 0.919  & 0.948  & 0.936  & 0.916  & 0.924  & 0.876  & 0.936  & 0.975  & 0.926  &  \\
\cmidrule{3-23}          &       & \multirow{2}[4]{*}{R2R\cite{pang2021recorrupted}} & -     & 34.78$^\ast$     & 36.32  & 38.11  & 38.47$^\ast$  & 38.71  & -     & -     & -     & -     & -     & -     & -     & -     & -     & -     & -     & -     & -     & \multirow{2}[4]{*}{4481.98} \\
\cmidrule{4-22}          &       &       & -     & 0.844$^\ast$     & 0.937  & 0.961  & 0.965$^\ast$  & 0.958  & -     & -     & -     & -     & -     & -     & -     & -     & -     & -     & -     & -     & -     &  \\
\cmidrule{3-23}          &       & \multirow{2}[4]{*}{SASL\cite{li2023spatially}} & 38.00  & 37.41     & 34.93  & -     & 37.13  & 38.24  & 40.04  & 41.86  & 43.07  & 40.29  & 37.22  & 37.38  & 36.45  & 37.64  & 40.85  & 41.64  & 39.10  & 42.89  & 34.25  & \multirow{2}[4]{*}{0.06} \\
\cmidrule{4-22}          &       &       & 0.936  & 0.934     & 0.936  & -     & 0.954  & 0.964  & 0.966  & 0.982  & 0.987  & 0.963  & 0.947  & 0.965  & 0.933  & 0.958  & 0.977  & 0.976  & 0.965  & 0.980  & 0.930  &  \\
\cmidrule{3-23}          &       & \multirow{2}[4]{*}{Self2Self\cite{quan2020self2self}} & -     & -     & 36.26  & 38.23  & -     & 39.49  & -     & -     & -     & -     & -     & -     & -     & -     & -     & -     & -     & -     & -     & \multirow{2}[4]{*}{4100.76} \\
\cmidrule{4-22}          &       &       & -     & -     & 0.947  & 0.969  & -     & 0.963  & -     & -     & -     & -     & -     & -     & -     & -     & -     & -     & -     & -     & -     &  \\
    \bottomrule
    \end{tabular}}%
  \label{Table_Results_Color_image_all}%
\end{table*}%

\vspace{-0.2cm}
\subsubsection{Objective Results}
Detailed denoising results are presented in Table \ref{Table_Results_Color_image_all}. Briefly, it is observed that many DNN methods show outstanding performance on both DnD and SIDD datasets, but they do not always demonstrate evident advantages over traditional denoisers on other datasets when training or validation data are not available. \\
\indent More specifically, for traditional denoisers, three effective transform-domain approaches namely CBM3D, CMSt-SVD, and NLHCC show comparable denoising performance on almost all datasets. The effectiveness of CBM3D1 and CMSt-SVD indicates that a one-step implementation with a small number of local patches and the hard-thresholding technique is able to produce very competitive results. In addition, it is noticed that the matrix-based methods such as MCWNNM and the proposed M-SVD present similar denoising capability with several tensor-based methods such as LLRT and 4DHOSVD, which indicates that the use of tensor representation may not significantly boost structural information retrieval in realistic cases. \\
\indent For DNN methods, we can see that with the aid of training or validation process, models targeting real-world denoising such as AINDNet, DIDN, NAFNet and Restormer produce much better performance on SIDD and DND datasets. For example, compared to CBM3D, Restormer provides the PSNR improvements of 1.99 dB on SIDD and 4.79 dB on DnD, respectively. However, once the testing data is no longer compatible with the training conditions, they can exhibit poor generalization and lead to overfitting or degrading performance. As a result, none of the pretrained models can significantly advance benchmark traditional denoisers on other real-world datasets. Nevertheless, we also notice that several network frameworks such as DBF, DIDN, DRUNet, FCCF and PNGAN show competitive and robust performance in the absence of training data. Furthermore, it is noteworthy that DBF, DRUNet and FCCF utilize Gaussian noise modeling and denoisers, which supports the effectiveness of incorporating Gaussian noise modeling. \\
\indent From the perspective of denoising speed, DNN methods are normally much faster than traditional denoisers in the test phase thanks to the power of advanced GPU devices. Specifically, efficient DNN models are able to handle images of size $512 \times 512 \times 3$ within 0.2 seconds. We also notice that representative self-supervised DNN methods such as Self2Self and R2R bear high computational burden, since for each noisy input, the training process involves thousands of iterations. For traditional denoisers, the time complexity lies mainly in the iterative local patch search and learning. For example, M-SVD spends 26 and 42 seconds on grouping and performing local SVD transforms, respectively. But it is slightly faster than its tensor counterpart 4DHOSVD, because it avoids folding and unfolding operations of high dimensional data along different modes. Among all the traditional denoisers, the state-of-the-art CBM3D is the most efficient because its grouping step is performed only on the luminance channel, and it does not need to train local transforms or solve optimization problems.
\vspace{-6pt}
\subsubsection{Visual Evaluation}
\begin{figure*}[htbp]
\graphicspath{{Figs/Figs_Image(sRGB)/result_images_sample/selected_for_comparison/Combined/DND_plus_SIDD/}}
\centering
\subfigure[Noisy]{
\label{Fig4}
\includegraphics[width=0.728in]{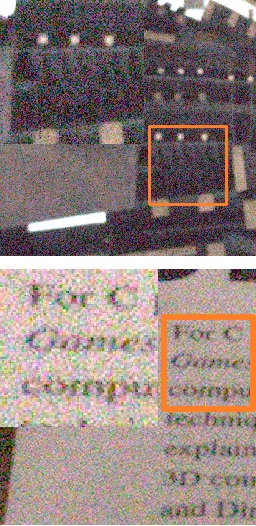}}
\subfigure[Bitonic]{
\label{Fig4}
\includegraphics[width=0.728in]{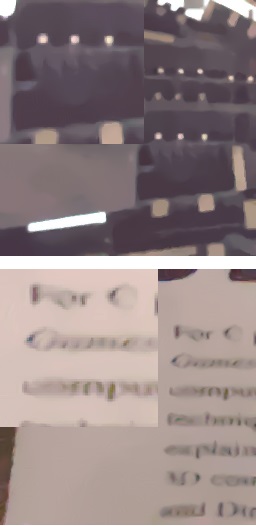}}
\subfigure[CBM3D2]{
\label{Fig4}
\includegraphics[width=0.728in]{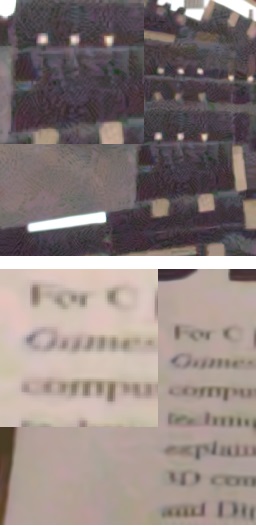}}
\subfigure[CMSt-SVD]{
\label{Fig4}
\includegraphics[width=0.728in]{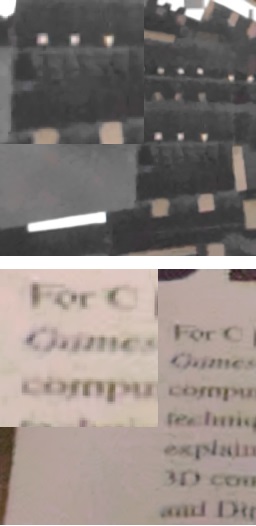}}
\subfigure[NLHCC]{
\label{Fig4}
\includegraphics[width=0.728in]{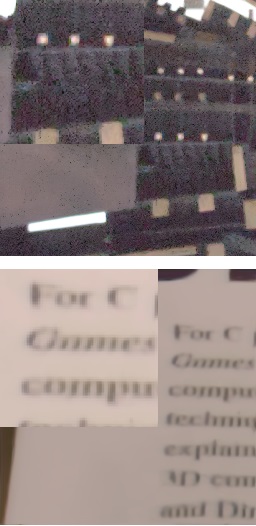}}
\subfigure[AINDNet]{
\label{Fig4}
\includegraphics[width=0.728in]{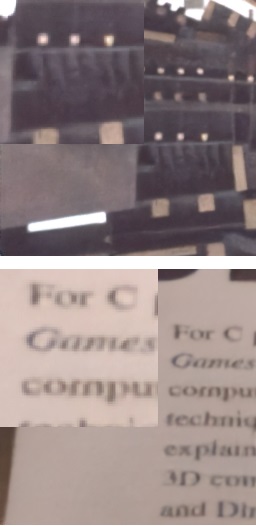}}
\subfigure[NAFNet]{
\label{Fig4}
\includegraphics[width=0.728in]{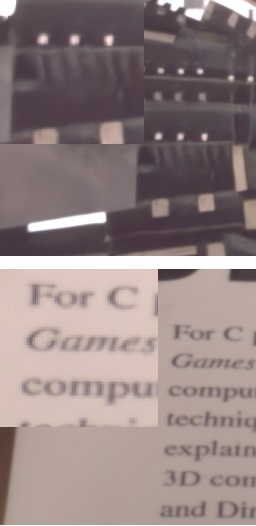}}
\subfigure[PNGAN]{
\label{Fig4}
\includegraphics[width=0.728in]{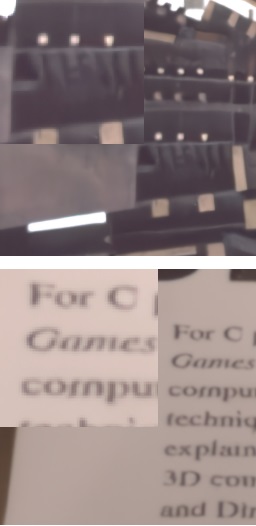}}
\subfigure[Restormer]{
\label{Fig4}
\includegraphics[width=0.728in]{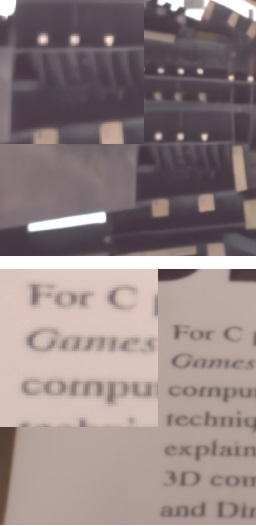}}

\caption{Visual evaluation of compared methods on DnD (first row) and SIDD (second row) datsets.}
\label{Fig_DND_plus_SIDD}
\end{figure*}

\begin{figure*}[htbp]
\graphicspath{{Figs/Figs_Image(sRGB)/result_images_sample/selected_for_comparison/Combined/PolyU/}}
\centering
\subfigure[Clean]{
\label{Fig4}
\includegraphics[width=0.829in]{Clean_combined_marked}}
\subfigure[Noisy]{
\label{Fig4}
\includegraphics[width=0.829in]{Noisy_combined}}
\subfigure[Bitonic]{
\label{Fig4}
\includegraphics[width=0.829in]{Bitonic_combined}}
\subfigure[CBM3D2]{
\label{Fig4}
\includegraphics[width=0.829in]{CBM3D2_combined}}
\subfigure[CMSt-SVD]{
\label{Fig4}
\includegraphics[width=0.829in]{CMSt-SVD_combined}}
\subfigure[MCWNNM]{
\label{Fig4}
\includegraphics[width=0.829in]{MCWNNM_combined}}
\subfigure[NLHCC]{
\label{Fig4}
\includegraphics[width=0.829in]{NLHCC_combined}}
\subfigure[TWSC]{
\label{Fig4}
\includegraphics[width=0.829in]{TWSC_combined}}\\

\subfigure[AINDNet]{
\label{Fig4}
\includegraphics[width=0.829in]{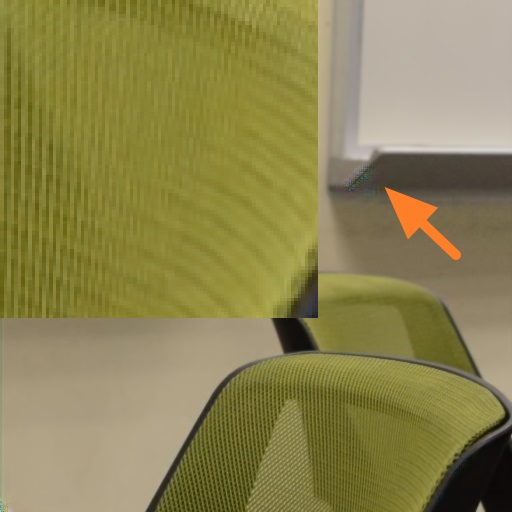}}
\subfigure[DIDN]{
\label{Fig4}
\includegraphics[width=0.829in]{DIDN_combined}}
\subfigure[DRUNet]{
\label{Fig4}
\includegraphics[width=0.829in]{DRUNet_combined}}
\subfigure[FCCF]{
\label{Fig4}
\includegraphics[width=0.829in]{FCCF_combined}}
\subfigure[NAFNet]{
\label{Fig4}
\includegraphics[width=0.829in]{NAFNet_combined}}
\subfigure[PNGAN]{
\label{Fig4}
\includegraphics[width=0.829in]{PNGAN_combined}}
\subfigure[Restormer]{
\label{Fig4}
\includegraphics[width=0.829in]{Restormer_combined}}
\subfigure[VDIR]{
\label{Fig4}
\includegraphics[width=0.829in]{VDIR_combined}}

\caption{Visual evaluation of compared methods on the real-world PolyU dataset. The camera device is NIKON D800.}
\label{Fig_PolyU_new}
\end{figure*}

\begin{figure*}[htbp]
\graphicspath{{Figs/Figs_Image(sRGB)/result_images_sample/selected_for_comparison/Combined/My_dataset/IPHONE13/}}
\centering
\subfigure[Clean]{
\label{Fig4}
\includegraphics[width=0.829in]{Clean_combined_marked}}
\subfigure[Noisy]{
\label{Fig4}
\includegraphics[width=0.829in]{Noisy_combined}}
\subfigure[Bitonic]{
\label{Fig4}
\includegraphics[width=0.829in]{Bitonic_combined}}
\subfigure[CBM3D2]{
\label{Fig4}
\includegraphics[width=0.829in]{CBM3D2_combined}}
\subfigure[CMSt-SVD]{
\label{Fig4}
\includegraphics[width=0.829in]{CMSt-SVD_combined}}
\subfigure[MCWNNM]{
\label{Fig4}
\includegraphics[width=0.829in]{MCWNNM_combined}}
\subfigure[NLHCC]{
\label{Fig4}
\includegraphics[width=0.829in]{NLHCC_combined}}
\subfigure[TWSC]{
\label{Fig4}
\includegraphics[width=0.829in]{TWSC_combined}}\\

\subfigure[AINDNet]{
\label{Fig4}
\includegraphics[width=0.829in]{AINDNet_combined}}
\subfigure[DIDN]{
\label{Fig4}
\includegraphics[width=0.829in]{DIDN_combined}}
\subfigure[DRUNet]{
\label{Fig4}
\includegraphics[width=0.829in]{DRUNet_combined}}
\subfigure[FCCF]{
\label{Fig4}
\includegraphics[width=0.829in]{FCCF_combined}}
\subfigure[NAFNet]{
\label{Fig4}
\includegraphics[width=0.829in]{NAFNet_combined}}
\subfigure[PNGAN]{
\label{Fig4}
\includegraphics[width=0.829in]{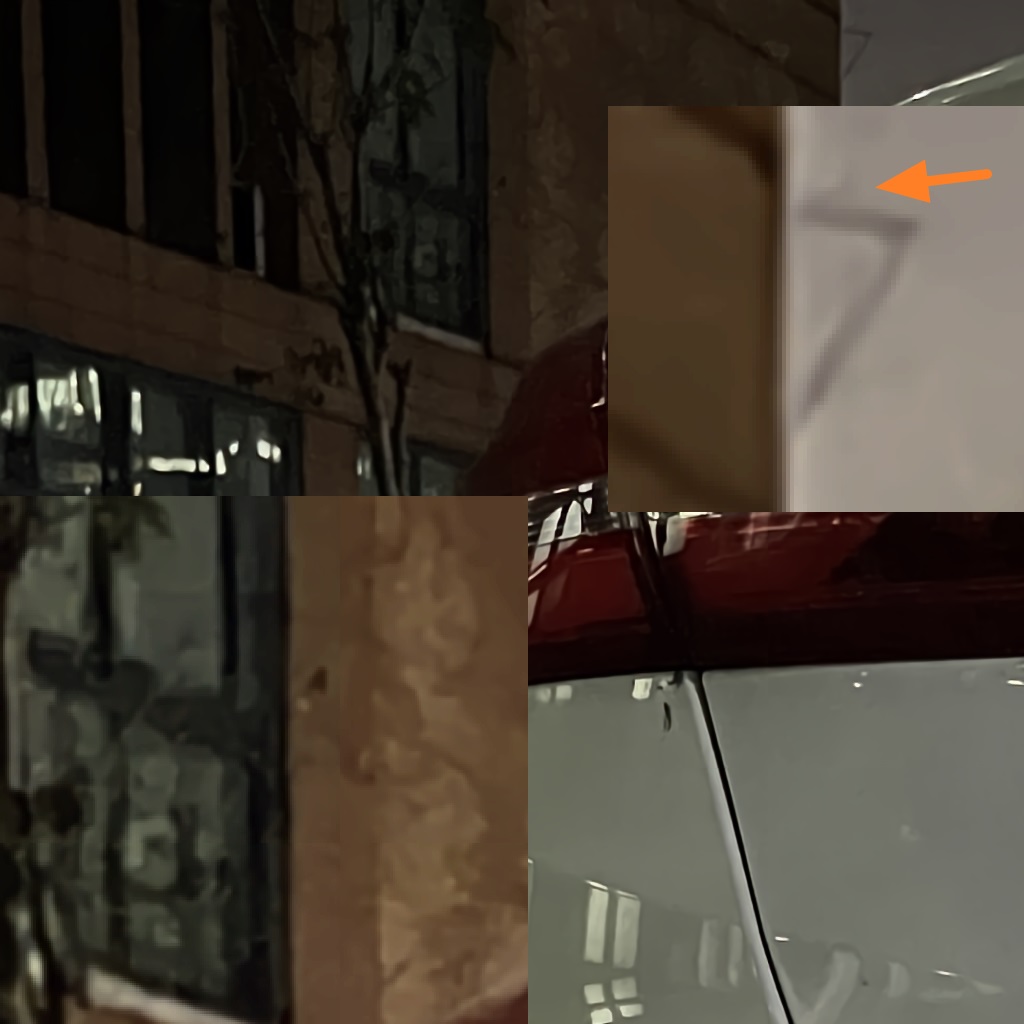}}
\subfigure[Restormer]{
\label{Fig4}
\includegraphics[width=0.829in]{Restormer_combined}}
\subfigure[VDIR]{
\label{Fig4}
\includegraphics[width=0.829in]{VDIR_combined}}

\caption{Visual evaluation of compared methods on the real-world IOCI dataset. The camera device is IPHONE 13.}
\label{Fig_My_dataset_IPHONE13}
\end{figure*}

Denoised results of compared methods on the DND and SIDD datasets are illustrated in Fig. \ref{Fig_DND_plus_SIDD}. The drawback of traditional patch-based denoisers is obvious when dealing with severely corrupted images, since the patch search and local transform learning steps are adversely affected by the presence of noise. Consequently, it is not difficult to see the distortion, artifacts, loss of true color and details. By comparison, fine-tuned supervised DNN models show clear advantages in terms of both noise removal and detail recovery, which demonstrates their powerful feature learning and extraction capability. Interestingly, the difference among the sophisticated and well-trained networks is barely noticeable in many cases of the two benchmark datasets. Therefore, we present visual evaluations of the PolyU and the IOCI datasets in Fig. \ref{Fig_PolyU_new} and Fig. \ref{Fig_My_dataset_IPHONE13}, respectively. When facing different and unseen noise patterns, the pretrained DNN models may also leave unwanted artifacts and produce over-smooth effects to varying degrees, while traditional denoisers show their strengths by exploiting nonlocal information of the noisy image, which renders certain robustness and adaptability. Nevertheless, we notice that several DNN models such as DRUNet, FCCF and PNGAN show impressive generalizability and achieve good balance between noise removal and detail preservation. The observation is consistent with results reported in Table \ref{Table_Results_Color_image_all}.

\vspace{-0.269cm}
\subsection{Results for Video Datasets}
\subsubsection{Experimental Settings}
Compared to images, videos are more engaging and informative by recording and displaying dynamic objects. In our evaluations, four benchmark datasets are included for video denoising. Briefly, DAVIS-2017-test-dev-480p and Set8 are used for synthetic experiments with Gaussian noise, and the entire CRVD and IOCV for real-world experiments. Our evaluations mainly focus on video denoising in the sRGB space, and the parameters and models of compared methods are carefully chosen in the same way as section \ref{color_image_settings}. Similar to \cite{Tassano_2020_CVPR}, the PSNR and SSIM of a sequence are computed as their average values of each frame.
\vspace{-0.2cm}
\subsubsection{Objective Results}
Table \ref{Table_Results_Color_video_all} lists the average PSNR and SSIM results of compared methods. For traditional denoisers, it is noticed
that CVMSt-SVD and VBM4D are able to achieve state-of-the-art PSNR results on the CRVD and IOCV datasets, respectively. This demonstrates the effectiveness of patch-based paradigm on capturing NLSS features among different frames. For DNN methods, representative models such as FastDVDNet, FloRNN, RVRT and VNLNet show dominating performance in synthetic cases, especially when dealing with severe Gaussian noise, which manifests their ability to extract deep features and exploit spatio-temporal information. When it comes to real-world experiments, although DNN methods are no longer significantly superior in the absence of corresponding training data, certain Gaussian-based models exhibit impressive generalization ability. For experiments with the CRVD dataset, by fusing information from multiple frames and benefiting from recurrent design, RVRT and FloRNN provide the PSNR and SSIM improvements of 0.28dB and 0.0142, respectively. Besides, by integrating the NLSS prior into CNN models, VNLNet produces state-of-the-art results on the IOCV dataset. \\
\indent In terms of denoising efficiency, FastDVDNet achieves almost real-time noise removal by getting rid of the time-consuming patch search and explicit flow estimation steps. More specifically, it takes FastDVDNet less than 0.1 second to process a single video frame of size $960 \times 540\times 3$, which is 8 times faster than FloRNN, and at least 100 times faster than the benchmark traditional denoisers such as VBM4D and CVMSt-SVD. The denoising speed of FastDVDNet is remarkable considering its competitive performance in different cases, making it an exceedingly appealing denoising algorithm in handling high definition videos.

\begin{table*}[htbp]
\tiny
  \centering
  \caption{Average PSNR and SSIM values of compared methods on four video datasets (sRGB space). The average computational time (s) is calculated based on the Set8 dataset with noise level $\sigma = 50$. `-' means the results are not available due to high computational cost or other issues.}
  \renewcommand{\arraystretch}{0.668}
  \scalebox{0.938}{
    \begin{tabular}{ccccccccccccccccc}
    \toprule
    \multirow{3}[4]{*}{Dataset} & \multirow{3}[4]{*}{$\sigma$} & \multicolumn{6}{c|}{Traditional denoisers}   & \multicolumn{9}{c}{DNN methods} \\
    \cmidrule{3-17}          &       & CVMSt-SVD & RTA-LSM & VBM4D1 & VBM4D2 & VIDOSAT & DVDNet & FastDVDNet & FloRNN & MAP-VDNet & MMNet & RVRT  & RFR   & UDVD  & ViDeNN & VNLNet \\
              &     &  \cite{kong2019color}   &   \cite{dong2018robust}  &     \cite{maggioni2012video}   &   \cite{maggioni2012video}    &    \cite{wen2018vidosat}   &  \cite{tassano2019dvdnet}   &  \cite{Tassano_2020_CVPR}   &  \cite{li2022unidirectional}     &  \cite{sun2021deep}     &    \cite{chen2021multiframe}   &  \cite{liang2022recurrent} &  \cite{lee2021restore}   &    \cite{sheth2021unsupervised}   &  \cite{claus2019videnn}    &  \cite{davy2019non}  \\
    \midrule
    \multirow{10}[20]{*}{Set8} & \multirow{2}[4]{*}{10} & 36.25  & 36.36  & 35.69  & 35.96  & 34.75  & 36.04  & 36.38  & 37.55  & -     & 37.28  & {\textbf{37.58}} & -  & 32.30  & 34.80  & 37.28  \\
\cmidrule{3-17}          &       & 0.9453  & 0.9474  & 0.9342  & 0.9405  & 0.9244  & 0.9487  & 0.9513  & 0.9617  & -     & 0.9595  & {\textbf{0.9626}} & - & 0.9081  & 0.9114  & 0.9584  \\
\cmidrule{2-17}          & \multirow{2}[4]{*}{20} & 32.50  & 32.87  & 31.56  & 32.10  & 31.12  & 33.42  & 33.35  & 34.65  & -     & 34.11  & {\textbf{34.85}} & 30.95  & 32.24  & 28.34  & 34.02  \\
\cmidrule{3-17}          &       & 0.8891  & 0.8977  & 0.8485  & 0.8731  & 0.8571  & 0.9151  & 0.9159  & 0.9347  & -     & 0.9270  & {\textbf{0.9383}} & 0.8321  & 0.8937  & 0.7619  & 0.9243  \\
\cmidrule{2-17}          & \multirow{2}[4]{*}{30} & 30.38  & 30.84  & 29.20  & 29.90  & -     & 31.70  & 31.59  & 32.94  & -     & 32.27  & {\textbf{33.31}} & 30.18  & 31.02  & 23.07  & - \\
\cmidrule{3-17}          &       & 0.8369  & 0.8542  & 0.7626  & 0.8081  & -     & 0.8825  & 0.8848  & 0.9100  & -     & 0.8973  & {\textbf{0.9173}} & 0.8299  & 0.8653  & 0.5719  & - \\
\cmidrule{2-17}          & \multirow{2}[4]{*}{40} & 28.92  & 29.41  & 27.56  & 28.37  & 27.67  & 30.45  & 30.36  & 31.71  & -     & 30.98  & {\textbf{32.21}} & 28.89  & 29.46  & 18.89  & 30.72  \\
\cmidrule{3-17}          &       & 0.7871  & 0.8170  & 0.6820  & 0.7482  & 0.7526  & 0.8528  & 0.8569  & 0.8869  & -     & 0.8702  & {\textbf{0.8981}} & 0.7891  & 0.8221  & 0.4070  & 0.8596  \\
\cmidrule{2-17}          & \multirow{2}[4]{*}{50} & 27.80  & 28.34  & 26.29  & 27.20  & -     & 29.45  & 29.42  & 30.75  & 26.48  & 29.99  & {\textbf{31.32}} & 26.57  & 27.89  & 15.42  & - \\
\cmidrule{3-17}          &       & 0.7392  & 0.7860  & 0.6086  & 0.6935  & -     & 0.8264  & 0.8317  & 0.8653  & 0.7220  & 0.8452  & {\textbf{0.8800}} & 0.6461  & 0.7654  & 0.2379  & - \\
    \midrule
    \multirow{10}[20]{*}{DAVIS} & \multirow{2}[4]{*}{10} & 38.49  & -     & 37.57  & 37.97  & -     & 37.70  & 38.33  & 38.92  & -     & 38.25  & {\textbf{39.07}} & -  & 33.65  & 35.90  & 38.68  \\
\cmidrule{3-17}          &       & 0.9611  & -     & 0.9475  & 0.9555  & -     & 0.9600  & 0.9631  & 0.9680  & -     & 0.9639  & {\textbf{0.9691}} & -  & 0.9204  & 0.9253  & 0.9653  \\
\cmidrule{2-17}          & \multirow{2}[4]{*}{20} & 34.73  & -     & 33.33  & 34.03  & -     & 35.34  & 35.52  & 36.43  & -     & 35.63  & {\textbf{36.69}} & 33.09  & 33.95  & 30.20  & 35.84  \\
\cmidrule{3-17}          &       & 0.9147  & -     & 0.8709  & 0.9000  & -     & 0.9361  & 0.9364  & 0.9493  & -     & 0.9384  & {\textbf{0.9520}} & 0.8765  & 0.9154  & 0.8124  & 0.9398  \\
\cmidrule{2-17}          & \multirow{2}[4]{*}{30} & 32.53  & -     & 30.82  & 31.71  & -     & 33.84  & 33.88  & 35.06  & -     & 34.06  & {\textbf{35.47}} & 31.93  & 32.85  & 24.75  & - \\
\cmidrule{3-17}          &       & 0.8665  & -     & 0.7866  & 0.8415  & -     & 0.9140  & 0.9134  & 0.9341  & -     & 0.9166  & {\textbf{0.9394}} & 0.8564  & 0.8918  & 0.6586  & - \\
\cmidrule{2-17}          & \multirow{2}[4]{*}{40} & 30.96  & -     & 29.02  & 30.06  & -     & 32.70  & 32.71  &       & -     & 32.93  & {\textbf{34.58}} & 30.22  & 31.29  & 20.60  & 33.01  \\
\cmidrule{3-17}          &       & 0.8182  & -     & 0.7036  & 0.7838  & -     & 0.8931  & 0.8926  &       & -     & 0.8969  & {\textbf{0.9280}} & 0.7997  & 0.8545  & 0.5121  & 0.8959  \\
\cmidrule{2-17}          & \multirow{2}[4]{*}{50} & 29.72  & -     & 27.61  & 28.78  & -     & 31.76  & 31.79  & 33.18  & 28.35  & 32.04  & {\textbf{33.85}} & 27.22  & 29.65  & 17.01  & - \\
\cmidrule{3-17}          &       & 0.7701  & -     & 0.6270  & 0.7287  & -     & 0.8730  & 0.8735  & 0.9054  & 0.7911  & 0.8788  & {\textbf{0.9170}} & 0.6350  & 0.8033  & 0.3429  & - \\
    \midrule
    \multirow{2}[4]{*}{CRVD} & \multirow{2}[4]{*}{-} & 36.66  & -     & 34.66  & 34.14  & 34.16  & 34.50  & 35.84  & 36.66  & -  & -  & {\textbf{36.94}} & 31.30  & -     & 32.31  & 36.11  \\
\cmidrule{3-17}          &       & 0.9463  & -     & 0.9224  & 0.9079  & 0.9384  & 0.9493  & 0.9306  &  {\textbf{0.9605}} & -  & -  & 0.9559  & 0.7785  & -     & 0.8449  & 0.9449  \\
    \midrule
    \multirow{2}[4]{*}{IOCV} & \multirow{2}[4]{*}{-} & 38.22  & -     & 38.65  & {\textbf{38.76}} & -     & 38.53  & 37.57  & 38.64  & 35.52  & -  & 38.50  & 31.46  & 35.02  & 36.13  & {\textbf{38.76}} \\
\cmidrule{3-17}          &       & 0.9736  & -     & 0.9763  & {\textbf{0.9765}} & -     & 0.9754  & 0.9699  & 0.9743  & 0.9313  & -  & 0.9672  & 0.8346  & 0.9660  & 0.9506  & {\textbf{0.9765}} \\
    \midrule
    Implementation & -     & MEX & MATLAB &  MEX &  MEX & C++   & Python & Python & Python & Python & Python & Python & Python & Python & Python & Python
 \\
    \midrule
    Time (s) & -     & 888.0  & $>1$h  & 920.9  & 2076.0  & 4313.7  & 818.8  & {\textbf{6.6}} & 53.9  & 77.2  & 30.1  & 93.3  & 37.8  & 163.4  & 19.0  & 217.1  \\
    \bottomrule
    \end{tabular}}%
  \label{Table_Results_Color_video_all}%
\end{table*}%

\vspace{-0.1cm}
\subsubsection{Visual Evaluation}
\begin{figure*}[htbp]
\graphicspath{{Figs/Figs_Video/result_images_sample/selected_for_comparison/Combined/CRVD_plus_Set8/}}
\centering
\subfigure[Clean]{
\label{Fig4}
\includegraphics[width=0.828in]{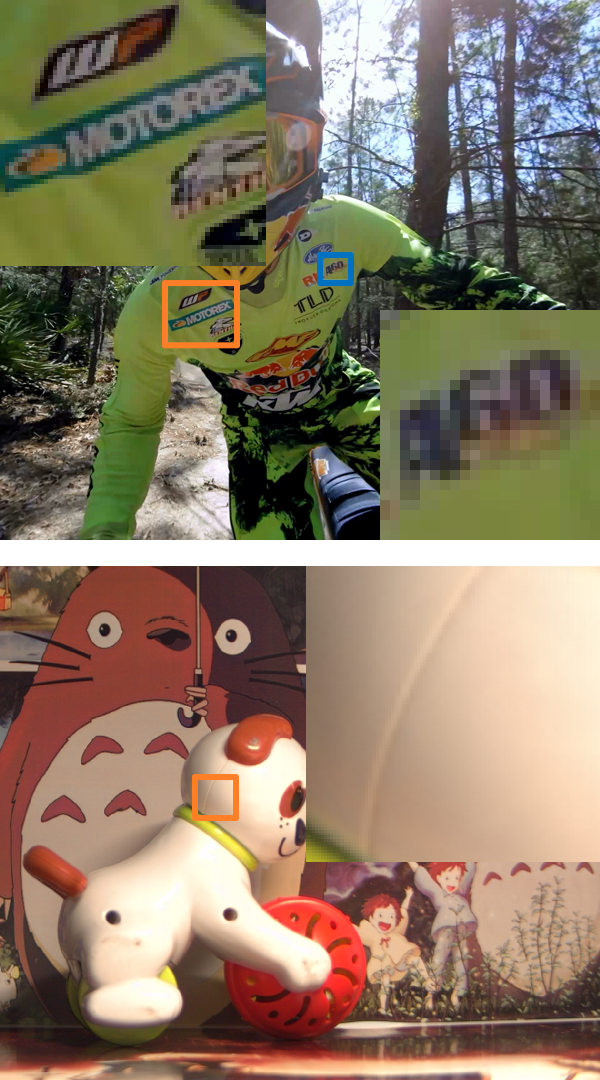}}
\subfigure[Noisy]{
\label{Fig4}
\includegraphics[width=0.828in]{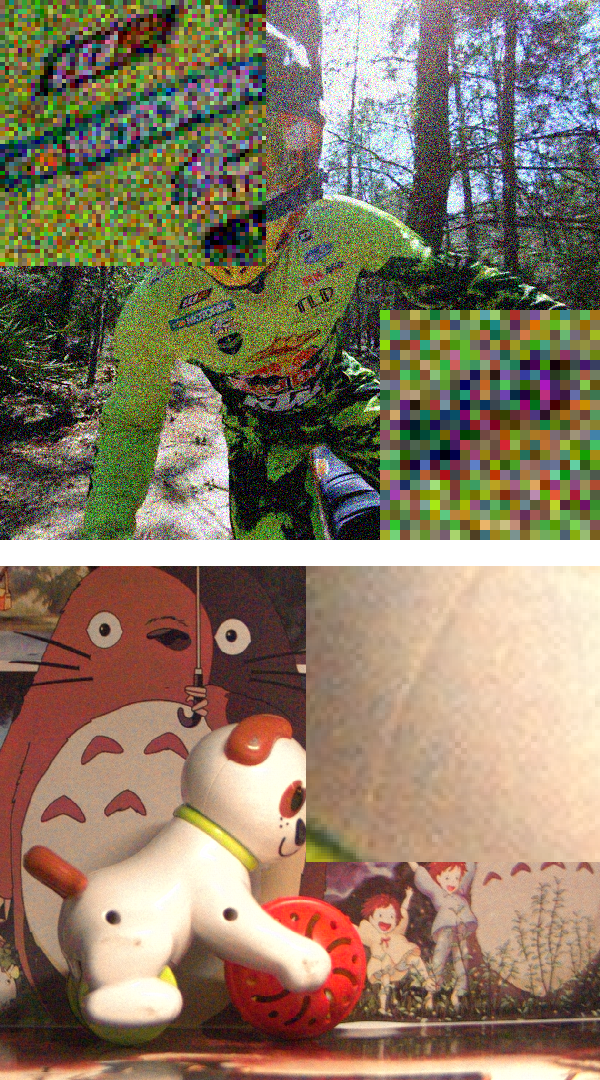}}
\subfigure[CVMSt-SVD]{
\label{Fig4}
\includegraphics[width=0.828in]{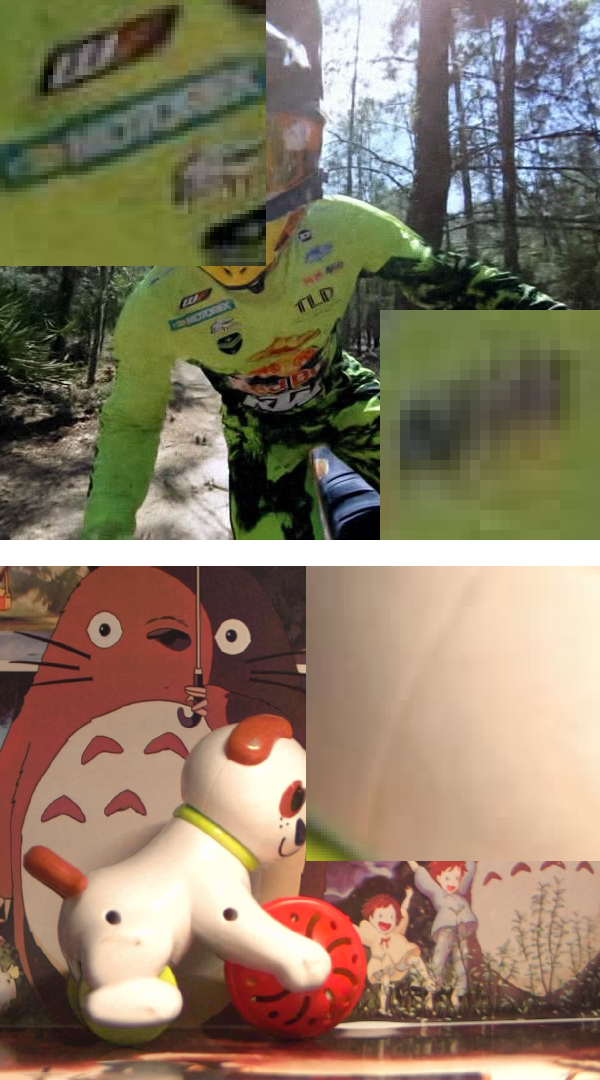}}
\subfigure[VBM4D2]{
\label{Fig4}
\includegraphics[width=0.828in]{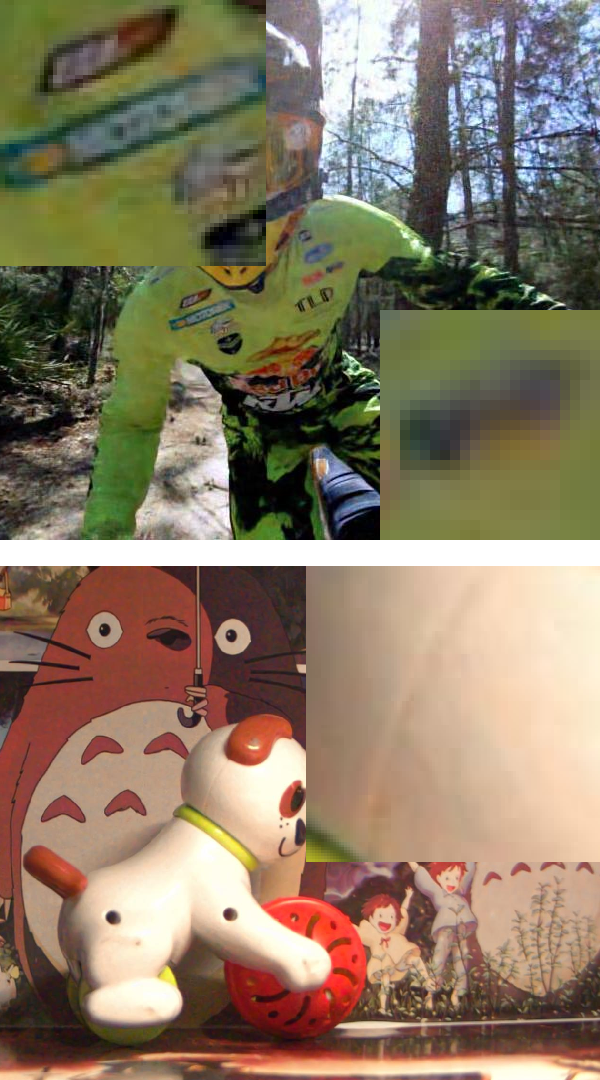}}
\subfigure[FastDVDNet]{
\label{Fig4}
\includegraphics[width=0.828in]{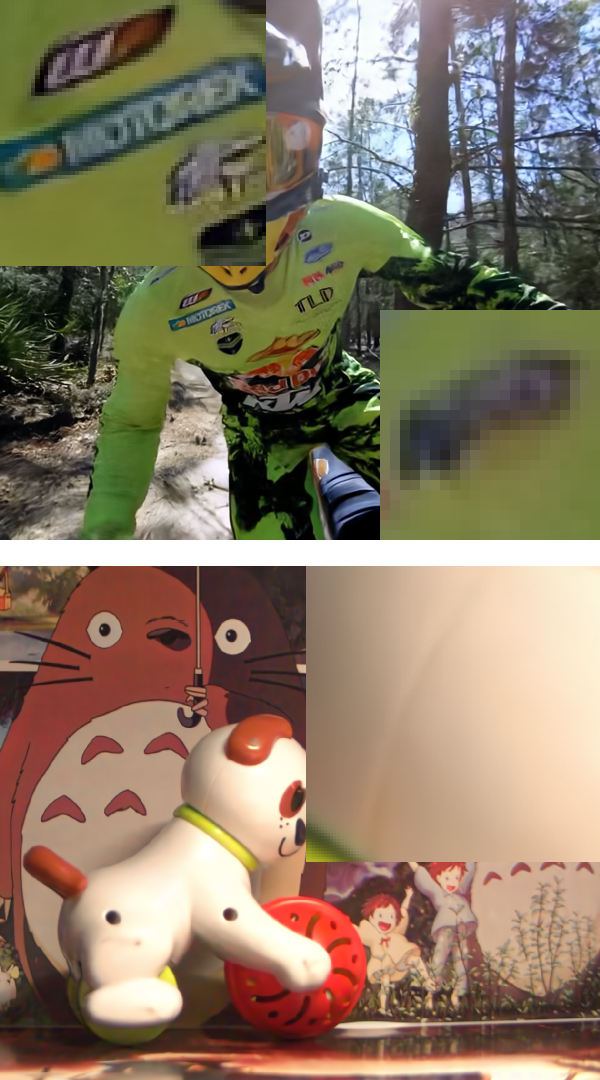}}
\subfigure[FloRNN]{
\label{Fig4}
\includegraphics[width=0.828in]{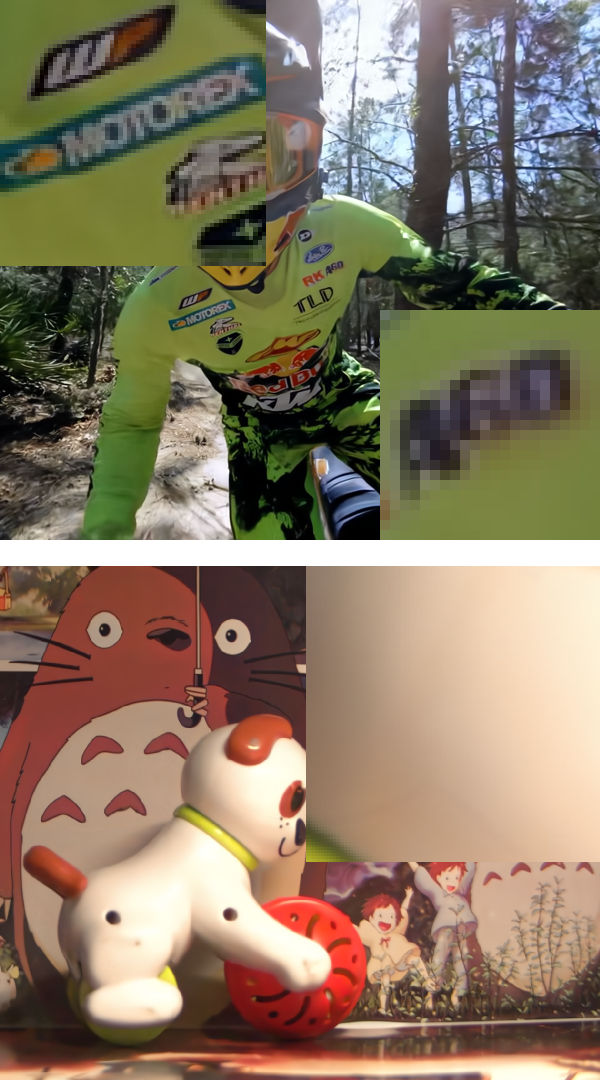}}
\subfigure[RVRT]{
\label{Fig4}
\includegraphics[width=0.828in]{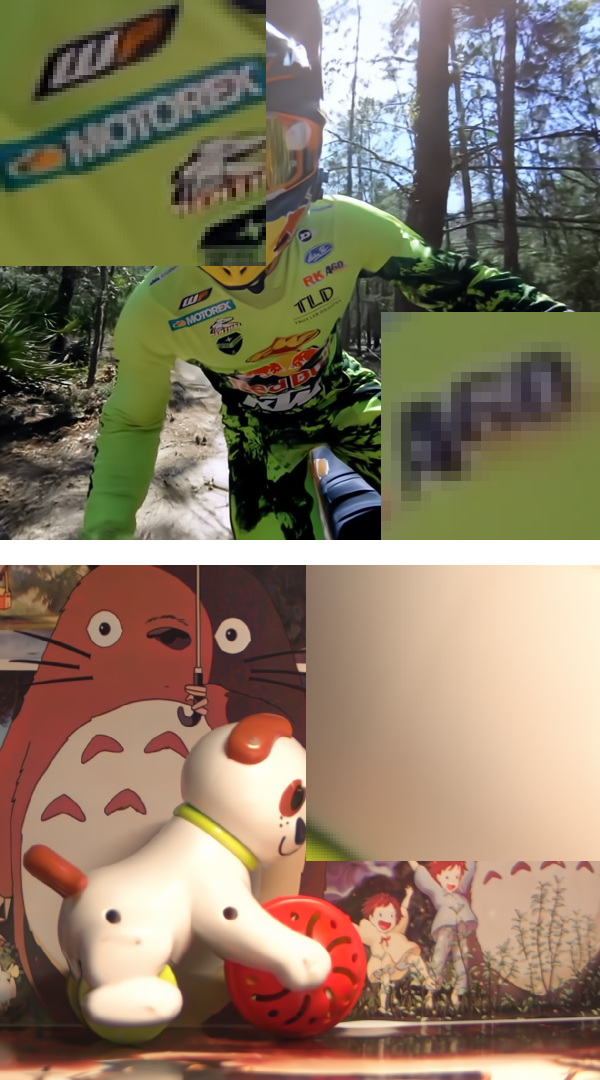}}
\subfigure[VNLNet]{
\label{Fig4}
\includegraphics[width=0.828in]{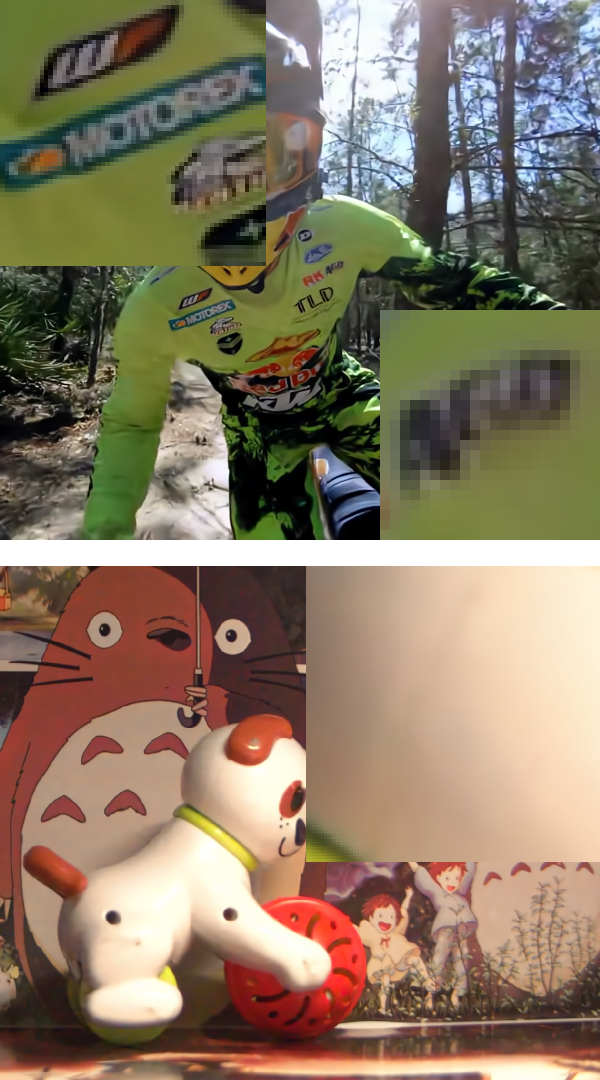}}

\caption{Visual evaluation of compared methods on synthetic and real-world video data. First row: Set8 ($\sigma = 40$), second row: CRVD (ISO = 6400).}
\label{Fig_CRVD_plus_Set8}
\end{figure*}

Visual comparison is presented in Fig. \ref{Fig_CRVD_plus_Set8}. From the results of synthetic experiments, we can observe that Gaussian-based DNN models such as FloRNN and VNLNet produce pleasant visual effects by suppressing noise and restoring true colors and details, while traditional denoisers CVMSt-SVD and VBM4D struggle to remove Gaussian noise and therefore generate unwanted color artifacts. Interestingly, from the results of CRVD data, it can be seen that the powerful DNN models may lead to more obvious over-smooth effects in real-world experiments. In this case, we notice that the background is static, the toy dog on the wheels is dynamic and moves fast in more than one directions, thus some details and textures are present only in certain frames. Therefore, traditional denoisers are able to benefit from their NLSS framework to leverage spatial similarity and preserve more structural information.

\vspace{-0.2cm}
\subsubsection{Human Ratings}
Due to the limitations of hardware equipments and environment, the videos acquired by the motorized slider also inevitably exhibit some noise, flickering, staircase effects, motion blur and misalignment that will undermine the accuracy of objective evaluations. In addition, the quality of videos may not be assessed frame-by-frame. Therefore, we conduct additional qualitative evaluations by collecting human opinions \cite{fang2019perceptual}. Specifically, we randomly select 10 videos from our IOCV dataset and invite 10 volunteers to rate the mean, noisy and denoised sequences of four compared methods. The invited volunteers have very little background knowledge of denoising, and they are not aware of how the presented video sequences are processed. For each of the 10 videos, the volunteers are asked to choose at least 2 best sequences, which then earn 1 point for the corresponding methods. The detailed human rating results are reported in Table \ref{Table_color_video_rating}. First, we observe that the mean video obtains the highest score on 9 of 10 videos, which suggests that our video averaging strategy may provide an alternative to the approach of generating reference videos frame-by-frame \cite{yue2020supervised}. Second, the human rating results show that DNN methods produce videos of higher quality than the benchmark traditional denoisers on our IOCV dataset, since both CVMSt-SVD and VBM4D leave behind medium-to-low-frequency noise, resulting in noticeable flickering. Last but not least, for the comparison of two traditional denoisers, VBM4D looks more visually pleasant than CVMSt-SVD because CVMSt-SVD presents more temporally decorrelated low-frequency noise in flat areas, which will appear as particularly bothersome for the viewers.

\begin{table}[htbp]
\vspace{-0.1cm}
\scriptsize
  \centering
  \caption{Human rating results of sequences generated by different methods based on our IOCV dataset. The top two results are bolded.}
  \scalebox{0.89}{
  \renewcommand{\arraystretch}{0.909}
    \begin{tabular}{ccccccc}
    \multicolumn{7}{c}{Ratings} \\
    \midrule
    \# Image & Mean  & Noisy & CVMSt-SVD & VBM4D1 & FastDVDNet & VNLNet \\
    \midrule
    1     & {\textbf{9}} & 0     & 1     & 2     & {\textbf{5}} & 3 \\
    \midrule
    2     & {\textbf{8}} & 1     & 0     & 4     & {\textbf{7}} & 6 \\
    \midrule
    3     & {\textbf{10}} & 0     & 0     & {\textbf{8}} & 3     & 5 \\
    \midrule
    4     & {\textbf{8}} & 0     & 0     & 4     & 7     & {\textbf{8}} \\
    \midrule
    5     & {\textbf{10}} & 0     & 0     & 4     & 4     & {\textbf{7}} \\
    \midrule
    6     & {\textbf{10}} & 0     & 0     & 5     & 6     & {\textbf{7}} \\
    \midrule
    7     & {\textbf{10}} & 0     & 0     & 5     & {\textbf{8}} & 5 \\
    \midrule
    8     & {\textbf{10}} & 2     & 2     & {\textbf{8}} & 5     & 6 \\
    \midrule
    9     & {\textbf{9}} & 0     & 2     & 4     & 2     & {\textbf{7}} \\
    \midrule
    10    & 4     & 4     & 4     & 4     & {\textbf{5}}     & {\textbf{8}} \\
    \midrule
    Average & {\textbf{8.80}} & 0.70  & 0.90  & 4.80  & 5.20  & {\textbf{6.20}} \\
    \bottomrule
    \end{tabular}%
    }
  \label{Table_color_video_rating}%
\end{table}%

\vspace{-0.218cm}
\subsection{Results for MSI/HSI Datasets}
\subsubsection{Experimental Settings} \label{MSI-HSI_settings}
MSI/HSI play an important role in a variety of remote sensing applications\cite{song2020unsupervised}. In this subsection, we evaluate the performance of various related denoising methods on synthetic and real noisy data. For synthetic experiments, due to the high computational cost, we mainly use the CAVE dataset for comparison. We assume that entries in all slices of noisy data are corrupted by zero-mean i.i.d Gaussian noise. In addition to the classical spatial-based quality indices PSNR and SSIM, we adopt two widely used spectral-based quality indicators for MSI/HSI data, namely spectral angle mapper (SAM) \cite{yuhas1990determination} and relative dimensionless global error in synthesis (ERGAS) \cite{wald2002data}. Different from PSNR and SSIM, recovered data with lower SAM and ERGAS are considered of better quality.
\vspace{-0.18cm}
\subsubsection{Objective Results}
\begin{table*}[htbp]
\tiny \tabcolsep2.08pt
  \centering
  \caption{Comparison of quantitative results on CAVE (with i.i.d Gaussian noise $\sigma = \{10, 30, 50, 100\}$) and Real-HSI datasets. The average denoising time (minutes) is calculated based on the CAVE dataset when $\sigma = 100$.}
  \scalebox{0.806}{
    \begin{tabular}{cccccccccccccccccccccccccccccc}
    \toprule
    \multirow{3}[4]{*}{Datasets} & \multirow{3}[4]{*}{$\sigma$} & \multirow{3}[4]{*}{Metrics} & \multicolumn{20}{c}{Traditonal denoisers}                                                                                                                     & \multicolumn{1}{c|}{} & \multicolumn{6}{c}{DNN methods} \\
\cmidrule{4-30}          &       &       & 4DHOSVD & ANLM  & BM4D1 & BM4D2 & FastHyDe & ISTReg & LLRGTV & LLRT  & LRMR  & LRTA  & LTDL  & LRTDGS & MSt-SVD & MTSNMF & NGMeet & OLRT  & PARAFAC & RTA-LSM & SDS   & SSTV  & \multicolumn{1}{c|}{TDL} & GRN   & HSI-DeNet & MAN   & QRNN3D & sDeCNN & SST \\
          &       &       & \cite{rajwade2012image} & \cite{manjon2010adaptive}  & \cite{maggioni2012nonlocal} & \cite{maggioni2012nonlocal} & \cite{zhuang2018fast} & \cite{xie2016multispectral} & \cite{He2018LLRGTV} & \cite{chang2017hyper} & \cite{zhang2013hyperspectral} & \cite{renard2008denoising} & \cite{gong2020low} & \cite{chen2019hyperspectral} & \cite{kong2017new} & \cite{ye2014multitask} & \cite{he2019non} & \cite{chang2020hyperspectral} & \cite{liu2012denoising} & \cite{dong2018robust} & \cite{lam2012denoising} & \cite{zeng2021hyperspectral} & \multicolumn{1}{c}{\cite{peng2014decomposable}} & \cite{cao2021deep} & \cite{chang2018hsi} & \cite{lai2023mixed} & \cite{wei20203} & \cite{maffei2019single} & \cite{li2022spatial} \\
    \midrule
    \multirow{19}[32]{*}{CAVE} & \multirow{4}[8]{*}{10} & PSNR  & 45.43  & 41.52  & 43.04  & 44.61  & 42.87  & 45.14  & 39.62  & 46.60  & 39.47  & 41.36  & 45.90  & 35.99  & 45.20  & 43.04  & 46.59  & {\textbf{47.07}} & 35.43  & 46.43  & 39.69  & 38.91  & 44.39  & 30.98  & 37.01  & 43.31  & 40.36  & 41.40  & 39.75  \\
\cmidrule{3-30}          &       & SSIM  & 0.9812  & 0.9576  & 0.9700  & 0.9784  & 0.9763  & 0.9792  & 0.9274  & 0.9868  & 0.9125  & 0.9499  & 0.9855  & 0.8976  & 0.9814  & 0.9706  & {\textbf{0.9880}} & 0.9877  & 0.8762  & 0.9857  & 0.9483  & 0.9153  & 0.9797  & 0.8622  & 0.8936  & 0.9804  & 0.9672  & 0.9624  & 0.9730  \\
\cmidrule{3-30}          &       & SAM   & 0.1084  & 0.2183  & 0.1844  & 0.1289  & 0.1398  & 0.1311  & 0.2676  & 0.0842  & 0.3282  & 0.1718  & 0.0923  & 0.1764  & 0.1064  & 0.1342  & 0.0853  & {\textbf{0.0840}} & 0.2407  & 0.0926  & 0.2159  & 0.2669  & 0.1055  & 0.4971  & 0.3236  & 0.1273  & 0.1926  & 0.1524  & 0.1632  \\
\cmidrule{3-30}          &       & EGRAS & 30.83  & 47.79  & 40.07  & 33.32  & 42.90  & 31.71  & 63.65  & 26.75  & 63.52  & 49.53  & 29.06  & 103.69  & 32.05  & 40.29  & 27.26  & {\textbf{25.53}} & 108.34  & 27.15  & 61.88  & 65.86  & 34.39  & 209.61  & 85.92  & 44.09  & 65.31  & 52.04  & 66.63  \\
\cmidrule{2-30}          & \multirow{4}[8]{*}{30} & PSNR  & 39.78  & 34.70  & 37.05  & 38.80  & 37.22  & 40.37  & 32.26  & 41.49  & 31.43  & 36.06  & 41.21  & 35.12  & 40.23  & 37.02  & 41.60  & {\textbf{41.79}} & 33.46  & 41.32  & 32.07  & 31.35  & 39.06  & 30.94  & 30.83  & 40.06  & 38.48  & 37.58  & 37.82  \\
\cmidrule{3-30}          &       & SSIM  & 0.9336  & 0.8068  & 0.9236  & 0.9283  & 0.9151  & 0.9462  & 0.7150  & 0.9681  & 0.6517  & 0.8775  & 0.9656  & 0.8795  & 0.9530  & 0.8758  & {\textbf{0.9693}} & 0.9680  & 0.8255  & 0.9598  & 0.6723  & 0.6078  & 0.9493  & 0.8605  & 0.6467  & 0.9620  & 0.9496  & 0.9247  & 0.9564  \\
\cmidrule{3-30}          &       & SAM   & 0.2272  & 0.4302  & 0.2898  & 0.2579  & 0.3123  & 0.2254  & 0.5538  & 0.1221  & 0.5899  & 0.2446  & 0.1299  & 0.2104  & 0.1737  & 0.2267  & 0.1238  & {\textbf{0.1180}} & 0.3362  & 0.1556  & 0.4959  & 0.5190  & 0.1496  & 0.4970  & 0.5105  & 0.1835  & 0.2320  & 0.1959  & 0.2144  \\
\cmidrule{3-30}          &       & EGRAS & 59.12  & 103.96  & 83.28  & 65.23  & 81.44  & 54.82  & 141.67  & 48.50  & 154.15  & 90.50  & 49.68  & 109.98  & 56.26  & 79.30  & 48.15  & {\textbf{46.65}} & 126.99  & 49.42  & 143.67  & 155.14  & 63.24  & 209.90  & 177.76  & 61.45  & 76.21  & 78.28  & 80.25  \\
\cmidrule{2-30}          & \multirow{4}[8]{*}{50} & PSNR  & 36.82  & 30.76  & 34.51  & 35.90  & 34.56  & 37.88  & 28.53  & 38.65  & 27.44  & 33.52  & 38.62  & 33.84  & 37.73  & 33.50  & 38.87  & {\textbf{39.13}} & 31.25  & 38.60  & 26.09  & 25.48  & 36.44  & 30.82  & -     & 38.11  & 37.02  & 35.49  & 36.43  \\
\cmidrule{3-30}          &       & SSIM  & 0.8722  & 0.6064  & 0.8823  & 0.8685  & 0.8471  & 0.9168  & 0.5368  & 0.9482  & 0.4410  & 0.8201  & 0.9464  & 0.8470  & 0.9285  & 0.7585  & {\textbf{0.9507}} & 0.9481  & 0.7307  & 0.9301  & 0.3529  & 0.3019  & 0.9171  & 0.8564  & -     & 0.9444  & 0.9314  & 0.8947  & 0.9408  \\
\cmidrule{3-30}          &       & SAM   & 0.3385  & 0.5801  & 0.3540  & 0.3557  & 0.4395  & 0.2737  & 0.7113  & 0.1551  & 0.7273  & 0.2897  & 0.1630  & 0.2674  & 0.2231  & 0.2956  & 0.1617  & {\textbf{0.1436}} & 0.4541  & 0.2199  & 0.7170  & 0.7473  & 0.2007  & 0.4966  & -     & 0.2309  & 0.2741  & 0.2141  & 0.2524  \\
\cmidrule{3-30}          &       & EGRAS & 83.36  & 164.17  & 110.14  & 91.19  & 108.74  & 73.32  & 211.99  & 67.56  & 241.87  & 121.15  & 66.73  & 122.70  & 75.03  & 118.36  & 65.72  & {\textbf{63.41}} & 158.53  & 67.27  & 290.93  & 306.21  & 85.24  & 210.86  & -     & 76.14  & 87.15  & 98.54  & 92.86  \\
\cmidrule{2-30}          & \multirow{4}[8]{*}{100} & PSNR  & 32.66  & 24.92  & 31.21  & 31.84  & 30.62  & 31.85  & 23.36  & {\textbf{35.39}} & 21.91  & 30.06  & 34.86  & 30.34  & 34.20  & 28.04  & 34.98  & 35.18  & 26.77  & 34.66  & 18.48  & 16.75  & 32.89  & 29.45  & -     & 35.35  & 30.33  & 32.51  & 33.96  \\
\cmidrule{3-30}          &       & SSIM  & 0.7307  & 0.2828  & 0.7854  & 0.7197  & 0.6717  & 0.8176  & 0.2859  & {\textbf{0.9154}} & 0.1962  & 0.7138  & 0.9002  & 0.6852  & 0.8800  & 0.4988  & 0.9073  & 0.8944  & 0.4481  & 0.8418  & 0.1067  & 0.0720  & 0.8283  & 0.7814  & -     & 0.9053  & 0.7128  & 0.8309  & 0.9016  \\
\cmidrule{3-30}          &       & SAM   & 0.5600  & 0.7968  & 0.4638  & 0.5160  & 0.6097  & 0.3605  & 0.9229  & {\textbf{0.1962}} & 0.9175  & 0.3649  & 0.2349  & 0.4707  & 0.3142  & 0.4180  & 0.2504  & 0.2100  & 0.6942  & 0.3714  & 0.9868  & 1.0569  & 0.3138  & 0.4991  & -     & 0.3161  & 0.5161  & 0.2748  & 0.3225  \\
\cmidrule{3-30}          &       & EGRAS & 134.34  & 324.49  & 158.47  & 144.91  & 168.92  & 148.29  & 376.10  & {\textbf{99.37}} & 456.12  & 180.03  & 102.61  & 174.87  & 115.24  & 221.84  & 101.86  & 99.93  & 260.85  & 105.99  & 718.40  & 834.58  & 128.22  & 225.29  & -     & 103.68  & 172.68  & 136.17  & 120.41  \\
    \midrule
    \multirow{4}[8]{*}{Real-HSI} & \multirow{4}[8]{*}{-} & PSNR  & 25.86  & -     & 25.84  & 25.88  & 23.99  & 25.91  & 24.63  & 25.90  & 24.81  & 24.27  & 25.80  & 25.12  & 25.86  & 25.51  & 25.87  & {\textbf{25.94}} & 25.08  & -     & 25.58  & 25.31  & 25.33  & 24.91  & 25.63  & 25.82  & 25.82  & 25.70  & - \\
\cmidrule{3-30}          &       & SSIM  & 0.8664  & -     & 0.8574  & 0.8653  & 0.5484  & 0.8688  & 0.7344  & 0.8610  & 0.7753  & 0.5876  & 0.8413  & 0.8181  & 0.8665  & 0.7819  & 0.8659  & {\textbf{0.8695}} & 0.7668  & -     & 0.8126  & 0.7923  & 0.7634  & 0.8006  & 0.8534  & 0.8694  & 0.8691  & 0.8597  & - \\
\cmidrule{3-30}          &       & SAM   & 0.0632  & -     & 0.0698  & 0.0660  & 0.2275  & 0.0551  & 0.1057  & 0.0604  & 0.1542  & 0.2198  & 0.0742  & 0.0669  & 0.0635  & 0.1020  & {\textbf{0.0509}} & 0.0538  & 0.1084  & -     & 0.0891  & 0.0816  & 0.1232  & 0.1635  & 0.0921  & 0.0596  & 0.0638  & 0.0928  & - \\
\cmidrule{3-30}          &       & EGRAS & 222.73  & -     & 224.51  & 222.67  & 252.24  & 222.18  & 243.68  & 224.05  & 246.39  & 248.57  & 223.32  & 234.02  & 222.64  & 226.58  & 222.69  & {\textbf{221.94}} & 233.43  & -     & 226.23  & 229.18  & 229.32  & 248.41  & 232.73  & 224.41  & 225.32  & 227.88  & - \\
    \midrule
    Time  & -     & minutes & 2.5   & 1.7   & 1.5   & 4.1   & {\textbf{0.1}} & 41.9  & 2.9   & 16.5  & 4.9   & {\textbf{0.1}} & 35.0  & 0.6   & 2.8   & 0.5   & 4.1   & 24.3  & 1.5   & 12.5  & {\textbf{0.1}} & 3.0   & 0.9   & {\textbf{0.1}} & 0.8   & 0.2   & {\textbf{0.1}} & 0.7   & {\textbf{0.1}} \\
    \bottomrule
    \end{tabular}}%
  \label{Table_Results_MSI-HSI_all}%
\end{table*}%

Since the ground-truth data of the HHD dataset is not available, we report the objective results of compared methods on CAVE and Real-HSI datasets in Table \ref{Table_Results_MSI-HSI_all}. \\
\indent For traditional denoisers, tensor-based methods such as NGMeet, LLRT, LTDL and OLRT demonstrate outstanding performance by exploiting both spatial and spectral correlation, which consistently outperform other compared approaches at all noise levels by more than 1 dB on the CAVE dataset. However, their iterative denoising strategy with a large number of local similar patches significantly increases computational burden. By comparison, BM4D and MSt-SVD are more efficient, in that BM4D does not need to learn local patch transforms, and MSt-SVD is a one-step approach that utilizes global patch representation. In addition, they also produce very competitive performance on the Real-HSI dataset. \\
\indent For DNN methods, we can observe that their results on the CAVE dataset are not satisfactory, since the networks are trained with predefined and also a limited range of noise levels, distributions and bandwidths on other datasets such as ICVL. Nevertheless, it is noticed that MAN and QRNN3D are able to gain state-of-the-art performance on the Real-HSI dataset, which indicates the success of CNN-based frameworks in capturing spatial and spectral correlations in real-world cases. Besides, they are potentially effective and efficient tools for handling large data in practice due to much faster denoising speed.
\vspace{-0.2cm}
\subsubsection{Visual Evaluation}
\begin{figure}[htbp]
\graphicspath{{Figs/combine_CAVE/}}
\centering
\subfigure[Clean]{
\label{Fig4}
\includegraphics[width=0.78in]{Clean_combined_marked}}
\subfigure[Noisy]{
\label{Fig4}
\includegraphics[width=0.78in]{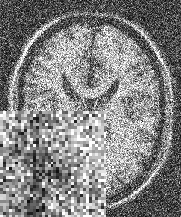}}
\subfigure[BM4D2]{
\label{Fig4}
\includegraphics[width=0.78in]{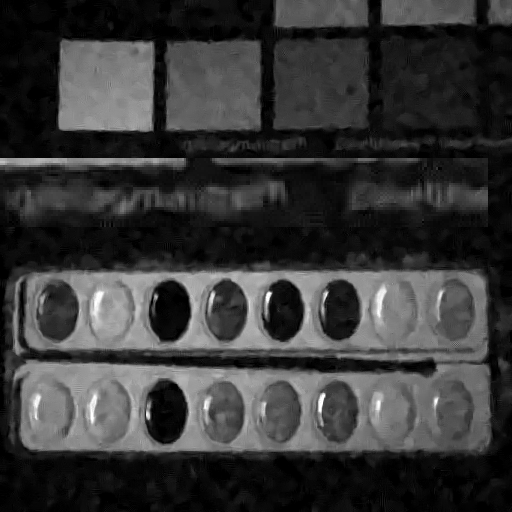}}
\subfigure[LLRT]{
\label{Fig4}
\includegraphics[width=0.78in]{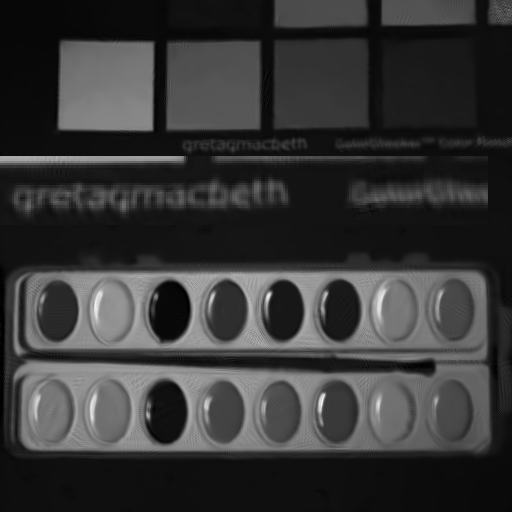}} \\
\subfigure[LTDL]{
\label{Fig4}
\includegraphics[width=0.78in]{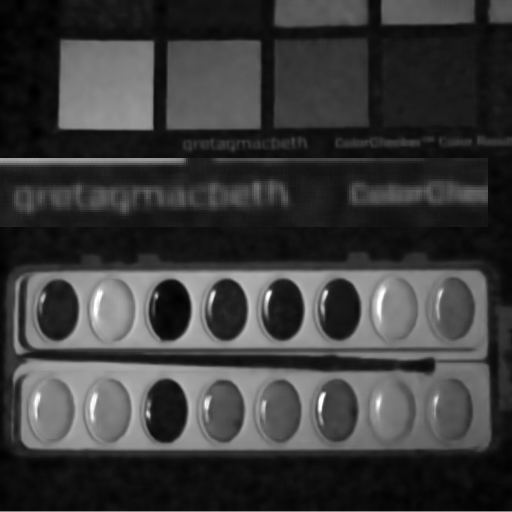}}
\subfigure[MSt-SVD]{
\label{Fig4}
\includegraphics[width=0.78in]{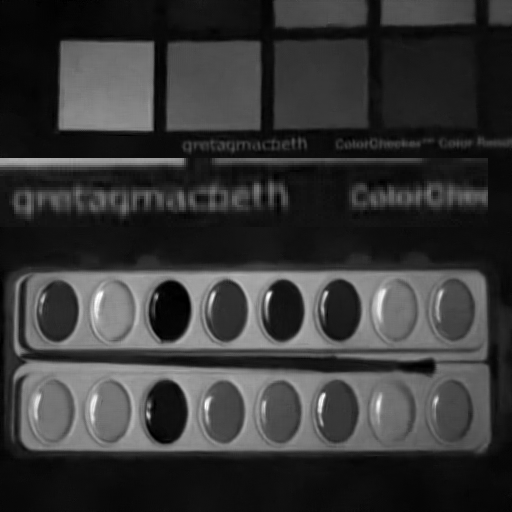}}
\subfigure[NGMeet]{
\label{Fig4}
\includegraphics[width=0.78in]{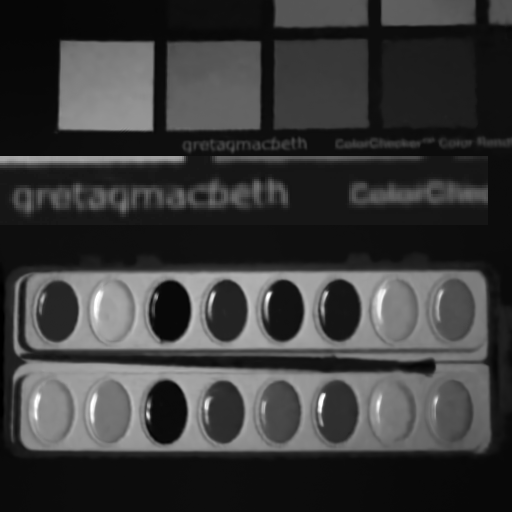}}
\subfigure[OLRT]{
\label{Fig4}
\includegraphics[width=0.78in]{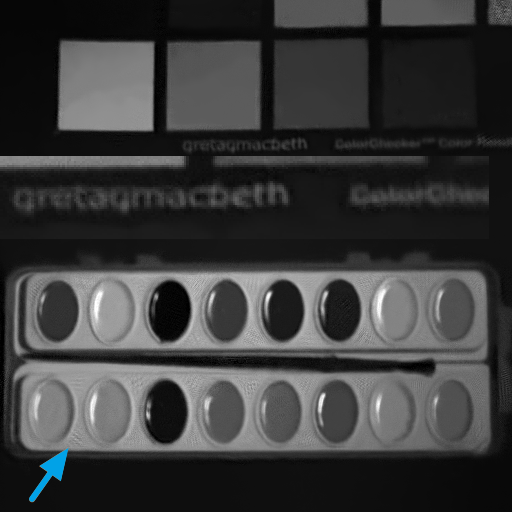}}

\caption{Visual evaluation of the CAVE dataset with noise level $\sigma = 100$.}
\label{Fig_CAVE}
\end{figure}

\begin{figure}[htbp]
\graphicspath{{Figs/combined_Real_HSI/}}
\centering
\subfigure[Clean]{
\label{Fig4}
\includegraphics[width=0.8in]{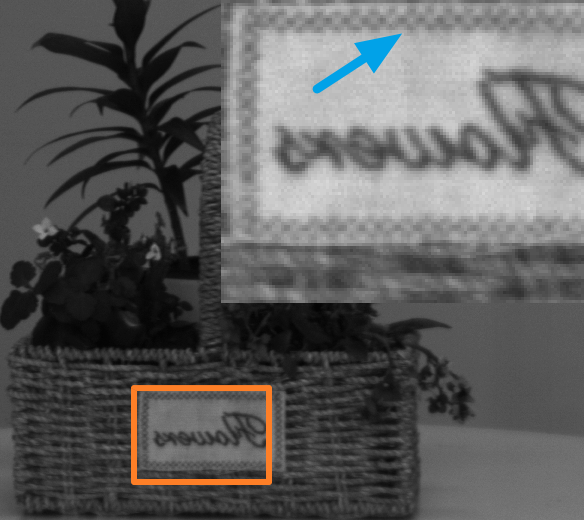}}
\subfigure[Noisy]{
\label{Fig4}
\includegraphics[width=0.8in]{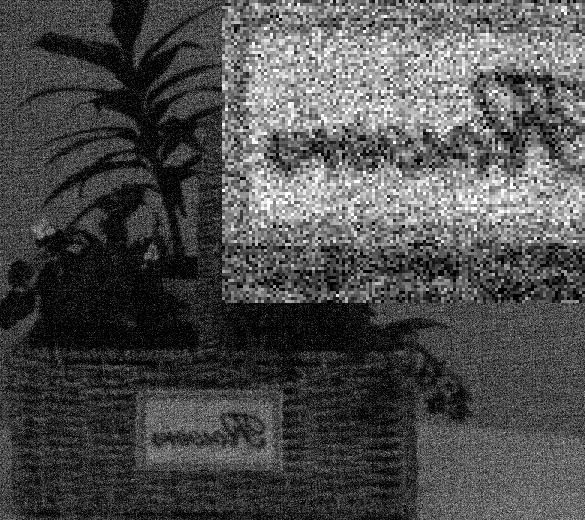}}
\subfigure[BM4D2]{
\label{Fig4}
\includegraphics[width=0.8in]{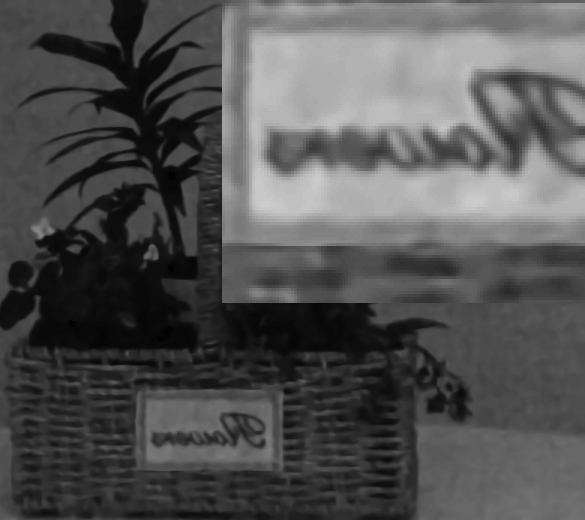}}
\subfigure[LTDL]{
\label{Fig4}
\includegraphics[width=0.8in]{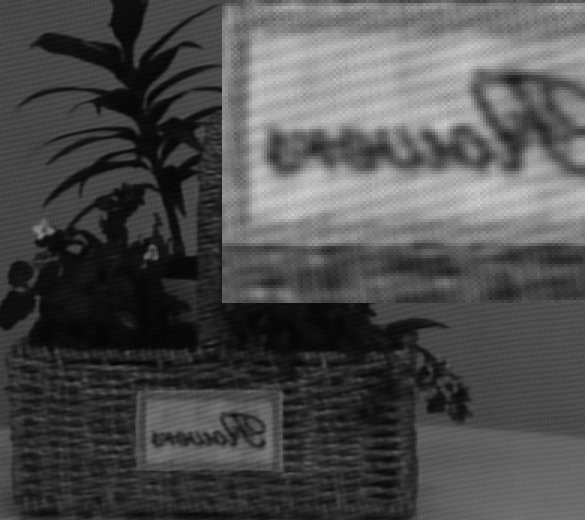}} \\
\subfigure[MSt-SVD]{
\label{Fig4}
\includegraphics[width=0.8in]{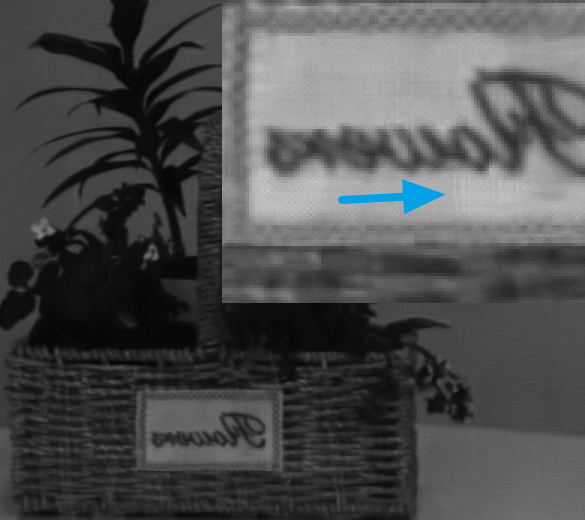}}
\subfigure[NGMeet]{
\label{Fig4}
\includegraphics[width=0.8in]{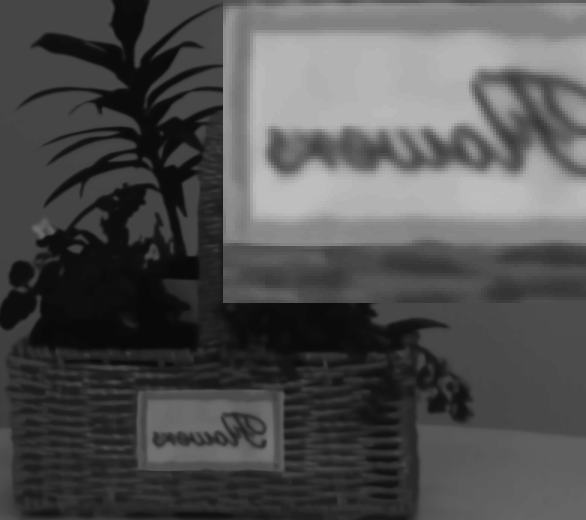}}
\subfigure[OLRT]{
\label{Fig4}
\includegraphics[width=0.8in]{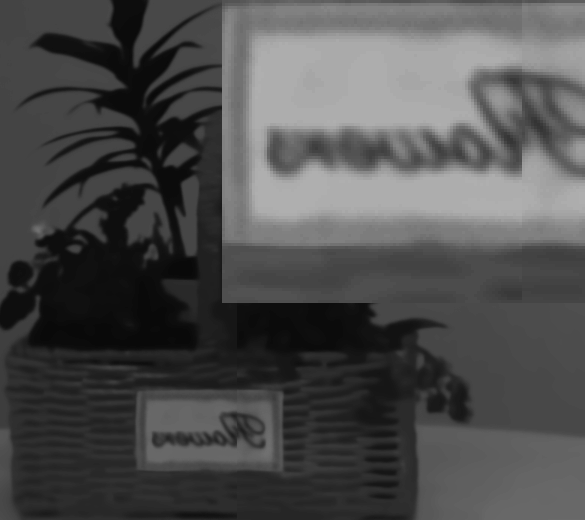}}
\subfigure[QRNN3D]{
\label{Fig4}
\includegraphics[width=0.8in]{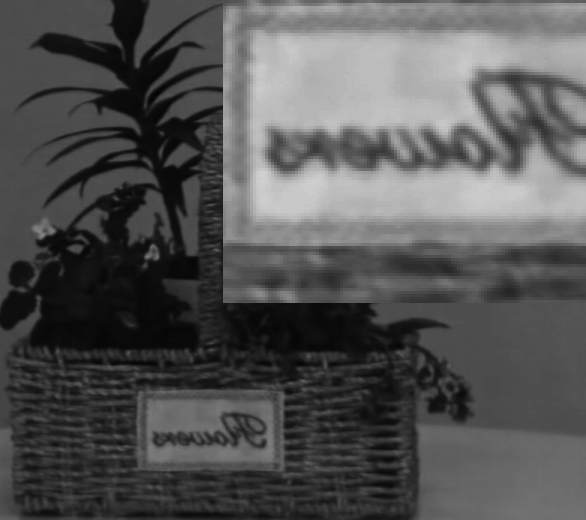}}

\caption{Visual evaluation of the real-world Real-HSI dataset.}
\label{Fig_Real_HSI}
\end{figure}

\begin{figure}[htbp]
\graphicspath{{Figs/combine_HHD/Combine_new/}}
\centering


\subfigure[Noisy]{
\label{Fig4}
\includegraphics[width=1.08in]{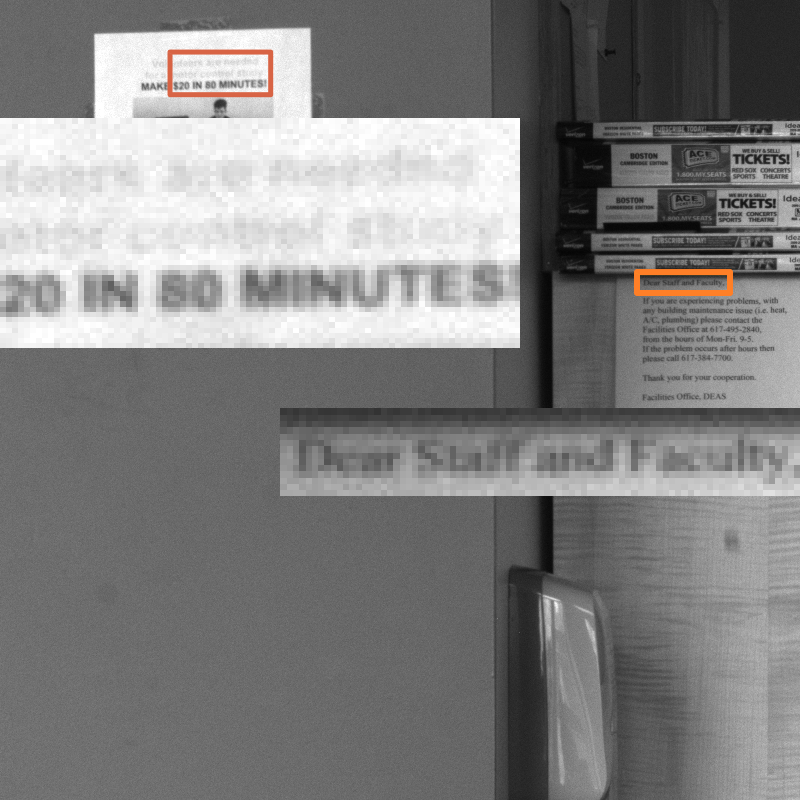}}
\subfigure[BM4D1]{
\label{Fig4}
\includegraphics[width=1.08in]{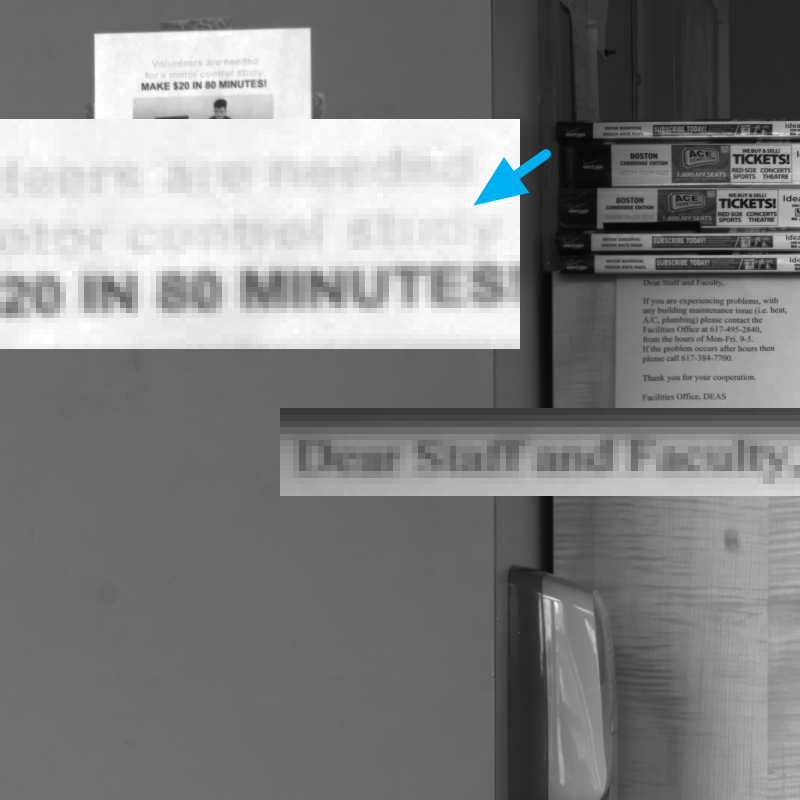}}
\subfigure[TDL]{
\label{Fig4}
\includegraphics[width=1.08in]{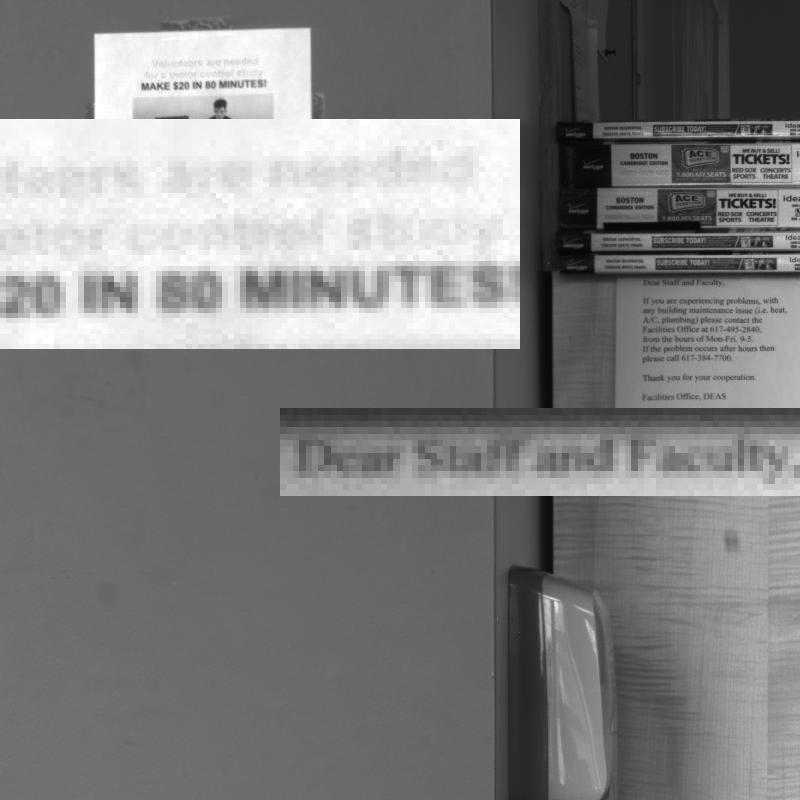}}\\
\subfigure[QRNN3D]{
\label{Fig4}
\includegraphics[width=1.08in]{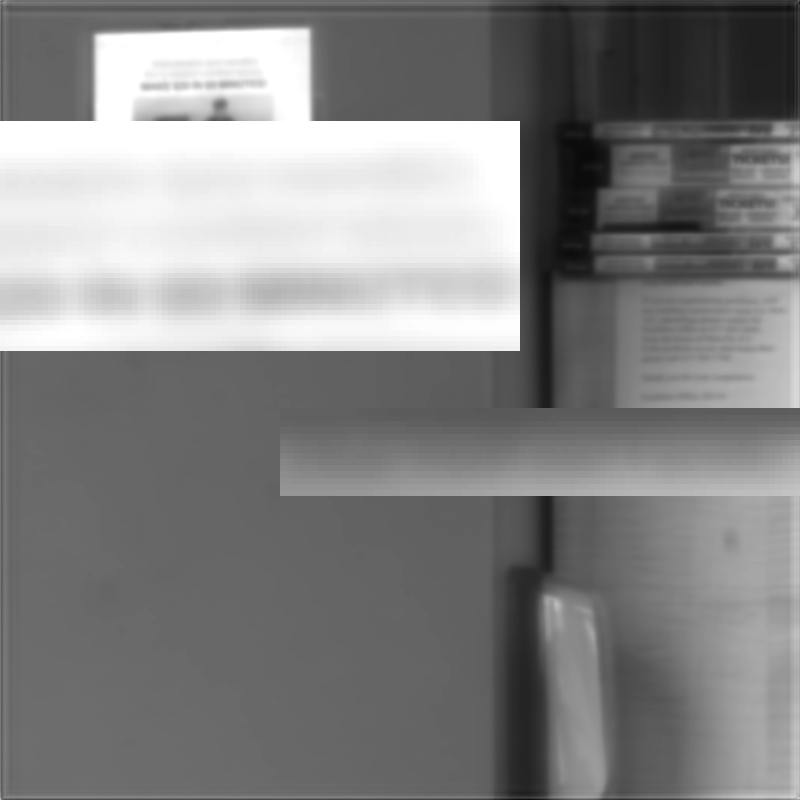}}
\subfigure[NGMeet]{
\label{Fig4}
\includegraphics[width=1.08in]{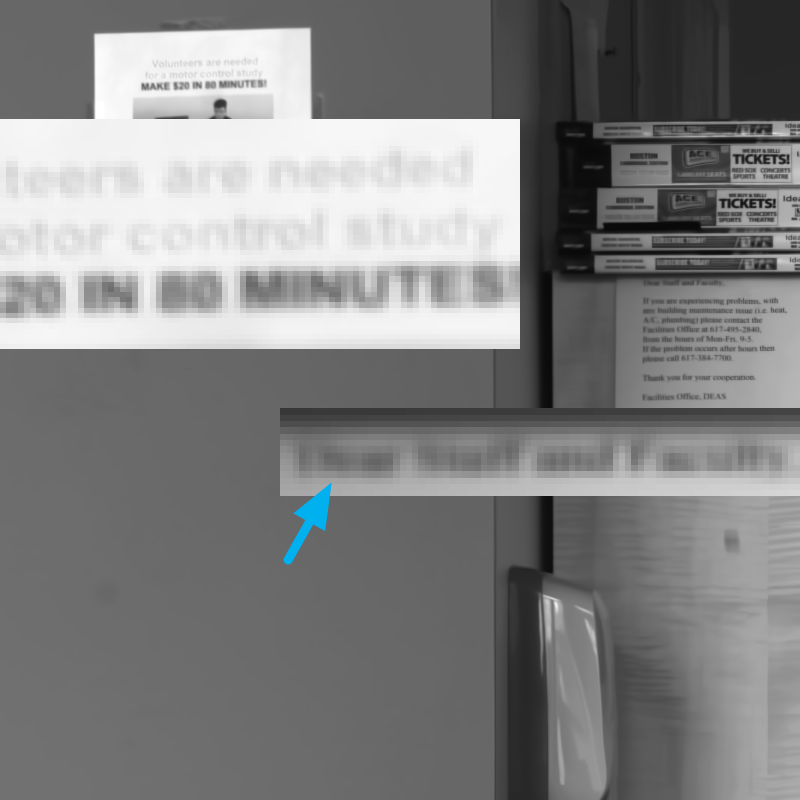}}
\subfigure[MSt-SVD]{
\label{Fig4}
\includegraphics[width=1.08in]{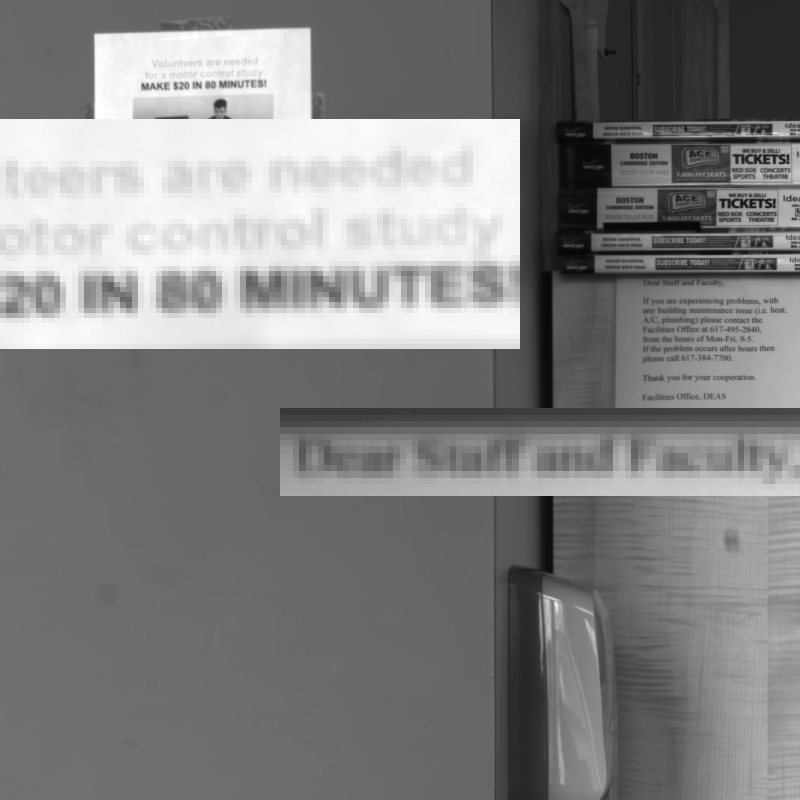}}

\caption{Visual evaluation of the real-world HHD dataset.}
\label{Fig_HHD}
\end{figure}

\noindent Fig. \ref{Fig_CAVE}, Fig. \ref{Fig_Real_HSI} and Fig. \ref{Fig_HHD} illustrate denoised results of competitive methods on the CAVE, Real-HSI and HHD datasets, respectively. For the HHD dataset, due to the large size of noisy observation, we mainly examine the effectiveness of five efficient methods. \\
\indent First, we notice that all compared methods can effectively remove noise to certain extents, but their performance are far from satisfactory. Specifically, BM4D and MSt-SVD tend to produce unexpected artifacts at high noise levels since the predefined transforms may not fully exploit the correlation among all the spectral bands. Besides, although NGMeet, LLRT and LTDL produce outstanding quantitative results, they fail to preserve high-frequency components such as edges and textures, as can be seen from Fig. \ref{Fig_CAVE} and Fig. \ref{Fig_Real_HSI}. This observation indicates that increasing the number of iterations and local similar patches may not help preserve fine details and structure of MSI/HSI data. In addition, the representative CNN-based method QRNN3D shows impressive noise removal ability at the cost of over-smooth and blurry effects on the edges and details. To conclude, all state-of-the-art methods struggle to adapt to different local image contents to balance smoothness and sharp details. For MSI/HSI data, each spectral band signal has unique spectroscopic characteristics \cite{song2020unsupervised}, applying a same set of parameters or transforms to multiple spectral bands may result in similar denoising patterns across the whole data. Therefore, apart from the high computational burden, another challenge of filtering MSI/HSI data lies in dealing with the spatial and spectral variation of noise levels and distributions.

\vspace{-0.568cm}
\subsection{Results for MRI Datasets}  \label{MR_image_experiments}
\subsubsection{Experimental Settings} \label{MRI_settings}
MRI is a non-invasive imaging technology that produces three dimensional detailed anatomical images, which has been shown to be helpful in disease detection, diagnosis, and treatment monitoring \cite{vlaardingerbroek2013magnetic}. Different from the AWGN noise modeling, MRI data are often corrupted by Rician noise \cite{awate2007feature}. Specifically, Let $\mathcal{X}$ be the original noise-free signal, the noisy Rician MRI data $\mathcal{Y}$ is defined by
\begin{equation}\label{rician_noise}
  \mathcal{Y} = \sqrt{(\mathcal{X} + \sigma \eta_r)^2 + ( \sigma \eta_i)^2}
\end{equation}
where $\eta_r$,$\eta_i$ $\sim$ $N(0,1)$ are i.i.d. random vectors following the standard normal distribution, $\sigma$ is the standard deviation in both real and imaginary channels of the noisy 3D MR images. To handle Rician noise, the technique of forward and inverse variance stabilizing transform (VST) \cite{foi2011noise} is often adopted by Gaussian denoisers via
\begin{equation}\label{equ_vst_gaussian}
    \hat{\mathcal{X}} = \text{VST}^{-1}(denoise(\text{VST}(\mathcal{Y}, \sigma), \sigma_{\text{VST}}), \sigma)
\end{equation}
where $\text{VST}^{-1}$ denotes the inverse of VST, $\sigma_{\text{VST}}$ is the stabilized standard deviation after VST, and $\sigma$ is the standard deviation of the noise in Eq. (\ref{rician_noise}). According to Eq. (\ref{equ_vst_gaussian}), the noisy data $\mathcal{Y}$ is first stabilized by the VST and then filtered by certain Gaussian denoisers using a constant noise level $\sigma_{\text{VST}}$, and the final estimate is obtained by applying the inverse VST to the output of the denoising result \cite{maggioni2012nonlocal}. \\
 \indent In our experiments, we use PSNR and SSIM as the objective metrics, and similar to \cite{maggioni2012nonlocal}, \cite{manjon2012new}, the PSNR value is computed on the foreground defined by $\mathcal{X}_f = \{x \in \mathcal{X}: x > 10 \cdot D/255\}$, where $D$ is the peak of clean data $\mathcal{X}$.
\subsubsection{Synthetic Experiments}
\begin{figure}[htbp]
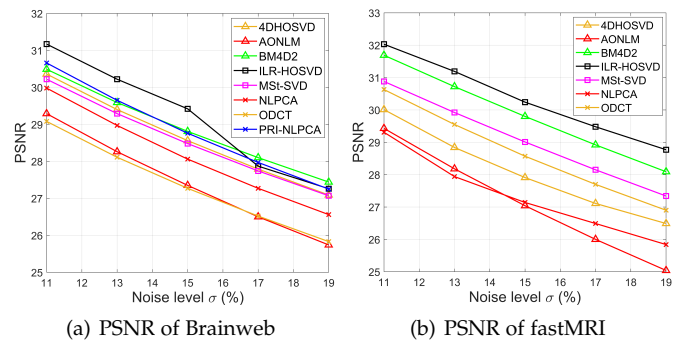

\graphicspath{{Figs/Fig_MRI_PSNR_SSIM/}}
\centering
\subfigure[PSNR of Brainweb]{
\label{Fig_MRI_results_a}
\includegraphics[width=1.69in]{Fig_Brainweb_PSNR}}
\subfigure[PSNR of fastMRI]{
\label{Fig_MRI_results_b}
\includegraphics[width=1.69in]{Fig_FastMRI-Knee_PSNR}}

\caption{Average PSNR values of compared methods on the Brainweb and fastMRI data at high noise levels ($\sigma \geq 11\%$).}
\label{Fig_illus_PSNR_SSIM_MRI}
\end{figure}

The volume data of Brainweb and fastMRI are added with varying levels of stationary Rician noise from 1$\%$ to 19$\%$ of the maximum intensity with an increase of 2$\%$. In real-world applications, the noise level is usually lower than 19$\%$, but we are also interested in the denoising capability of compared methods under extreme conditions. Table \ref{Table_Results_MRI_all} lists detailed quantitative results, and Fig. \ref{Fig_illus_PSNR_SSIM_MRI} compares the denoising performance at high noise levels when $\sigma \geq 11\%$. We notice that at lower noise levels, PRI-NLPCA is able to take advantage of the high-quality initial estimate of NLPCA, and thus shows outstanding performance when $\sigma \leq 9\%$. As noise level increases, the tensor-based method ILR-HOSVD demonstrates advantages of extracting latent features, which produces state-of-the-art results on the fastMRI dataset when $\sigma \geq 9\%$. However, from Fig. \ref{Fig_MRI_results_a} we observe that the iterative learning strategy of ILR-HOSVD is also subject to the presence of severe noise and may not generalize well to other datasets, while BM4D benefits from its predefined transforms and outperforms other methods on the Brainweb dataset when $\sigma \geq 17\%$.
\begin{table*}[htbp]
\scriptsize \tabcolsep4.1pt
  \centering
    \caption{Average PSNR-SSIM and computational time (s) of different methods on Brainweb and fastMRI data corrupted by Rician noise. The standard deviations $\sigma$ of the noise is expressed as percentage relative to the maximum intensity value of the noise-free data \cite{maggioni2012nonlocal}.}
     \scalebox{0.828}{
    \begin{tabular}{cccccccccccccc}
    \toprule
    \multirow{2}[1]{*}{Datasets} & \multirow{2}[1]{*}{Noise level} & 4DHOSVD & ANLM  & BM4D1 & BM4D2 & ILR-HOSVD & OLMMSE & ROLMMSE & MStSVD & NLPCA & ODCT  & PRINLM & PRI-NLPCA \\
          &       & \cite{rajwade2012image} & \cite{manjon2010adaptive} &  \cite{maggioni2012nonlocal}  &  \cite{maggioni2012nonlocal}  & \cite{lv2019denoising} & \cite{aja2008noise} & \cite{aja2008noise} & \cite{kong2019color} &  \cite{manjon2015mri}  & \cite{manjon2012new} & \cite{manjon2012new} &  \cite{manjon2015mri}  \\
    \toprule
    \multirow{10}[20]{*}{Brainweb} & 1     & 43.86/0.993 & 41.87/0.991 & 43.18/0.992 & 43.43/0.992 & {\textbf{44.55/0.994}} & 42.37/0.991 & 42.39/0.991 & 43.99/0.993 & 44.13/0.994 & 43.25/0.992 & 43.33/0.993 & 44.28/0.994 \\
\cmidrule{2-14}          & 3     & 37.49/0.976 & 36.52/0.971 & 36.59/0.971 & 37.21/0.975 & {\textbf{38.22/0.981}} & 35.99/0.968 & 35.98/0.969 & 37.46/0.975 & 37.74/0.977 & 36.43/0.970 & 36.87/0.975 & 38.12/0.981 \\
\cmidrule{2-14}          & 5     & 34.64/0.959 & 33.78/0.947 & 33.71/0.950 & 34.56/0.959 & {\textbf{35.46}}/0.966 & 33.37/0.949 & 33.36/0.949 & 34.65/0.958 & 33.84/0.940 & 33.57/0.949 & 33.97/0.955 & 35.29/{\textbf{0.967}} \\
\cmidrule{2-14}          & 7     & 32.85/0.941 & 31.91/0.921 & 31.80/0.926 & 32.84/0.943 & {\textbf{33.62/0.951}} & 31.58/0.925 & 31.61/0.928 & 32.83/0.940 & 31.87/0.912 & 31.66/0.927 & 32.01/0.934 & 33.36/0.951 \\
\cmidrule{2-14}          & 9     & 31.49/0.922 & 30.47/0.894 & 30.35/0.901 & 31.54/0.928 & {\textbf{32.31/0.938}} & 30.16/0.898 & 30.23/0.904 & 31.31/0.924 & 31.15/0.908 & 30.23/0.905 & 30.49/0.911 & 31.86/0.934 \\
\cmidrule{2-14}          & 11    & 30.36/0.903 & 29.29/0.865 & 29.15/0.873 & 30.49/0.912 & {\textbf{31.17/0.922}} & 28.93/0.868 & 29.09/0.880 & 30.22/0.906 & 29.98/0.883 & 29.08/0.882 & 29.28/0.888 & 30.66/0.916 \\
\cmidrule{2-14}          & 13    & 29.40/0.883 & 28.26/0.836 & 28.14/0.844 & 29.59/0.895 & {\textbf{30.22/0.909}} & 27.82/0.834 & 28.08/0.856 & 29.29/0.888 & 28.97/0.855 & 28.11/0.861 & 28.24/0.864 & 29.65/0.898 \\
\cmidrule{2-14}          & 15    & 28.56/0.862 & 27.35/0.807 & 27.26/0.814 & 28.81/0.879 & {\textbf{29.42/0.894}} & 26.82/0.799 & 27.26/0.835 & 28.48/0.870 & 28.06/0.826 & 27.27/0.841 & 27.36/0.841 & 28.76/0.879 \\
\cmidrule{2-14}          & 17    & 27.79/0.842 & 26.50/0.775 & 26.46/0.785 & {\textbf{28.10/0.862}} & 27.87/0.845 & 25.91/0.762 & 26.54/0.815 & 27.74/0.853 & 27.27/0.798 & 26.52/0.819 & 26.59/0.817 & 27.97/0.860 \\
\cmidrule{2-14}          & 19    & 27.09/0.822 & 25.74/0.744 & 25.73/0.755 & {\textbf{27.44/0.845}} & 27.26/0.828 & 25.05/0.725 & 25.86/0.797 & 27.07/0.835 & 26.56/0.772 & 25.83/0.799 & 25.92/0.797 & 27.25/0.841 \\
    \midrule
    \multirow{10}[20]{*}{fastMRI} & 1     & 41.50/0.975 & 39.36/0.956 & 42.07/0.977 & {\textbf{42.53/0.979}} & 42.30/0.977 & -     & -     & 42.09/0.977 & 41.69/0.976 & 40.96/0.968 & -     & - \\
\cmidrule{2-14}          & 3     & 36.58/0.931 & 36.12/0.921 & 36.68/0.933 & 37.07/{\textbf{0.937}} & {\textbf{37.09}}/0.933 & -     & -     & 36.74/0.935 & 37.01/0.932 & 36.68/0.930 & -     & - \\
\cmidrule{2-14}          & 5     & 34.13/0.901 & 34.01/0.884 & 34.57/0.901 & 35.12/0.911 & {\textbf{35.34/0.911}} & -     & -     & 34.61/0.906 & 34.15/0.902 & 34.62/0.899 & -     & - \\
\cmidrule{2-14}          & 7     & 32.58/0.871 & 32.29/0.842 & 33.01/0.867 & 33.82/0.889 & \textbf{34.01/0.888} & -     & -     & 33.10/0.878 & 32.08/0.866 & 33.11/0.869 & -     & - \\
\cmidrule{2-14}          & 9     & 31.27/0.845 & 30.79/0.793 & 31.61/0.829 & 32.71/0.866 & {\textbf{33.04/0.868}} & -     & -     & 31.92/0.854 & 30.73/0.838 & 31.81/0.836 & -     & - \\
\cmidrule{2-14}          & 11    & 30.01/0.814 & 29.43/0.741 & 30.29/0.785 & 31.69/0.841 & {\textbf{32.03/0.844}} & -     & -     & 30.88/0.827 & 29.31/0.804 & 30.63/0.801 & -     & - \\
\cmidrule{2-14}          & 13    & 28.84/0.785 & 28.18/0.683 & 29.05/0.737 & 30.72/0.814 & {\textbf{31.19/0.822}} & -     & -     & 29.92/0.799 & 27.94/0.768 & 29.55/0.764 & -     & - \\
\cmidrule{2-14}          & 15    & 27.91/0.754 & 27.04/0.626 & 27.87/0.686 & 29.80/0.785 & {\textbf{30.24/0.794}} & -     & -     & 29.01/0.770 & 27.12/0.711 & 28.57/0.727 & -     & - \\
\cmidrule{2-14}          & 17    & 27.11/0.708 & 26.01/0.571 & 26.77/0.634 & 28.92/0.755 & {\textbf{29.48/0.769}} & -     & -     & 28.15/0.741 & 26.49/0.668 & 27.70/0.693 & -     & - \\
\cmidrule{2-14}          & 19    & 26.49/0.671 & 25.04/0.519 & 25.76/0.583 & 28.09/0.725 & {\textbf{28.77/0.744}} & -     & -     & 27.34/0.711 & 25.84/0.606 & 26.90/0.660 & -     & - \\
    \midrule
    Time (s) &  -    & 614.1  & 426.7  & 57.1  & 171.2  & 1552.9  & $>1$h   & $>1$h   & 141.4  & 517.8  & {\textbf{4.5}} & 11.0  & 986.9  \\
    \bottomrule
    \end{tabular}}%
  \label{Table_Results_MRI_all}%
\end{table*}%

%
%

\begin{figure*}[htbp]
\graphicspath{{Figs/combined_Brainweb/}}
\centering
\subfigure[Clean]{
\label{Fig4}
\includegraphics[width=0.81in]{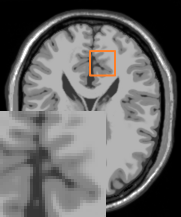}}
\subfigure[Noisy]{
\label{Fig4}
\includegraphics[width=0.81in]{Noisy}}
\subfigure[NLPCA]{
\label{Fig4}
\includegraphics[width=0.81in]{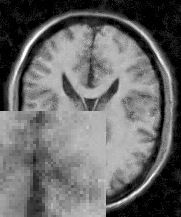}}
\subfigure[PRINLM]{
\label{Fig4}
\includegraphics[width=0.81in]{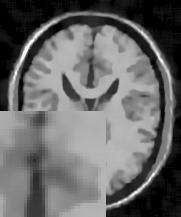}}
\subfigure[PRI-NLPCA]{
\label{Fig4}
\includegraphics[width=0.81in]{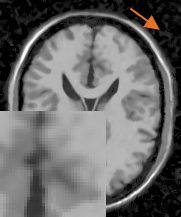}}
\subfigure[ILR-HOSVD]{
\label{Fig4}
\includegraphics[width=0.81in]{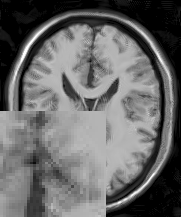}}
\subfigure[MSt-SVD]{
\label{Fig4}
\includegraphics[width=0.81in]{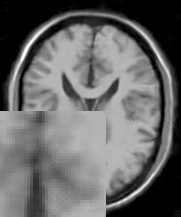}}
\subfigure[BM4D2]{
\label{Fig4}
\includegraphics[width=0.81in]{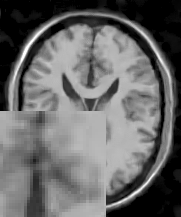}}

\caption{Visual evaluation of compared methods on the Brainweb T1w data with noise level $\sigma$ = 19\%.}
\label{Fig_Brainweb}
\end{figure*}

\indent Visual evaluation of compare methods on T1w data ($\sigma = 19\%$) is illustrated in Fig. \ref{Fig_Brainweb}. BM4D is successful in both noise suppression and detail preservation. Compared with ILR-HOSVD, another tensor-based filter MSt-SVD removes noise to a greater extent but blurs the edges and textures. The state-of-the-art NLM methods, namely PRINLM and PRI-NLPCA exhibit pleasant visual effects in homogeneous areas at the cost of slight over-smoothness along the edges. Furthermore, PRI-NLPCA does not remove noise in the background sufficiently since the local PCA transforms are severely degraded.

\subsubsection{Real-world MRI Dataset}
\begin{figure}[htbp]
\graphicspath{{Figs/combine_real_MRI/}}
\centering
\subfigure[MRI\_0112]{
\label{Fig4}
\includegraphics[width=0.78in]{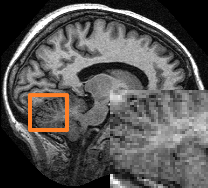}}
\subfigure[NLPCA]{
\label{Fig4}
\includegraphics[width=0.78in]{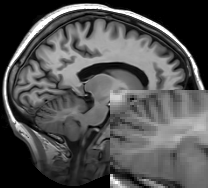}}
\subfigure[PRINLM]{
\label{Fig4}
\includegraphics[width=0.78in]{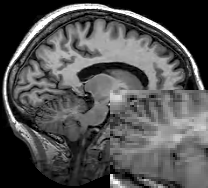}}
\subfigure[PRI-NLPCA]{
\label{Fig4}
\includegraphics[width=0.78in]{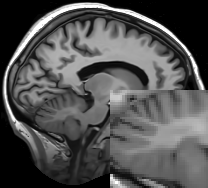}}\\

\subfigure[MSt-SVD]{
\label{Fig4}
\includegraphics[width=0.78in]{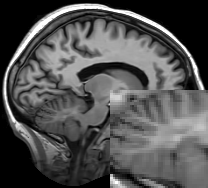}}
\subfigure[ILR-HOSVD]{
\label{Fig4}
\includegraphics[width=0.78in]{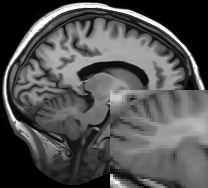}}
\subfigure[BM4D1]{
\label{Fig4}
\includegraphics[width=0.78in]{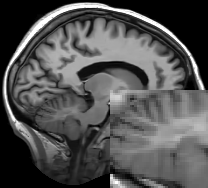}}
\subfigure[BM4D2]{
\label{Fig4}
\includegraphics[width=0.78in]{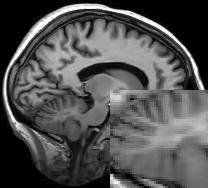}}

\caption{Visual evaluation of compared methods on the real OAS1\_0092 T1w data with estimated noise level $\sigma$ = 3\%.}
\label{Fig_real_MRI_0112}
\end{figure}

\begin{figure}[htbp]
\graphicspath{{Figs/combine_real_MRI/}}
\centering
\subfigure[MRI\_0092]{
\label{Fig4}
\includegraphics[width=0.78in]{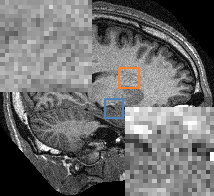}}
\subfigure[NLPCA]{
\label{Fig4}
\includegraphics[width=0.78in]{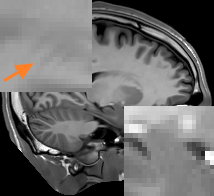}}
\subfigure[PRINLM]{
\label{Fig4}
\includegraphics[width=0.78in]{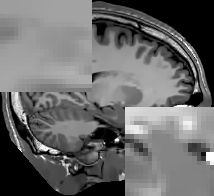}}
\subfigure[PRI-NLPCA]{
\label{Fig4}
\includegraphics[width=0.78in]{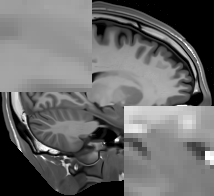}}\\

\subfigure[MSt-SVD]{
\label{Fig4}
\includegraphics[width=0.78in]{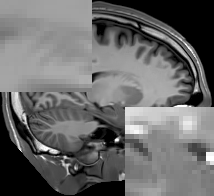}}
\subfigure[ILR-HOSVD]{
\label{Fig4}
\includegraphics[width=0.78in]{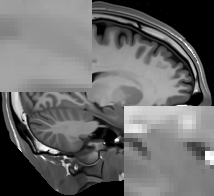}}
\subfigure[BM4D1]{
\label{Fig4}
\includegraphics[width=0.78in]{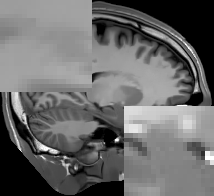}}
\subfigure[BM4D2]{
\label{Fig4}
\includegraphics[width=0.78in]{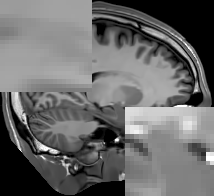}}

\caption{Visual evaluation of compared methods on the real OAS1\_0092 T1w data with estimated noise level $\sigma$ = 4.5\%.}
\label{Fig_real_MRI}
\end{figure}

To evaluate the performance of compared methods on real-world 3D MRI data, we carry out experiments on T1w MR images of the OASIS dataset \cite{marcus2007open}. The Rician noise levels of two selected T1w data, namely OAS1\_0112 and OAS1\_0092 are estimated to be 3$\%$ and 4.5$\%$ of the maximum intensity, respectively \cite{zhang2015denoising}. The filtered images of different methods are compared in Fig. \ref{Fig_real_MRI_0112} and Fig. \ref{Fig_real_MRI}, respectively. \\
\indent According to synthetic experimental results in Table \ref{Table_Results_MRI_all}, we notice that in most cases all methods can achieve promising denoising performance at relatively low noise levels, as can also be seen from the visual effects in Fig. \ref{Fig_real_MRI_0112}. Therefore, from the perspective of real-world denoising, BM4D1 and PRINLM are more competitive with low computational cost. Despite the similar denoising results of compared methods, we observe that for certain slices of OAS1\_0092, NLPCA restores more details and textures, while other methods suffer from over-smoothness to varying degrees, as illustrated in Fig. \ref{Fig_real_MRI}. This interesting observation is another vivid example to show that the use of tensor representation and transforms do not always help preserve more structural information of the underlying clean images.
\vspace{-0.16cm}
\subsection{Discussion}
\begin{figure}[htbp]

\graphicspath{{Figs/Fig_discussion/}}

\centering
\subfigure[PSNR]{
\label{Fig4}
\includegraphics[width=1.66in]{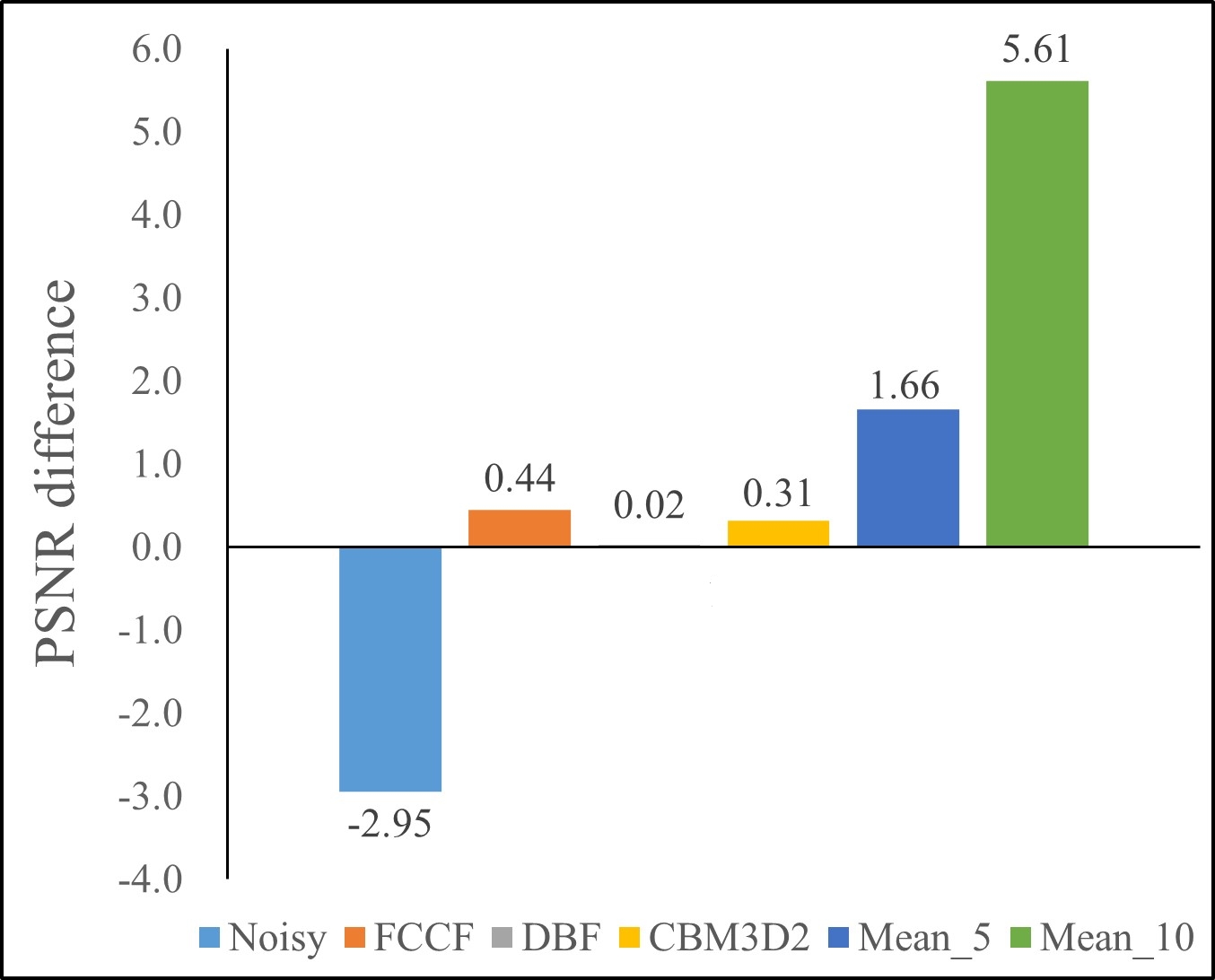}}
\subfigure[SSIM]{
\label{Fig4}
\includegraphics[width=1.66in]{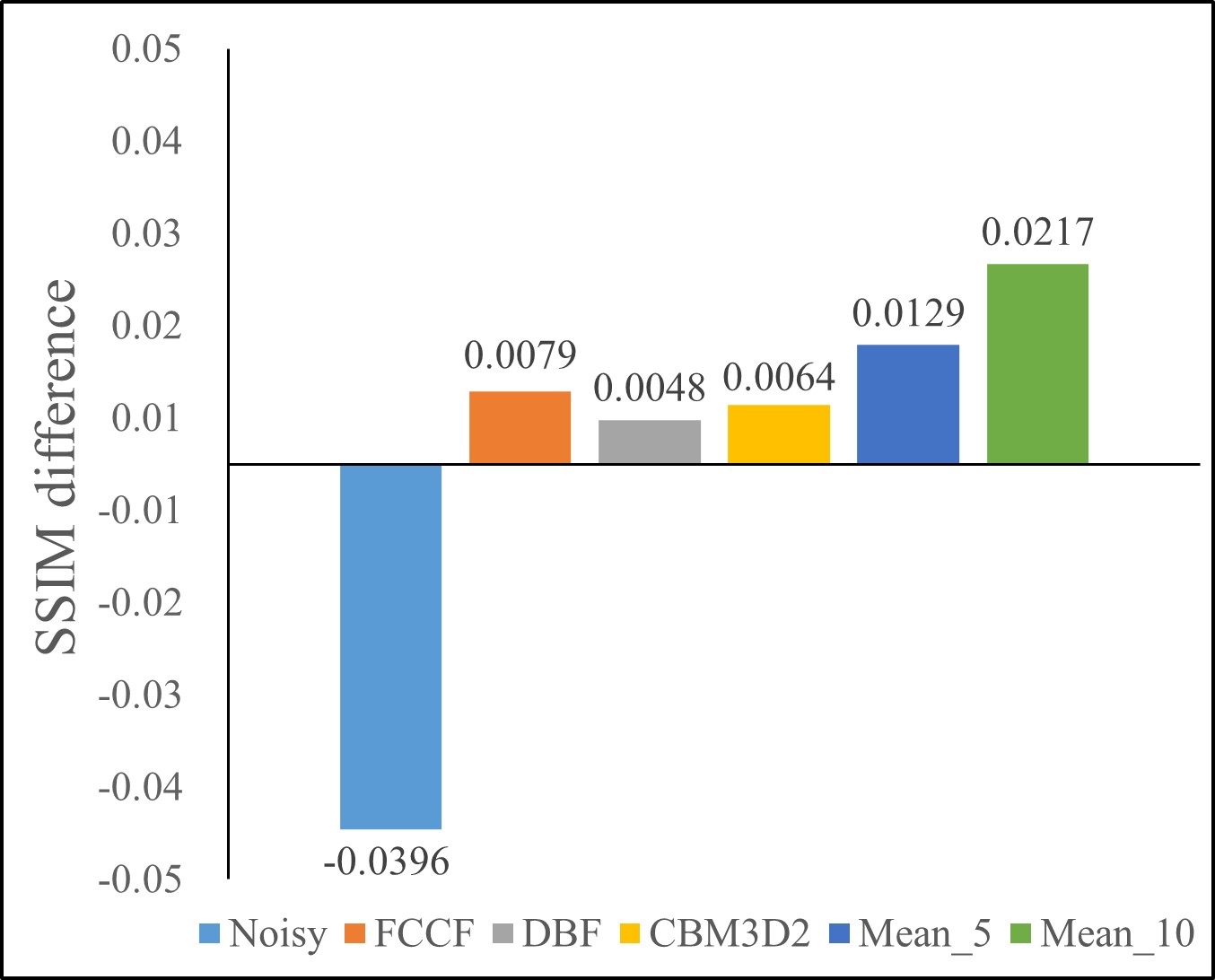}}

\caption{PSNR and SSIM difference of six implementations compared with 'Mean\_3' on the FUJI dataset.}

\label{Fig_PSNR_diff_Fujifilm_test}
\end{figure}

With numerous denoising techniques in hand, it is natural to ask how well we may expect a denoiser to perform. The theoretical bound of compared denoising methods is hard to obtain, but it is interesting to investigate the denoising capability under the challenging practical cases when prior information, such as training data and camera settings are unavailable. We use data captured by FUJIFILM X100T camera as an example, and for each scene, we generate three new mean images by averaging 3, 5 and 10 noisy images, and they are named Mean\_3, Mean\_5 and Mean\_10, respectively. They can be regarded as images captured via different continuous shooting modes with \textit{high}, \textit{medium} and \textit{low} noise levels. Fig. \ref{Fig_PSNR_diff_Fujifilm_test} illustrates the average PSNR differences of six implementations compared with Mean\_3 on the FUJI data. Interestingly, Fig. \ref{Fig_PSNR_diff_Fujifilm_test} shows that state-of-the-art denoising methods only produce marginal improvements compared with Mean\_3, which indicates that their denoising performance is similar to obtaining the mean image by averaging 3 consecutive noisy observations.

\vspace{-3.6pt}
\vspace{-9.9pt}
\section{Conclusion} \label{section_conclusion}
Over the past several decades, the rapid development of imaging systems and techniques has fostered the emergence of numerous denoising approaches, varying from traditional Gaussian denoisers to advanced DNN methods. Driven by the curiosity about the applicability and generalization ability of different methods, we conduct both synthetic and real-world experiments. A new dataset is also introduced to enrich benchmarking. Our experiments demonstrate: (i) the effectiveness of traditional denoisers for various denoising tasks, (ii) a simple matrix-based algorithm may be able to produce similar results compared with its tensor counterparts, and (iii) the substantial achievements of DNN models, which exhibit impressive generalization ability and show state-of-the-art performance on different datasets.

\indent Nowadays, image denoising also serves as a perfect test-bed for assessing new ideas and techniques \cite{elad2023image}. Many recent studies focus on the extention of image denoising methods to other computer vision tasks such as dehazing \cite{wu2019learning} and demoisacking \cite{liu2020joint}. It is therefore interesting to further explore the potential of related works to satisfy needs beyond denoising.

\vspace{-8.8pt}


\ifCLASSOPTIONcaptionsoff
  \newpage
\fi

\bibliographystyle{IEEEtran}
\bibliography{IEEEabrv,ms}

\end{document}